\documentclass[twocolumn,showpacs,superscriptaddress,floatfix,prl]{revtex4-2}
\usepackage{graphicx,amsfonts,amssymb,amsmath,hyperref,hypcap,enumerate}
\usepackage{braket}

\usepackage{amsthm}
\usepackage{multirow}
\usepackage{graphicx}
\usepackage{dcolumn}   
\usepackage{bm}        
\usepackage{amssymb}   
\usepackage{amsmath}
\usepackage{braket}
\usepackage{epstopdf}
\usepackage{color}
\usepackage{subfigure}
\usepackage{diagbox}

\usepackage{multirow}
\usepackage{makecell}
\usepackage{bbold}

\DeclareMathAlphabet{\mathitb}{OT1}{cmr}{bx}{sl}

\begin{document}
	\title{Unifying the Anderson Transitions in Hermitian and Non-Hermitian Systems}
	
	\author{Xunlong Luo}
	\email{luoxunlong@pku.edu.cn}
	\affiliation{Science and Technology on Surface Physics and Chemistry Laboratory, Mianyang 621907, China}
	
	\author{Zhenyu Xiao}
	\email{wjkxzy@pku.edu.cn}
	\affiliation{International Center for Quantum Materials, Peking University, Beijing 100871, China}
	\affiliation{Collaborative Innovation Center of Quantum Matter, Beijing 100871, China}
	
	\author{Kohei Kawabata}
	\email{kohei.kawabata@princeton.edu}
	\affiliation{Department of Physics, University of Tokyo, 7-3-1 Hongo, Bunkyo-ku, Tokyo 113-0033, Japan}
	\affiliation{Department of Physics, Princeton University, Princeton, New Jersey, 08540, USA}
	
	\author{Tomi Ohtsuki}
	\email{ohtsuki@sophia.ac.jp}
	\affiliation{Physics Division, Sophia University, Chiyoda-ku, Tokyo 102-8554, Japan}
	
	\author{Ryuichi Shindou}
	\email{rshindou@pku.edu.cn}
	\affiliation{International Center for Quantum Materials, Peking University, Beijing 100871, China}
	\affiliation{Collaborative Innovation Center of Quantum Matter, Beijing 100871, China}
	
	\date{\today}
	\begin{abstract}
		Non-Hermiticity enriches the 10-fold Altland-Zirnbauer symmetry class into the 38-fold symmetry class, 
		where critical behavior of the Anderson transitions (ATs) has been extensively studied recently. 
		Here, we propose a correspondence of the universality classes of the ATs between 
        Hermitian and non-Hermitian systems. We illustrate that the critical exponents of the length scale 
        in non-Hermitian systems coincide with the critical exponents in the corresponding Hermitian systems 
        with additional chiral symmetry. A remarkable consequence of the correspondence is superuniversality, 
        i.e., the ATs in some different symmetry classes of non-Hermitian systems are characterized 
        by the same critical exponent. In addition to the comparisons between the known critical exponents for  
		non-Hermitian systems and their Hermitian counterparts,
		we obtain the critical exponents in symmetry classes AI, AII, AII$^{\dagger}$, CII$^{\dagger}$, and DIII 
		in two and three dimensions. Estimated critical exponents are consistent 
        with the proposed correspondence. According to the correspondence, some of 
       the exponents also give useful information of the unknown critical exponents in 
        Hermitian systems, paving a way to study the ATs of Hermitian systems by the corresponding non-Hermitian systems.
	\end{abstract}
	
	\maketitle
	
	\textit{Introduction.}---
	Scattering, transmission, and interference of 
	waves in dissipative media lead to a rich variety 
	of physical phenomena. A prime example is localization, where a 
	propagating wave and its counter-propagating wave caused by scattering form a standing wave. 
	After Anderson's seminal work~\cite{Anderson58},  which predicted delocalization-localization 
	transitions of electron wavefunctions in disordered solids, a general scaling theory of localization 
    was introduced~\cite{Wegner76,Abrahams79}.  Subsequent development of field theory descriptions,  
	as well as renormalization-group analyses, clarified the universality classes of the Anderson transitions 
    (ATs) in three fundamental symmetry classes of time-reversal symmetry: Wigner-Dyson 
	classes~\cite{Efetov80,Hikami81}. Furthermore, chiral symmetry~\cite{Gade91, Gade93} and 
    particle-hole symmetry enrich the universality classes into the ten-fold symmetry classification~\cite{Altland97}.
	
	Like other continuous phase transitions, a universality class of the ATs is characterized by scaling 
    properties of an effective theory. Based on the single-parameter-scaling hypothesis, 
	the critical exponents of the ATs in the ten symmetry classes have been numerically  studied~\cite{Slevin14,Slevin16,Slevin18,Slevin09,Asada05,Asada04,Bitan17,Luo18QMCT,Wang21,Luo20,Medvedyeva10,Gruzberg99,Beamond02,Kagalovsky99,Kagalovsky99,Ortuno09,Fulga12,Roy17}.
	It is commonly believed that the universality classes are determined solely by spatial dimension and symmetry, 
	being independent from details of Hamiltonians. 
	In some cases, two distinct symmetry classes share the same scaling property, 
	which is called superuniversality~\cite{blackschtein84, Kivelson92,Lutken93,Fradkin96, oshikawa00, Pruisken07,Goswami17}. 
	Superuniversality can be numerically observed by precisely determining critical exponents 
    and other universal scaling properties. In this paper, we show that superuniversality 
	emerges also in non-Hermitian disordered systems.
	
	Recently, the ATs in non-Hermitian disordered systems attract considerable research interest~\cite{Hatano96, Feinberg97,Efetov97,Feinberg99,Hatano16,Xu16,Gong18,Tzortzakakis20,Wang20,Huang20,Huang20SR,Kawabata20,Luo21,Luo21TM}. 
	Non-Hermitian disordered systems describe random media with amplification or dissipation, which include 
	open classical systems~\cite{Konotop16,Feng17,El-Ganainy18,Ozdemir19,Miri19}, 
	as well as quantum systems of quasi-particles with finite lifetime~\cite{Kozii17,Shen18,Papaj19,Sun21}.
	In contrast to Hermitian systems, non-Hermitian systems are 
	classified into 38 symmetry classes~\cite{BL02,Kawabata19,Zhou19}.
	However, universality classes of the ATs in these 38 symmetry classes have yet to be understood clearly. 
	
\begin{table*}[bt]
\centering
\caption{Critical exponents $\nu$ and normalized localization lengths $\Lambda_c$ at the 
 Anderson transitions 
 for non-Hermitian symmetry classes (NHSCes) 
 in three dimensions (3D) and two dimensions (2D). 
Non-Hermitian Hamiltonians ${\cal H}$ are classified by time-reversal symmetry (TRS) 
${\cal U} {\cal H}^{*} {\cal U}^{\dagger}={\cal H}$, 
particle-hole symmetry (PHS) ${\cal U} {\cal H}^{T} {\cal U}^{\dagger}=-{\cal H}$, 
time-reversal symmetry$^{\dagger}$ (TRS$^{\dagger}$) 
${\cal U} {\cal H}^{T} {\cal U}^{\dagger}={\cal H}$, 
particle-hole symmetry$^{\dagger}$ (PHS$^{\dagger}$) 
${\cal U} {\cal H}^{*} {\cal U}^{\dagger}=-{\cal H}$, chiral symmetry (CS) 
${\cal U} {\cal H}^{\dagger} {\cal U}^{\dagger}=-{\cal H}$,
and 
sublattice symmetry (SLS) ${\cal U} {\cal H} {\cal U}^{\dagger}=-{\cal H}$. 
with unitary matrices ${\cal U}$.
In classes AI, 
AI$^{\dagger}$, AII, AII$^{\dagger}$, and AIII, ${\cal H}$ respects TRS with ${\cal U}{\cal U}^*=+1$, 
TRS$^{\dagger}$ with ${\cal U}{\cal U}^*=+1$, TRS with ${\cal U}{\cal U}^*=-1$, 
TRS$^{\dagger}$ with ${\cal U}{\cal U}^*=-1$, and CS, respectively. 
In classes 
CII$^{\dagger}$, ${\cal H}$  respects TRS$^{\dagger}$ with ${\cal U}{\cal U}^*=-1$, 
PHS$^{\dagger}$ with ${\cal U}{\cal U}^*=-1$, and CS. 
In class DIII, ${\cal H}$ respects TRS with ${\cal U}{\cal U}^*=-1$, PHS with 
${\cal U}{\cal U}^*=1$, and CS. The symmetry class depends not only 
on the symmetry of ${\cal H}$ but also on the eigenvalue $E$. For example, when the symmetry 
of ${\cal H}$ is that of the non-Hermitian symmetry class AI and $E\ne E^{*}$, the symmetry class is the non-Hermitian class A. 
 For comparison, critical exponents for the corresponding Hermitian symmetry classes
 (HSCes) are also listed. The square brackets denote 
 the 95\% confidence intervals 
 estimated by the Monte Carlo simulation. 
 For the literature values, the error bars 
 with * and ** denote the standard deviation $\sigma$, and 
 its double $2\sigma$, respectively. Note that we found a significant discrepancy in critical 
 exponents between 3D Hermitian class BDI and 3D non-Hermitian class AI.}
		\begin{tabular}{c|ccccc|cc}
\hline
\hline
& ~ ${\cal H}$~ & Energy & ~NHSC~ &$\nu$ &$\Lambda_c$ & ~HSC~ & $\nu$  \\
\hline
\multirow{7}{*}{~3D~} & A & $E$ & A & 1.00$\pm$0.04$^{**}$~\cite{Luo21TM} 
& ~0.598[0.593, 0.605]~\cite{Luo21TM}~ & AIII&  ~1.06$\pm$0.02$^{*}$~\cite{Wang21}~ 
\\ 
 &	AI$^{\dagger}$ & $E$ & AI$^{\dagger}$ & 1.19$\pm$0.01$^{**}$~\cite{Luo21TM}& 0.837[0.835, 0.839]~\cite{Luo21TM}& CI &
1.17$\pm$0.02$^{*}$~\cite{Wang21}, 1.16$\pm$0.02$^{**}$~\cite{Luo20} 
\\
 &	AI&$E=E^*$ & AI & 0.933[0.799, 1.041]\footnotemark[1] & 0.269[0.259, 0.293]\footnotemark[1]
 &BDI&  1.12$\pm$0.06$^{*}$~\cite{Wang21}, 0.80$\pm$0.02$^{**}$~\cite{Luo20} 
 \\ 
 &	AII& $E=E^*$ & AII &~0.8745[0.8710, 0.8783]\footnotemark[1]~ &0.936[0.935, 0.937]\footnotemark[1]&CII &unknown \\ 
	& AII$^{\dagger}$ & $E$& AII$^{\dagger}$ & 0.903[0.896, 0.908]\footnote{this paper} 
	& 0.581[0.576, 0.586]\footnotemark[1]&DIII &  0.85$\pm$0.05~\cite{Bitan17} \\ 
\hline
\multirow{6}{*}{2D}	&~AII~& $E\ne E^*$ & A &1.562[1.524, 1.609]\footnotemark[1]  &1.290[1.276, 1.303]\footnotemark[1]&AIII & unknown \\ 
 &~AII~& $E=E^*$ & AII & no AT found\footnotemark[1]& no AT found\footnotemark[1] &CII& unknown 
 \\ 
 &	~AII$^{\dagger}$~ & $E$ & AII$^{\dagger}$ & 1.377[1.331, 1.439]\footnotemark[1]  & 0.48[0.29, 0.61]\footnotemark[1]
 &DIII & 
 1.5$\pm$0.1~\cite{Yoshioka18}, \!\ $\approx$2.0~\cite{Fulga12} 
 \\ 
 & ~AIII~& $E=-E^{*}$ & AIII & 2.7$\pm$0.1$^{*}$~\cite{Xu16} & unknown &A& 2.59$\pm$0.01$^{**}$~\cite{Slevin09}\footnote{$\Lambda_c=1.284 [1.268, 1.305]$ for class A; $\Lambda_c=1.844\pm 0.004$ for class AII} \\
 &~CII$^{\dagger}$~& $E=0$& CII$^{\dagger}$ & 2.740[2.706, 2.773]\footnotemark[1]  &1.852[1.848, 1.855]\footnotemark[1]&AII& 2.75$\pm$0.04$^{**}$~\cite{Asada04}\footnotemark[2] \\ 
 &~DIII~& $E=0$& DIII & 2.757[2.726, 2.788]\footnotemark[1] &1.852[1.847, 1.855]\footnotemark[1]&AII  & 2.75$\pm$0.04$^{**}$~\cite{Asada04}\footnotemark[2] \\ 
\hline
\hline
		\end{tabular}
		\label{table_nu}
\end{table*} 
	
	In this paper, we propose 
    a correspondence between the ATs in Hermitian systems and those in 
	non-Hermitian systems and develop a unified understanding about the ATs. 
	We argue that the critical behavior of the length scale in 
    non-Hermitian systems is identical to the critical behavior in the corresponding Hermitian system 
    with additional chiral symmetry.
	To examine the proposed correspondence, we carry out extensive numerical studies of 
    critical exponents in non-Hermitian disordered systems. In particular, 
	we study in this paper the universal critical behavior of the ATs for 
	non-Hermitian models 
	in classes AI, AII, AII$^{\dagger}$, CII$^{\dagger}$, and DIII in two dimensions (2D) and three dimensions (3D). 
	We calculate the localization lengths of these models by the transfer matrix method,
	analyze them by the finite size scaling~\cite{Luo21TM},
	and determine 
    values of the critical exponents of the ATs, as summarized in Table~\ref{table_nu}. 
	Combining with the critical exponents for 
	classes A and AI$^{\dag}$ previously obtained in Refs.~\cite{Luo21TM,Luo21},
	we show that the critical exponents in these non-Hermitian symmetry classes 
	are consistent with the known critical exponents in the corresponding Hermitian 
	symmetry classes, supporting 
	the correspondence of the ATs between Hermitian and non-Hermitian systems. 
	Notably, estimated critical exponents in some non-Hermitian systems also provide useful information of   
    critical behaviour in Hermitian symmetry classes with chiral or particle-hole 
	symmetry, where 
	the critical exponents were previously difficult to estimate.
	
	\textit{Unified universality classes.}---
	Our correspondence of the ATs between Hermitian and 
	non-Hermitian systems is based on Hermitization~\cite{Feinberg97,Efetov97,Gong18,Kawabata19,Okuma20}.
	A non-Hermitian Hamiltonian $\cal H$ with complex energy $E\in \mathbb{C}$ is mapped to the Hermitian Hamiltonian 
	$\tilde{\cal H}$ by 
	\begin{equation}
		\tilde{\cal H}=
		\begin{pmatrix}
			0& {\cal H}-E\\
			{\cal H}^{\dagger}-E^{*}&0\\
		\end{pmatrix}.
		\label{eq:Hermitization}
	\end{equation}
	By construction,
	the Hermitian Hamiltonian $\tilde{\cal H}$ respects additional chiral symmetry 
	$\tau_z \tilde{\cal H} \tau_z = - \tilde{\cal H}$. 
	Let $\ket{\phi_r}$ and $\ket{\phi_l}$ be a right 
	eigenmode and a left eigenmode of the non-Hermitian Hamiltonian ${\cal H}$ with eigenenergy $E$, respectively: ${\cal H} \ket{\phi_r} = E \ket{\phi_r}$ and ${\cal H}^{\dag} \ket{\phi_l} = E^{*} \ket{\phi_l}$.
	Then, $(0~~|\phi_r\rangle)^{\rm T}$ and $(|\phi_l\rangle~~0)^{\rm T}$ comprise doubly degenerate zero modes of the Hermitian Hamiltonian $\tilde{\cal H}$ 
	(i.e., $\tilde{\cal H} \left( 0~~|\phi_r\rangle \right)^{\rm T} = \tilde{\cal H} \left( |\phi_l\rangle~~0 \right)^{\rm T} = 0$).
	This is the Hermitization, which associates the non-Hermitian Hamiltonian ${\cal H}$ with the Hermitian Hamiltonian $\tilde{\cal H}$ with 
	chiral symmetry. 
	Hermitization is relevant to non-Hermitian random matrices~\cite{Feinberg97} and  
	topological phases~\cite{Gong18, Kawabata19}, as well as topological characterization~\cite{Zhang20, Okuma20} of the anomalous boundary physics due to non-Hermiticity (i.e., non-Hermitian skin effect~\cite{Lee16, Yao18, Kunst18}).
	However, the significance of Hermitization has been unclear for the ATs.

	We argue that
	Hermitization unifies the ATs in  Hermitian and non-Hermitian systems.
	The ATs are continuous phase transitions that are characterized 
	by the universal scaling properties of the  
	localization lengths. As shown above, eigenmodes of $\cal{H}$ and the corresponding zero modes of $\tilde{\cal H}$  
	share the same spatial profiles, including the localization lengths. 
	Therefore, the universal scaling properties of the localization lengths in non-Hermitian systems, 
    as well as the absence or presence of the ATs, are generally the same as those in the Hermitian counterparts. 
	Notably, the right eigenmode $\ket{\phi_r}$ and the corresponding left eigenmode $\ket{\phi_l}$ exhibit similar localization properties with the same localization length, since they correspond to zero modes in the Hermitized Hamiltonian $\tilde{\cal H}$ with opposite chiralities. Notably, although the Hermitization procedure always maps non-Hermitian Hamiltonians to Hermitian Hamiltonians with chiral symmetry, nonchiral symmetry classes can appear in the Hermitized Hamiltonians. Even if the Hermitized Hamiltonians 
respect chiral symmetry, they can respect additional unitary symmetry and then be block diagonalized. In such a case, 
the relevant symmetry classes (or equivalently, classifying spaces) are not necessarily chiral classes~\cite{supplemental}.

For several non-Hermitian symmetry classes, 
we summarize the correspondence in Table~\ref{table_nu} (see Ref.~\cite{Kawabata19} and the Supplemental Material~\cite{supplemental} for the correspondence of all the 38 symmetry classes).
	For these classes in 2D and 3D in Table~\ref{table_nu}, we illustrate
   the correspondence by numerical evaluations of the critical exponents, as shown below.

	\textit{Model and symmetry class.}---
To study the AT in class AI, we introduce the following  
O(1) tight-binding model on 3D cubic lattice:
	\begin{align}\label{model_AI}
		{\cal H}=\sum_i \varepsilon_i c^{\dagger}_i c_i+\sum_{\langle i,j\rangle}
		V_{i,j}c^{\dagger}_i c_j,
	\end{align}
where $\varepsilon_i$ is the random potential characterized by 
the uniform distribution in $[-W/2, W/2]$ with the disorder strength $W$.
Here, $\langle i, j \rangle$ denotes 
nearest-neighbor lattice sites.
$V_{i,j}$ is set to either $-1$ or $+1$ randomly with the
equal probability, and $V_{i,j}$ and $V_{j,i}$ are treated as 
independent random numbers. 
Hermiticity is broken because of $V_{i, j}^{*} \neq V_{j, i}$, and 
reciprocity is absent in each 
disorder realization (${\cal H}^{\rm T} \ne {\cal H}$). 
Still, ${\cal H}$ is statistically reciprocal in a sense that 
${\cal H}$ and ${\cal H}^{\rm T}$ appear with the equal probability in the ensemble.
Eigenstates of ${\cal H}$ at real and complex energy $E$ belong to non-Hermitian 
symmetry classes AI and A respectively. For the real and complex $E$, the Hermitized
 Hamiltonian $\tilde{\cal H}$ belongs to symmetry classes BDI and AIII, respectively.

	
	To study the ATs in 
	classes AII, AII$^{\dagger}$, CII$^{\dagger}$, and DIII, 
	we introduce the following 
	non-Hermitian extension of the 
	SU(2) model~\cite{Asada02,Asada04,Asada05} on 2D square and 3D cubic lattices,
	\begin{align}
		{\cal H}=\sum_{i,\sigma} \varepsilon_{i,\sigma} c_{i,\sigma}^{\dagger}c_{i,\sigma}
		+\sum_{\langle i, j\rangle,\sigma,\sigma'}R(i,j)_{\sigma,\sigma'}c_{i,\sigma}^{\dagger}c_{j,\sigma'} \label{su2a}
	\end{align}
	with $\sigma=\uparrow,\downarrow$. 
	The spin-dependent nearest-neighbor hoppings are parametrized by 
	the SU(2) matrix 
	\begin{align}
		R(i,j)=
		\begin{pmatrix}
			e^{\text{i}\alpha_{i,j}}\cos(\beta_{i,j}) & e^{\text{i}\gamma_{i,j}}\sin(\beta_{i,j}) \\
			-e^{-\text{i}\gamma_{i,j}}\sin(\beta_{i,j}) & e^{-\text{i}\alpha_{i,j}}\cos(\beta_{i,j}) \\
		\end{pmatrix},\label{su2b}
	\end{align}
	where $\text{i}$ is the imaginary unit,
	$\alpha_{i,j}$ and 
	$\gamma_{i,j}$ are uniformly distributed in 
	$[0, 2\pi)$, and 
	$\beta_{i,j}$ is distributed in 
	$[0,\pi/2]$ 
	according to 
	the probability density 
	$P(\beta)d\beta=\sin(2\beta)d\beta$.
The hopping terms satisfy $R^{\dag}(i,j)=R(j,i)$ for classes 
AII, AII$^{\dagger}$, and CII$^{\dagger}$ 
($\alpha_{i,j}=-\alpha_{j,i}$, $\gamma_{i,j}=\gamma_{j,i}+\pi$), while they satisfy 
$\sigma_z R^{\dag}(i,j) \sigma_z = - R (j,i)$ for class DIII  ($\alpha_{i,j}=-\alpha_{j,i}+\pi$,  $\gamma_{i,j}=\gamma_{j,i}+\pi$). 
The on-site potentials $\varepsilon_{j,\sigma}=\omega_{j,\sigma}^r + {\rm i} \omega_{j,\sigma}^i$ are complex-valued, 
letting ${\cal H}$ be non-Hermitian.
	The complex-valued potentials are realized 
	in classical optical systems with random amplification and  dissipation~\cite{Cao99,Wiersma08,Wiersma13}. 
	$\omega_{j,\sigma}^r$ and $\omega_{j,\sigma}^i$ are independent for 
	each site $j$, and are uniformly distributed in 
	$[-W_r/2,W_r/2]$ and $[-W_i/2,W_i/2]$, respectively. 
	A relation between $\varepsilon_{j,\uparrow}$ and 
	$\varepsilon_{j,\downarrow}$, as well as $W_r$ and $W_i$, 
	is chosen appropriately so that ${\cal H}$ will belong to 
	the different 
	symmetry classes among 
	classes AII, AII$^{\dagger}$, CII$^{\dagger}$, and DIII~\cite{supplemental}.
	The SU(2) models are reciprocal in classes AII$^{\dagger}$, 
	CII$^{\dagger}$, and DIII; 
	the SU(2) model in class AII is reciprocal only statistically, similarly to the O(1) model.

	\begin{figure}[bt]
		\centering
		\includegraphics[width=1\linewidth]{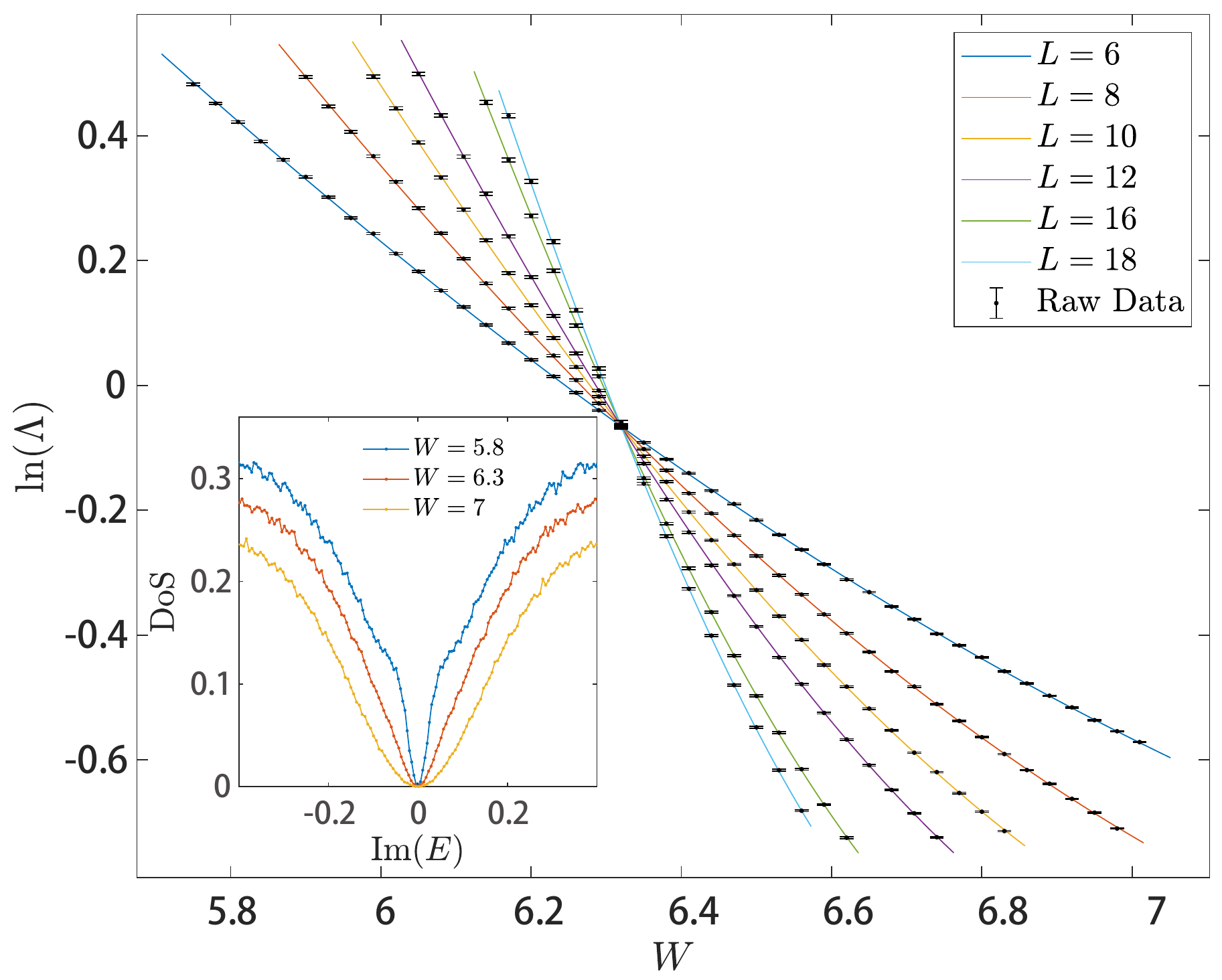}
		\caption{
			Normalized localization lengths $\Lambda$ as a function of the disorder strength $W \equiv W_r = W_i$ for 3D class AII at $E=0$.
			The points with the error bars are the 
			numerical data with the different system sizes $L$.
			The colored curves are the fitted curves.   
			Inset: density of states (DoS) for the imaginary part of eigenenergies. 
			Eigenenergies 
			in the $16 \times 16 \times 16$ cubic system 
			under the periodic boundary conditions are calculated, and the average over the $640$ samples 
			is taken.
		}
		\label{Lambda_3D_AII_E0}
	\end{figure}
	
	\textit{Transfer matrix study and polynomial fitting.}---
	Localization length and conductance of non-Hermitian systems were previously 
	calculated by the transfer matrix method~\cite{Luo21TM}. 
	Thereby, the critical exponents of the ATs in classes A and AI$^{\dagger}$  
	were determined precisely by the finite-size scaling analysis~\cite{MacKinnon94,Slevin99,Slevin14}. 
In this paper, the localization lengths for the five symmetry classes are calculated for different complex-valued energies in a quasi-one-dimensional 
geometry ($L\times L_z$ in 2D and $L\times L \times L_z$ in 3D with $L_z \gg L$). 
The 
quasi-one-dimensional localization length $\xi(L)$ along the $z$ direction is normalized by the system size $L$ along the transverse direction. Being dimensionless, the normalized length $\Lambda=\xi(W,L)/L$ shows 
scale-invariant behavior at the AT as a function of $L$. 
	
The single-parameter scaling~\cite{Wegner76,Abrahams79} 
has been demonstrated to be successful in analyses of the quantum 
criticality of the ATs in Hermitian 
systems~\cite{Slevin14,Slevin16,Slevin18,Slevin09,Asada05,Asada04,Bitan17,Luo18QMCT,Wang21,Luo20,Medvedyeva10,Gruzberg99,Beamond02,Kagalovsky99,Kagalovsky99,Ortuno09,Fulga12,Roy17} and in non-Hermitian systems~\cite{Huang20,Luo21,Luo21TM}. 
Apart from fine-tuned critical points such as multicritical points, critical properties of 
a generic continuous phase transition must be controlled by a saddle-point fixed point 
with only one relevant scaling variable. The scaling argument dictates that 
the dimensionless normalized localization length $\Lambda$ follows a  
scaling function that depends on the relevant scaling variable and possibly many other 
irrelevant scaling variables. The universal critical exponent $\nu$ 
associated with the relevant scaling variable can be estimated based on a polynomial  
expansion of the scaling function in terms of the scaling variables~\cite{Slevin14}.

	\textit{Numerical results.}---
The normalized 
quasi-one-dimensional localization lengths $\Lambda$ for 
classes AI, AII, AII$^{\dagger}$, CII$^{\dagger}$, and DIII in 2D or 3D are calculated at different complex energies~\cite{supplemental}. As an illustration,  Fig.~\ref{Lambda_3D_AII_E0} shows $\Lambda$ around the 
critical point at $E=0$ for 3D class AII with different system sizes $L$ and disorder strength $W$. 
As $L$ increases, $\Lambda$ increases below the critical point (delocalized phase) and decreases above the critical point (localized phase). 
In terms of numerical fitting based on the polynomial expansion~\cite{supplemental}, 
universal critical parameters of the ATs in the five non-Hermitian symmetry 
classes are obtained. The critical exponents $\nu$ as well as normalized localization lengths $\Lambda_c$
at the critical point are 
summarized in Table~\ref{table_nu}. 
Fitted critical parameters are confirmed to be stable against 
changing the system sizes and/or expansion orders~\cite{supplemental}. 
	
Universal critical exponents of the ATs in the non-Hermitian symmetry 
classes are mostly consistent with the known exponents 
in the corresponding Hermitian symmetry classes 
(Table~\ref{table_nu}). On the other hand, we also found discrepancies in the exponents 
between the 3D class AI model from Ref.~\cite{Wang21,Luo20} and the 2D class 
AII$^{\dagger}$ model from Ref.~\cite{Fulga12}. Causes of these deviations are 
currently under investigation, which will be discussed in a separate paper. 

		\textit{Superuniversality.}---
As a unique feature of non-Hermitian systems, our results show 
emergent superuniversality of the ATs:
two or more non-Hermitian disordered systems that belong to different symmetry classes in the 
38-fold symmetry classification can exhibit the same critical behavior of the length scale.
In fact, the critical exponent of the 2D SU(2) model in class CII$^{\dagger}$ is identical to that in class DIII (see Table~\ref{table_nu}).
In the Hermitian limit, for which parameters giving rise to non-Hermiticity vanish,
these two different symmetry classes fall into the two different universality classes with the different critical exponents. Hence, the superuniversality emerges as a consequence of non-Hermiticity.

Furthermore,  the correspondence of the ATs in Hermitian 
and non-Hermitian systems can also be regarded as superuniversality. 
Hermitian and non-Hermitian systems exhibit distinct transport phenomena, which 
implies that the underlying effective theories are different.  Nevertheless, our results in Table~\ref{table_nu} 
illustrate that such different effective theories share 
the same scaling  property of the length scale. 
	
	\textit{Implications on unknown critical exponents in Hermitian chiral classes}---
As a by-product of our correspondence, we can provide useful information for  
critical exponents for unexplored Hermitian symmetry classes that are difficult to estimate. 
To our best knowledge, the critical exponents for 3D class CII and 2D 
class AIII are unknown in the Hermitian case. The critical exponents obtained in 
3D class AII model with $E=E^*$ and 2D class AII model with $E\ne E^*$ 
respectively (see Table~\ref{table_nu}) can be the critical exponents of the ATs in these Hermitian chiral 
symmetry classes, given that the universality classes of the ATs are determined only by spatial dimension 
and symmetry. 
Importantly, calculations of non-Hermitian systems are much easier than the 
Hermitian counterparts because degrees of freedom of minimal non-Hermitian models 
are often half. We note that the critical localization lengths $\Lambda_c$ for Hermitian 
systems are also proposed by the non-Hermitian 
counterparts summarized in Table~\ref{table_nu}. 
	
The ATs of 2D Hermitian systems have remained elusive in chiral classes 
(AIII, BDI, and CII)~\cite{Asada2003chiral, Bocquet2003, Schweitzer2012} because of the 
vanishing $\beta$ functions~\cite{Gade91, Gade93, Konig2012}. We fail to find ATs for our 
2D O(1) models in class AI (not shown). Similarly, the 2D non-Hermitian models 
in Refs.~\cite{Tzortzakakis20,Huang20} exhibit no ATs.  On the other hand, for our 2D SU(2) model 
in class AII with $E \neq E^{*}$, which corresponds to class AIII in the Hermitian counterpart, we find 
the AT~\cite{supplemental} and estimate the critical exponent (Table~\ref{table_nu}).  It  could  
merit further study to investigate the 2D ATs on the basis of our correspondence.
	
	\textit{Density of states.}---
The density of states shows characteristic features and also contains relevant information about the ATs.
In our non-Hermitian systems, the density of states around the real axis exhibits a 
sharp peak in class AI~\cite{supplemental} and 
a soft gap in class AII (inset of Fig.~\ref{Lambda_3D_AII_E0}). 
	This behavior is consistent with the random-matrix behavior in classes AI  and AII~\cite{Ginibre65,Efetov97,Hamazaki20,supplemental}, and originates from the difference of time-reversal symmetry. 
	In class AI, time-reversal symmetry imposes a constraint on each real eigenenergy.
	Because of this constraint, real eigenenergies remain real unless they are mixed with other real eigenenergies.
	Consequently, some of them are stable even against non-Hermitian perturbations, forming the sharp peak of the density of states.
	In class AII, by contrast, time-reversal symmetry creates Kramers pairs with real eigenenergies. 
	In the presence of non-Hermitian perturbations,
	they are fragile and form complex-conjugate pairs~\cite{Kawabata19NC}. 
	Hence, eigenenergies tend to be away from the real axis, which leads to the soft gap of the density of states.
	We give other heuristic discussions in the Supplemental Material~\cite{supplemental}.

		\textit{Nonreciprocity.}---
In the numerical studies, we focused on statistically reciprocal models 
as illustrative examples. Nonreciprocity can give rise to unique non-Hermitian 
topology~\cite{Gong18, Kawabata19} and further change the universal critical properties.  Nevertheless, 
our correspondence of the ATs in Hermitian and non-Hermitian systems should remain valid even in the 
presence of nonreciprocity since it is based solely on Hermitization. We conjecture that  
even if nonreciprocity changes critical behavior of the ATs in non-Hermitian systems 
because of an additional mechanism such as topology, the critical behavior in the corresponding 
Hermitian systems should also change by the same mechanism and thus coincides with the non-Hermitian 
counterpart. As an example of this, the ATs in one-dimensional nonreciprocal systems are characterized by 
$\nu = 1$~\cite{Hatano96, Kawabata20}, which coincides with the critical behavior 
in the corresponding Hermitized systems~\cite{Altland14, Mondragon-Shem14}. The conjecture 
can be argued for the case of symmetry-conserving energy~\cite{supplemental}.  
It is worthwhile to further confirm our correspondence for higher-dimensional nonreciprocal systems.

	\textit{Summary and concluding remarks.}---
	In this paper, we propose a correspondence 
	of the ATs between Hermitian and non-Hermitian systems. 
	The 38-fold non-Hermitian symmetry class is mapped to the 10-fold 
	Hermitian symmetry class in terms of the universal scaling properties of the length scale. 
	Consequently, superuniversality emerges 
	in non-Hermitian systems: the ATs in several distinct symmetry classes  
	share the same universal scaling properties around their critical points. 
    To test this correspondence, we study the ATs in classes 
	AI, AII, AII$^{\dagger}$, CII$^{\dagger}$, and DIII 
	in 2D and 3D, and estimate the critical exponents 
	by the transfer matrix method. 
	The estimated critical exponents are consistent with the 
	correspondence and superuniversality in non-Hermitian disordered systems.
	From the correspondence, we also provide useful information of  
    the unknown critical exponents for 2D class AIII and 3D class CII in Hermitian systems.
	Investigating non-Hermitian systems is a new efficient way to study critical behavior of the ATs in Hermitian systems since 
	non-Hermitian matrices are often half the size of the corresponding Hermitian matrices.
	We note that conformal invariance~\cite{Cardy96,Janssen98,Obuse10}  
	should emerge at the ATs in 2D non-Hermitian systems from the correspondence.
	The multifractal properties at the ATs~\cite{Evers08} in 2D and 3D non-Hermitian
	systems should also be unified with the Hermitian counterparts.
	
	\textit{Acknowledgment.}---
	X.L., Z.X., and R.S. thank Zhiming Pan and Tong Wang for fruitful discussions and correspondences.
	K.K. thanks Shinsei Ryu for helpful discussions.
	T.O. thanks Matthias Stosiek and Masatoshi Imada for useful discussions. 
	X.L. was supported by National Natural Science Foundation of China of 
	Grant No.~51701190. K.K. was supported by JSPS KAKENHI Grant No. JP19J21927, JSPS Overseas Research Fellowship, 
        and Grant No. GBMF8685 from the Gordon and Betty Moore Foundation toward the Princeton theory program.
	T.O. was supported by JSPS KAKENHI Grants 19H00658.  
	Z.X. and R.S. were supported by the National Basic Research Programs of China (No.~2019YFA0308401) and by National 
	Natural Science Foundation of China (No.~11674011 and No.~12074008). 
	
	\bibliography{paper}
	
\clearpage
\begin{widetext}
\setcounter{equation}{0}
\setcounter{figure}{0}
\renewcommand{\theequation}{S.\arabic{equation}}
\renewcommand{\thefigure}{S.\arabic{figure}}
\section*{Supplementary Material for ``Unifying the Anderson Transitions in Hermitian and Non-Hermitian Systems"}

\section{Symmetry}
Non-Hermiticity ramifies and unifies symmetries in Hermitian 
physics~\cite{Kawabata19NC,Kawabata19}. In non-Hermitian systems, symmetries 
are defined by 
\begin{align}
	\text{time-reversal symmetry (TRS)}&:\quad
	\mathcal{U_{T_+}} {\cal H}^* \mathcal{U}^{\dagger}_{\mathcal{T}_+} = {\cal H},\quad
	\mathcal{U_{T_+}} \mathcal{U}^{*}_{\mathcal{T}_+} = \pm1,\\
	\text{particle-hole symmetry (PHS)}&:\quad
	\mathcal{U}_{\mathcal{P}_-}  {\cal H}^T \mathcal{U}^{\dagger}_{\mathcal{P}_-} = -{\cal H},\quad
	\mathcal{U}_{\mathcal{P}_-} \mathcal{U}_{\mathcal{P}_-}^{*} = \pm1,\\
	\text{time-reversal symmetry$^{\dag}$ (TRS$^{\dag}$)}&:\quad
	\mathcal{U_{P_+}} {\cal H}^T \mathcal{U}^{\dagger}_{\mathcal{P}_+} = {\cal H},\quad
	\mathcal{U_{P_+}} \mathcal{U}^{*}_{\mathcal{P}_+} = \pm1,\\
	\text{particle-hole symmetry$^{\dag}$ (PHS$^{\dag}$)}&:\quad
	\mathcal{U}_{\mathcal{T}_-}  {\cal H}^* \mathcal{U}^{\dagger}_{\mathcal{T}_-} = -{\cal H},\quad
	\mathcal{U}_{\mathcal{T}_-} \mathcal{U}^{*}_{\mathcal{T}_-} = \pm1,\\
	\text{chiral symmetry (CS)}&:\quad
	\mathcal{U_C} {\cal H}^{\dagger} \mathcal{U}_{\cal C}^{\dagger} = -{\cal H},\quad
	\mathcal{U_C}^2 = 1,\\
	\text{sublattice symmetry (SLS)}&:\quad
	\mathcal{U_S} {\cal H} \mathcal{U}_{\cal S}^{\dagger} = -{\cal H},\quad
	\mathcal{U_S}^2 = 1,
\end{align}
where $\mathcal{U_{T_{\pm}}}$, $\mathcal{U}_{\mathcal{P}_{\pm}}$, 
$\mathcal{U_C}$, and $\mathcal{U_S}$ are unitary matrices. TRS and PHS$^\dagger$ are unified:  
if ${\cal H}$ respects TRS, $\text{i}{\cal H}$ respects PHS$^\dagger$, and vice versa~\cite{Kawabata19NC}. 
It is useful to group the non-Hermitian symmetry classes according to 
the number $N$ of independent 
anti-unitary symmetries (TRS, PHS, TRS$^{\dagger}$, and PHS$^{\dagger}$)
(Table~\ref{tab: symmetry-supplement1})~\cite{Kawabata19}. 
In the simultaneous presence of multiple symmetries, the commutation or anticommutation 
relations between them are relevant.

Hermitization is a powerful tool to analyze non-Hermitian systems.
It is relevant to understand topological phases of non-Hermitian systems~\cite{Gong18, Kawabata19}.
In this work, we provide a universal understanding about the 
Anderson transitions (ATs) by Hermitization.
Here, we will explain in detail some of the mappings between 
the 38 non-Hermitian symmetry classes and 10 Hermitian symmetry classes, and relate   
the localization properties of non-Hermitian systems with those in Hermitian systems. 
The comprehensive correspondence between the 38-fold non-Hermitian symmetry class and 
10-fold symmetry class 
is summarized in Table~\ref{tab: symmetry-supplement1} 
and Table~\ref{tab: symmetry-supplement2}. 
Note also that the correspondence table  
was previously derived also in the context of the point-gap classification of non-Hermitian 
topological phases~\cite{Kawabata19}.

Hermitization is a local operation which can be regarded as 
a local introduction of a sublattice structure. When a non-Hermitian Hamiltonian 
only has short-range hoppings in $d$-dimensional space, so 
does 
the Hermitian 
Hamiltonian obtained from 
Hermitization.
Eigenmodes of the non-Hermitian 
Hamiltonian are the same as zero modes of the Hermitized Hamiltonian; 
both eigenmodes share the same localization properties. 
Thus, universal critical properties of disorder-driven quantum phase transitions 
are shared by the non-Hermitian Hamiltonians and Hermitian Hamiltonians that are 
related to each other through 
Hermitization. 

Hermitization maps a non-Hermitian Hamiltonian to a Hermitian Hamiltonian with CS (SLS). 
When the original non-Hermitian Hamiltonian respects CS or SLS,
the  
Hermitized Hamiltonian 
commutes with a 
unitary matrix and can be 
further block diagonalized. 
Even in the presence of disorder, the 
Hermitized Hamiltonian can 
be 
block diagonalized, given that 
disorder does not break any symmetries. 
The relevant symmetry class (classifying space) of the 
Hermitized Hamiltonian is determined by the symmetry class 
of its irreducible block, 
not by the symmetries of the Hermitized Hamiltonian itself.  

In the following, we summarize 
relevant symmetry classes of the 
Hermitized 
Hamiltonians
for several non-Hermitian symmetry classes in an elementary manner.  
For clarify of the following description, we always use $\tilde x$ for a Hermitian matrix 
and $x$ 
for a non-Hermitian matrix ($x$ can be different symbols). 

\begin{table}[t]
	\caption{
        38-fold non-Hermitian symmetry class (NHSC). 
		Any universality 
		class of disorder-driven quantum phase transitions in 
		a NHSC can be realized in the corresponding Hermitian symmetry class (HSC).
		The blank entries mean the absence of the symmetries. 
		For the anti-unitary symmetries (i.e., TRS, PHS, TRS$^{\dagger}$, and PHS$^{\dagger}$), 
        the entries $\pm 1$ denote the signs of the symmetry operations. $N$ is the number of the 
        independent anti-unitary symmetries. A symmetry class labelled as $...+{\cal S}_{\pm}$ or 
       $...+{\cal S}_{\pm\pm}$ has 
       sublattice symmetry (SLS) in addition to 
       the symmetries in 
       the 
       symmetry class specified by $``..."$. For $N=0$, the subscript of 
        ${\cal S}_{\pm}$ specifies the commutation ($+1$) or anti-commutation ($-1$) relation 
        between 
        SLS and chiral symmetry (CS). For $N=2$, the subscript 
        of $\mathcal{S}_{\pm}$ specifies the commutation ($+1$) or anti-commutation ($-1$) relation 
        between 
        SLS and TRS or PHS. For $N = 3$, only three anti-unitary symmetries 
       are independent of one another, where the first subscript of $\mathcal{S}_{\pm\pm}$ specifies 
       the 
       commutation or anti-commutation relation between SLS and TRS, and the second one specifies the 
       commutation or anti-commutation relation between SLS and PHS.
		In the presence of both CS and SLS, the commutation 
		or anti-commutation relation between CS and 
       SLS is respectively specified by $+$ 
       or $-$ 
       in the column of $[\mathcal{C},\mathcal{S}]_{\pm}=0$.}
	\begin{tabular}{c|c|c|c|c|c|c|c|c}
		\hline \hline
		~NHSC~ &
		~TRS ($\mathcal{T}_+$)~ &
		~PHS ($\mathcal{P}_-$)~ &
		~TRS$^{\dagger} (\mathcal{P}_+)$~ &
		~PHS$^{\dagger} (\mathcal{T}_-)$~ &
		~CS ($\mathcal{C}$)~ &
		~SLS ($\mathcal{S}$)~ &
		~$[\mathcal{C},\mathcal{S}]_{\pm}$=0~ &
		~HSC~ \\ \hline\hline
		$N$=0              &    &    &    &    &         &         &    &        \\\hline
		A             &    &    &    &    &         &         &    & AIII \\
		AIII          &    &    &    &    & $\surd$ &         &    & A    \\
		AIII$^{\dag}$ (A+$\mathcal{S}$)&    &    &    &    &         & $\surd$&    & AIII \\
		AIII + $\mathcal{S}_{+}$ &    &    &    &    & $\surd$ & $\surd$ & + & AIII \\
		AIII + $\mathcal{S}_{-}$ &    &    &    &    & $\surd$ & $\surd$ & - & A   \\
		\hline\hline
		$N$=1              &    &    &    &    &         &         &    &        \\\hline
		AI            & 1  &    &    &    &         &         &    & BDI  \\
		AII           & -1 &    &    &    &         &         &    & CII  \\
		D             &    & 1  &    &    &         &         &    & DIII \\
		C             &    & -1 &    &    &         &         &    & CI   \\
		AI$^{\dagger}$   &    &    & 1  &    &         &         &    & CI   \\
		AII$^{\dagger}$  &    &    & -1 &    &         &         &    & DIII \\
		\hline\hline
		$N$=2              &    &    &    &    &         &         &    &        \\ \hline
		BDI            & 1  & 1  &    &    & $\surd$ &         &    & D   \\
		CI             & 1  & -1 &    &    & $\surd$ &         &    & AI   \\
		DIII          & -1 & 1  &    &    & $\surd$ &         &    & AII  \\
		CII            & -1 & -1 &    &    & $\surd$ &         &    & C    \\
		BDI$^{\dagger}$  &    &    & 1  & 1  & $\surd$ &         &    & AI  \\
		CI$^{\dagger}$   &    &    & 1  & -1 & $\surd$ &         &    & C    \\
		DIII$^{\dagger}$ &    &    & -1 & 1  & $\surd$ &         &    & D    \\
		CII$^{\dagger}$  &    &    & -1 & -1 & $\surd$ &         &    & AII \\
		D + $\mathcal{S}_{+}$ &    & 1  & 1  &    &         & $\surd$ &    & AIII \\
		C + $\mathcal{S}_{-}$ &    & -1 & 1  &    &         & $\surd$ &    & CI   \\
		D + $\mathcal{S}_{-}$ &    & 1  & -1 &    &         & $\surd$ &    & DIII \\
		C + $\mathcal{S}_{+}$ &    & -1 & -1 &    &         & $\surd$ &    & AIII \\
		AI + $\mathcal{S}_{+}$ & 1  &    &    & 1  &         & $\surd$ &    & BDI  \\
		AI + $\mathcal{S}_{-}$ & 1  &    &    & -1 &         & $\surd$ &    & AIII \\
		AII + $\mathcal{S}_{+}$ & -1 &    &    & -1 &         & $\surd$ &    & CII  \\
		\hline\hline
		$N$=3              &    &    &    &    &         &         &    &        \\ \hline
		BDI + $\mathcal{S}_{++}$ & 1  & 1  & 1 & 1  & $\surd$ & $\surd$ & +  & BDI  \\
		BDI + $\mathcal{S}_{--}$ & 1  & 1  & -1   & -1 & $\surd$ & $\surd$ & +  & DIII \\
		DIII + $\mathcal{S}_{++}$ & -1 & 1  & 1 & -1 & $\surd$ & $\surd$ & + & CII  \\
		CI + $\mathcal{S}_{++}$ & 1  & -1 & -1   & 1  & $\surd$ & $\surd$ & +  & BDI  \\
		CI + $\mathcal{S}_{--}$ & 1  & -1 & 1 & -1 & $\surd$ & $\surd$ & + & CI   \\
		CII + $\mathcal{S}_{++}$ & -1 & -1 & -1 & -1 & $\surd$ & $\surd$ & +  & CII  \\
		BDI + $\mathcal{S}_{+-}$ & 1  & 1  & -1 & 1  & $\surd$ & $\surd$ & - & D    \\
		BDI + $\mathcal{S}_{-+}$ & 1  & 1  & 1 & -1 & $\surd$ & $\surd$ & - & A    \\
		DIII + $\mathcal{S}_{+-}$ & -1 & 1  & -1 & -1 & $\surd$ & $\surd$ & - & AII  \\
		CI + $\mathcal{S}_{+-}$ & 1  & -1 & 1 & 1  & $\surd$ & $\surd$ & - & AI   \\
		CI + $\mathcal{S}_{-+}$ & 1  & -1 & -1  & -1 & $\surd$ & $\surd$ & - & A    \\
		CII + $\mathcal{S}_{+-}$ & -1 & -1 & 1 & -1 & $\surd$ & $\surd$ & - & C \\
		\hline \hline
	\end{tabular}
	\label{tab: symmetry-supplement1}
\end{table}

\subsection{Class AIII}
Non-Hermitian Hamiltonians ${\cal H}$ in class AIII respect CS: 
$\mathcal{C} {\cal H}^{\dagger} \mathcal{C} = -{\cal H}$ with a unitary matrix 
$\mathcal{C}$ satisfying $\mathcal{C}^2 = 1$ and 
$\mathcal{C}^{\dagger} = \mathcal{C}$. 
A Hermitian Hamiltonian is introduced through 
Hermitization, 
\begin{equation}
	\tilde{\cal H}:=\left(\begin{array}{cc}
		0 & {\cal H} \\
		{\cal H}^{\dagger} & 0
	\end{array}\right), \label{nh-aiii-a}
\end{equation}
where the reference energy is set to $E = 0$ without loss of generality. 
CS of the  
non-Hermitian Hamiltonian ${\cal H}$ leads to CS of the Hermitized Hamiltonian $\tilde{\cal H}$:
\begin{equation}
	\left( \tau_x \otimes \mathcal{C} \right) \tilde{\cal H} \left( \tau_x \otimes \mathcal{C} \right)^{-1} 
= - \tilde{\cal H},
\end{equation}
 where $\tau_x$ is the Pauli matrix that exchanges ${\cal H}$ and ${\cal H}^{\dagger}$ in 
Eq.~(\ref{nh-aiii-a}). By the construction of Hermitization, $\tilde{\cal H}$ also respects another CS:
\begin{equation}
	\tau_z \tilde{\cal H} \tau_{z}^{-1} = - \tilde{\cal H}.
\end{equation}
The simultaneous presence of the two CSs allows $\tilde{\cal H}$ to commute with a 
unitary symmetry:
\begin{equation}
	\left( \tau_y \otimes \mathcal{C} \right) \tilde{\cal H} \left( \tau_y \otimes \mathcal{C} \right)^{-1} = 
\tilde{\cal H}.
\end{equation}
The unitary symmetry enables block diagonalization of $\tilde{\cal H}$.

The CS operator can be written as $\mathcal{C} = \mathcal{C}_+ - \mathcal{C}_-$ with the projection operators $\mathcal{C}_+$ and $\mathcal{C}_-$ on subspaces in which eigenenergies of $\mathcal{C}$ are $+1$ and $-1$, respectively.
Here are some useful properties of the projection operators:
\begin{equation}
	\mathcal{C}_{+}+\mathcal{C}_{-} = 1,\quad
	\mathcal{C}_{+}\mathcal{C}_{-} 
	= \mathcal{C}_{-}\mathcal{C}_{+} = 0,\quad
	\mathcal{C}_{+}^2 = \mathcal{C}_+,\quad
	\mathcal{C}_-^2 = \mathcal{C}_-,\quad
	\mathcal{C}_+\mathcal{C} = \mathcal{C} \mathcal{C}_+ = \mathcal{C}_+,\quad
	\mathcal{C}_-\mathcal{C} = \mathcal{C} \mathcal{C}_- = -\mathcal{C}_-.
\end{equation}
Then, we have
\begin{align}
	\mathcal{C}_{\pm} {\cal H}^{\dagger} \mathcal{C}_{\pm} 
	&=  -\mathcal{C}_{\pm} \left( \mathcal{C} {\cal H} \mathcal{C} \right) \mathcal{C}_{\pm} 
	= -\mathcal{C}_{\pm}  {\cal H}  \mathcal{C}_{\pm},\\
	(\mathcal{C}_-{\cal H}\mathcal{C}_+)^{\dagger}
	&=\mathcal{C}_+ {\cal H}^{\dagger} \mathcal{C}_- 
	=  -\mathcal{C}_+ \left( \mathcal{C} {\cal H} \mathcal{C} \right) \mathcal{C}_- 
	= \mathcal{C}_+ {\cal H}  \mathcal{C}_-.
\end{align}
In the basis that diagonalizes $\mathcal{C}$, the Hamiltonian $H$ can be written as
\begin{equation}
	{\cal H} = \left(\begin{array}{cc}
		\text{i} \tilde{h}_1 & h_{12} \\
		h_{12}^{\dagger} &\text{i} \tilde{h}_2
	\end{array}\right), \label{aiii-a}
\end{equation}
where the matrices $\tilde{h}_{1}$, $\tilde{h}_{2}$, and $h_{12}$ satisfy
\begin{equation}
	\tilde{h}_{1}^{\dag} = \tilde{h}_{1},\quad
	\tilde{h}_{2}^{\dag} = \tilde{h}_{2},\quad
	\mathcal{C}_+ {\cal H} \mathcal{C}_+ =\text{i} \tilde{h}_1,\quad
	\mathcal{C}_- {\cal H} \mathcal{C}_- =\text{i} \tilde{h}_2,\quad
	\mathcal{C}_+ {\cal H} \mathcal{C}_- = h_{12},\quad
	\mathcal{C}_- {\cal H} \mathcal{C}_+ = h_{21} = h^{\dagger}_{12}.
\end{equation}
The following argument holds true for arbitrary dimensions of $\mathcal{C}_{+}$ 
and $\mathcal{C}_{-}$. For simplicity, let us assume that $\mathcal{C}_{+}$ 
and $\mathcal{C}_{-}$ have the same dimension and take 
$\mathcal{C} = \sigma_z$. In this basis, we can block diagonalize the Hermitized 
Hamiltonian $\tilde{\cal H}$ by
\begin{equation}
	\tilde{\cal H} = \left(
	\begin{matrix}
		0 & 0 & \text{i} \tilde{h}_1& h_{12} \\
		0 & 0 & h_{12}^{\dagger} & \text{i} \tilde{h}_2\\
		- \text{i} \tilde{h}_1&  h_{12} & 0 & 0 \\
		h_{12}^{\dagger} & - \text{i} \tilde{h}_2 & 0 & 0
	\end{matrix}
	\right)
	\quad \rightarrow \quad
	\mathcal{U} \tilde{\cal H} \mathcal{U}^{
		\dagger}=\left(
	\begin{matrix}
		-\tilde{h}_1 & -\text{i}h_{12}  & 0 & 0 \\
		\text{i} h_{12}^{\dagger} &  \tilde{h}_2 & 0 & 0 \\
		0 & 0 &\tilde{h}_1 &\text{i}h_{12}\\
		0 & 0 & -\text{i} h_{12}^{\dagger} &- \tilde{h}_2
	\end{matrix}
	\right),  
\end{equation}
where $\mathcal{U}$ is the following unitary matrix:
\begin{equation}
	\mathcal{U} := \frac{1}{\sqrt{2}} \left(
	\begin{matrix}
		1 & 0  & - \text{i} &0 \\
		0 & 1  & 0 & \text{i}  \\
		1 &  0 & \text{i} & 0\\
		0 & 1 & 0 & -\text{i} 
	\end{matrix}
	\right).
\end{equation}
The two irreducible blocks 
have no symmetry except for 
Hermiticity,  
\begin{align}
\tilde{\cal H}_{\rm block} := \left(\begin{array}{cc}
	-\tilde{h}_1 & -\text{i} h_{12} \\
	\text{i}h_{12}^{\dagger} & \tilde{h}_2
\end{array}\right). \label{block}
\end{align}
Thus, the relevant symmetry class of the 
Hermitized 
Hamiltonian ${\cal H}$ obtained 
from 
the 
non-Hermitian Hamiltonian ${\cal H}$ in 
class AIII is class A. 

\subsection{Class {CII}$^{\dagger}$}

Non-Hermitian Hamiltonians ${\cal H}$ in class CII$^{\dagger}$ have  
TRS$^{\dag}$ $\mathcal{U_{P_+}} {\cal H}^{T} \mathcal{U}_{\mathcal{P}_+}^{\dagger} = {\cal H}$ 
with $\mathcal{U_{P_+}}\mathcal{U}_{\mathcal{P}_+}^* = -1$ and 
PHS$^{\dag}$ $\mathcal{U_{T_-}} {\cal H}^* \mathcal{U}_{\mathcal{T}_-}^{\dagger} = -{\cal H}$ 
with $\mathcal{U_{T_-}} \mathcal{U}_{\mathcal{T}_-}^* =-1$. 
The combination of TRS$^{\dag}$ and PHS$^{\dag}$ gives CS.
From the above argument for 
class AIII, we can write ${\cal H}$ into Eq.~(\ref{aiii-a}) 
in the basis that diagonalizes $\mathcal{C} = \sigma_z$. 
Here, we choose the symmetry operators to be $\mathcal{U_{P_+}}=\sigma_0 \otimes \mathcal{V}$ and $\mathcal{U_{T_-}}=\sigma_z \otimes \mathcal{V}$ with $\mathcal{V}\mathcal{V}^* = -1$. 
A unitary matrix $\mathcal{V}$ 
respects
\begin{equation}
	\mathcal{V} \tilde{h}^*_1 \mathcal{V}^{\dagger} = \tilde{h}_{1},\quad 
	\mathcal{V} \tilde{h}^*_2 \mathcal{V}^{\dagger} = \tilde{h}_{2},\quad
	\mathcal{V}  h_{12}^*\mathcal{V}^{\dagger} = h_{12},
\end{equation}
where $\tilde{h}_1$, $\tilde{h}_2$, and $h_{12}$ are defined by Eq.~(\ref{aiii-a}).
Thus, the Hermitian block given by Eq.~(\ref{block}) respects TRS,  
$\left( \sigma_z \otimes \mathcal{V} \right) \tilde{\cal H}_{\text{block}}^* \left( \sigma_z \otimes \mathcal{V} \right)^{-1} = \tilde{\cal H}_{\text{block}}$, and belongs to class AII. 

\subsection{Class DIII}
Non-Hermitian Hamiltonians ${\cal H}$ in class DIII have TRS $\ \mathcal{U_{T_+}} {\cal H}^{*} \mathcal{U}_{\mathcal{T}_+}^{\dagger} = {\cal H}$ with  $\mathcal{U_{T_+}}\mathcal{U}_{\mathcal{T}_+}^* = -1$ and 
PHS $\ \mathcal{U_{P_-}} {\cal H}^T \mathcal{U}_{\mathcal{P}_-}^{\dagger} = -{\cal H}$ with $\mathcal{U_{P_-}} \mathcal{U}_{\mathcal{P}_-}^* =1$.  The combination of TRS and PHS gives CS.  
From the above argument for class AIII, we can write ${\cal H}$ into Eq.~(\ref{aiii-a}) 
in the diagonal basis of $\mathcal{C} = \sigma_z$. Here, we choose the symmetry operators to 
be $\mathcal{U_{T_+}}=\sigma_y \otimes \mathcal{V}$ and 
$\mathcal{U_{P_-}}=\sigma_x \otimes \mathcal{V}$ with $\mathcal{V}\mathcal{V}^* = 1$. 
A unitary matrix $\mathcal{V}$ 
respects
\begin{equation}
	\mathcal{V} \tilde{h}^*_1 \mathcal{V}^{\dagger} = -\tilde{h}_{2},\quad 
	\mathcal{V} \tilde{h}^*_2 \mathcal{V}^{\dagger} = -\tilde{h}_{1},\quad
	\mathcal{V}  h_{12}^T\mathcal{V}^{\dagger} = -h_{12},
\end{equation}
where $\tilde{h}_1$, $\tilde{h}_2$, and $h_{12}$ are defined by Eq.~(\ref{aiii-a}).
Thus, the Hermitian block given in Eq.~(\ref{block}) respects TRS, 
$\left( \sigma_y \otimes \mathcal{V} \right) \tilde{\cal H}_{\text{block}} ^* \left( \sigma_y \otimes \mathcal{V} \right)^{-1} = \tilde{\cal H}_{\text{block}}$.  The irreducible 
block
of the Hermitized Hamiltonian belongs to class AII.

\subsection{Non-Hermitian Hamiltonians with anti-commutative CS and SLS: class AIII + ${\cal S}_{-}$}

Let us consider a generic non-Hermitian Hamiltonian with CS and SLS 
which anti-commute with each other. We can choose the symmetry operators to 
be $\mathcal{U_{C}} = \sigma_y$ and $\mathcal{U_{S}} = \sigma_z$ in a certain basis. 
In this basis, the non-Hermitian Hamiltonian takes a form of 
\begin{align}
{\cal H} = \left(\begin{array}{cc}
	0 & h_1 \\
	h_2 & 0
\end{array}\right) 
\end{align}
with $h_1 = h_1^{\dagger}$ and $h_2 = h_2^{\dagger}$. Then,  
the Hermitized Hamiltonian $\tilde{\cal H}$ can be block diagonalized 
\begin{equation}
	\tilde{H} = \left(
	\begin{matrix}
		0 & 0 & 0  & h_{1} \\
		0 & 0 & h_{2} & 0 \\
		0 &  h_{2} & 0 & 0\\
		h_1 & 0 & 0 & 0
	\end{matrix}
	\right)
	\quad \rightarrow \quad
	\mathcal{U} \tilde{H} \mathcal{U}^{
		\dagger} =
	\left(
	\begin{matrix}
		h_1 & 0  & 0 & 0 \\
		0 & -h_1 & 0 & 0 \\
		0 & 0 &h_2 & 0\\
		0 & 0 & 0 & -h_2
	\end{matrix}
	\right),
\end{equation}
by the 
unitary matrix 
\begin{equation}
	\mathcal{U} = \frac{1}{\sqrt{2}} \left(
	\begin{matrix}
		1 & 0  & 0 &1 \\
		-1 & 0  & 0 &   1 \\
		0&  1 & 1  & 0\\
		0 & -1 & 1& 0
	\end{matrix}
	\right).
\end{equation}
Note that $h_1$ and $h_2$ have no symmetry restrictions 
other than 
Hermiticity. Thus, 
the 
relevant symmetry class of 
the Hermitized Hamiltonian 
is class A.

\begin{table}[t]
	\caption{
     10-fold Hermitian symmetry class (HSC) and 
     38-fold non-Hermitian symmetry class (NHSC).
		Any universality class 
		of disorder-driven quantum phase transitions in a Hermitian symmetry class 
       can be realized in the corresponding non-Hermitian symmetry classes. TRS, PHS, and CS in the Hermitian 
       symmetry classes are time-reversal, particle-hole, and chiral symmetries, respectively.  ${\cal T}_{+}$, 
       ${\cal P}_{-}$, ${\cal P}_{+}$, ${\cal T}_{+}$, ${\cal C}$, and ${\cal S}$ in 
       the NHSCes are time-reversal, particle-hole, time-reversal$^{\dagger}$, particle-hole$^{\dagger}$, chiral, 
       and sublattice symmetries defined in Eqs.~(S.1)-(S.6). The blank entries, $\pm 1$ entries, $\pm$ entries, 
       and the names of the 38-fold 
       NHSC are defined in the caption of 
       Table~\ref{tab: symmetry-supplement1}. ``*" in the remark means that the dimension of random 
       Hermitian matrices must be even, which is always the case with 
       the
       symplectic class, three chiral classes, and four 
       BdG classes. ``**" in the remark means that the dimension of the Hermitian matrices must be a multiple of 4.}
	\begin{tabular}{c|c||c|c|c||c||c|c|c|c|c|c|c|c}
		\hline \hline
	 &	~HSC~ &
	 	~TRS~ &
		~PHS~ &
		~CS~ &
		~NHSC~ &
		~$\mathcal{T}_+$~ &
		~$\mathcal{P}_-$~ &
		~$\mathcal{P}_+$~ &
		~$\mathcal{T}_-$~ &
		~$\mathcal{C}$~ &
		~$\mathcal{S}$~ &
		~$[\mathcal{C},\mathcal{S}]_{\pm}$=0~ & remark \\
		\hline
	unitary &	A & & &	 & AIII          &    &    &    &    & $\surd$ &         &   & \\
	&	 & & &	 & AIII + $\mathcal{S}_{-}$ &    &    &    &    & $\surd$ & $\surd$ & - & \\      
	&	 & & &	 & BDI + $\mathcal{S}_{-+}$ & 1  & 1  & 1 & -1 & $\surd$ & $\surd$ & - & \\
	&	 & & &	 & CI + $\mathcal{S}_{-+}$ & 1  & -1 & -1  & -1 & $\surd$ & $\surd$ & - & * \\   
		\hline
	 orthogonal &	AI & 1 & &	 & CI             & 1  & -1 &    &    & $\surd$ &         &  &  * \\
	&	&  & &	 & BDI$^{\dagger}$  &    &    & 1  & 1  & $\surd$ &         &   &  \\
	&	 &  & & & CI + $\mathcal{S}_{+-}$ & 1  & -1 & 1 & 1  & $\surd$ & $\surd$ & -  & \\
		\hline
	 symplectic &	AII & -1 & &	 & DIII          & -1 & 1  &    &    & $\surd$ &         &  & * \\
	&	&  & & &  CII$^{\dagger}$  &    &    & -1 & -1 & $\surd$ &         &   &  * \\
	&	&  & & & DIII + $\mathcal{S}_{+-}$ & -1 & 1  & -1 & -1 & $\surd$ & $\surd$ & - &  * \\
		\hline \hline 
	 chiral unitary &	AIII & & & $\surd$ & A             &    &    &    &    &         &         &   & * \\
	&	 & & &  & AIII$^{\dagger}$ &    &    &    &    &  & $\surd$ &  & * \\
	&	 & & &  & AIII + $\mathcal{S}_{+}$ &    &    &    &    & $\surd$ & $\surd$ & + & * \\
	&	& & &  & 	D + $\mathcal{S}_{+}$ &    & 1  & 1  &    &         & $\surd$ &  &  * \\
	&	 & & & & C + $\mathcal{S}_{+}$ &    & -1 & -1 &    &         & $\surd$ &  & ** \\
	&	& & &  & AI + $\mathcal{S}_{-}$ & 1  &    &    & -1 &         & $\surd$ &  &  * \\
		\hline
	chiral orthogonal &	BDI & 1 & 1 &	$\surd$ & AI            & 1  &    &    &    &         &         &  &  * \\
	&	&  & & & AI + $\mathcal{S}_{+}$ & 1  &    &    & 1  &         & $\surd$ &   &  * \\
	&	&  & & &  BDI + $\mathcal{S}_{++}$ & 1  & 1  & 1 & 1  & $\surd$ & $\surd$ & + & * \\
	&	&  & & &  	CI + $\mathcal{S}_{++}$ & 1  & -1 & -1   & 1  & $\surd$ & $\surd$ & + &  ** \\
		\hline
	chiral symplectic &	CII  & -1 & -1 & $\surd$ &  AII  & -1 &    &    &    &         &         &  &  ** \\
	&	&  & & & AII + $\mathcal{S}_{+}$ & -1 &    &    & -1 &         & $\surd$ &  &  ** \\
	&	&  & & & DIII + $\mathcal{S}_{++}$ & -1 & 1  & 1 & -1 & $\surd$ & $\surd$ & + &  ** \\
	&	&  & & &  CII + $\mathcal{S}_{++}$ & -1 & -1 & -1 & -1 & $\surd$ & $\surd$ & + &  ** \\
		\hline \hline 
	BdG &	D  &  & 1 & & BDI            & 1  & 1  &    &    & $\surd$ &         &   &   \\
	&	&  & & &  DIII$^{\dagger}$ &    &    & -1 & 1  & $\surd$ &         &  &  * \\
	&	&  & & & BDI + $\mathcal{S}_{+-}$ & 1  & 1  & -1 & 1  & $\surd$ & $\surd$ & - &  \\
		\hline 
	&	DIII & -1 & 1 & $\surd$ & D             &    & 1  &    &    &         &         &  &   * \\
	&	&  & & & AII$^{\dagger}$  &    &    & -1 &    &         &         &   & ** \\
	&	&  & & & D + $\mathcal{S}_{-}$ &    & 1  & -1 &    &         & $\surd$ & &  * \\
	&	&  & & & BDI + $\mathcal{S}_{--}$ & 1  & 1  & -1   & -1 & $\surd$ & $\surd$ & + &  * \\
		\hline
	&	C &  & -1 & &  CII            & -1 & -1 &    &    & $\surd$ &         &  &  * \\
	&	&  & & & CI$^{\dagger}$   &    &    & 1  & -1 & $\surd$ &         &  &  * \\
	&	&  & & & CII + $\mathcal{S}_{+-}$ & -1 & -1 & 1 & -1 & $\surd$ & $\surd$ & - &  * \\
		\hline 
	&	CI  & 1 & -1 & $\surd$ & C             &    & -1 &    &    &         &         &   &  ** \\
	&	&  & & & AI$^{\dagger}$   &    &    & 1  &    &         &         &  &  * \\
	&	&  & & & C + $\mathcal{S}_{-}$ &    & -1 & 1  &    &         & $\surd$ &  &  * \\
	&	&  & & & CI + $\mathcal{S}_{--}$ & 1  & -1 & 1 & -1 & $\surd$ & $\surd$ & + &  ** \\
        \hline \hline
	\end{tabular}
	\label{tab: symmetry-supplement2}
\end{table}

\subsection{Properties of Hermitized Hamiltonians}
Though non-Hermitian Hamiltonian ${\cal H}$ in Eq.~(1) in the main text  
takes a general form of a non-Hermitian symmetry class,  it is not obvious whether  
irreducible blocks of the Hermitized Hamiltonian take a general Hamiltonian 
form in its corresponding Hermitian symmetry class. To complement this point partially  
(see a subsection of `symmetry-conserving energy and symmetry-breaking energy' 
below for why we use ``partially" here), in the following, we start from a general Hermitian 
Hamiltonian $\tilde{\cal H}$ of a Hermitian symmetry class, an energy $\tilde{E}$ that respects all 
the symmetries of $\tilde{\cal H}$, and eigenmodes $|\psi\rangle$ at 
$\tilde{E}$, 
\begin{align}
\tilde{\cal H} |\psi\rangle = \tilde{E} |\psi\rangle. 
\end{align} 
Then, for each of its corresponding non-Hermitian symmetry classes, 
we introduce a non-Hermitian Hamiltonian ${\cal H}$ of the symmetry class, an energy 
$E$ that respects all the symmetries of the symmetry class of ${\cal H}$ 
(`symmetry-conserving energy'), and right eigenmodes $|\phi_r\rangle$ at the 
symmetry-conserving energy $E$, 
\begin{align}
{\cal H} |\phi_r\rangle = E |\phi_r\rangle. 
\end{align}
We show that $|\phi_r\rangle$ is given by $|\psi\rangle$; $|\phi_r\rangle$ 
and $|\psi\rangle$ share the same localization properties. Importantly, 
the non-Hermitian Hamiltonian ${\cal H}$ thus introduced takes a generic Hamiltonian 
form under the symmetries of the symmetry class. Here, by `takes a generic Hamiltonian form', 
we mean that when $\tilde{\cal H}$ takes all the possible Hamiltonian defined by the symmetries of 
the Hermitian symmetry class, ${\cal H}$ takes all the possible Hamiltonian defined by the symmetries 
of the non-Hermitian symmetry class, e.g. compare Eq.~(\ref{nh-aiii}) with Eq.~(\ref{a}).

The 38-fold non-Hermitian symmetry class contains more symmetry 
classes than the 10-fold Hermitian symmetry class. Consequently, a set of different 
non-Hermitian symmetry classes can be mapped to the same Hermitian symmetry 
class by the Hermitization. 
Table~\ref{tab: symmetry-supplement2} summarizes all the corresponding 
non-Hermitian symmetry classes for the 10 Hermitian symmetry classes. A similar 
correspondence between classifying spaces was obtained on the basis of the 
$K$-theory~\cite{Kawabata19}. 

\subsection{Symmetry-conserving energy and symmetry-breaking energy}
Note that the following argument focuses {\it only} on those cases where the energy 
$E$ respects all the symmetries of ${\cal H}$ (symmetry-conserving energy case). 
The argument does not cover the other cases where $E$ breaks (some) symmetries of ${\cal H}$ 
(symmetry-breaking energy case). In the latter cases, ${\cal H}$ {\it together with} $E$ belongs to a given 
non-Hermitian symmetry class, while ${\cal H}$ itself has higher symmetries than the symmetry class and  
$E$ breaks some of the symmetries of ${\cal H}$ such that ${\cal H}$ with $E$ belongs to the symmetry class. 
For example, eigenmodes of non-Hermitian class-AI (or AII) Hamiltonian at complex energy $E \ne E^*$ 
belong to non-Hermitian class A. According to the Hermitization, the class-AI (or AII) Hamiltonian with the 
complex energy maps to a Hermitian Hamiltonian whose relevant symmetry class is Hermitian class AIII. 
The mapping is related to numerical results in the 2D class AII model with $E \ne E^*$. 
In spite of the presence of such mapping, the argument in this section does not discuss how to construct a generic 
non-Hermitian class-AI (or AII) Hamiltonian whose eigenmodes at the complex energy are given by zero 
modes of a general Hermitian class-AIII Hamiltonian. Due to the lack of arguments for symmetry-breaking energy 
cases, we regard the proposed correspondence as a conjecture instead of as a mathematically-proven fact.  

\bigskip
\bigskip
In the next subsection, we first begin with Hermitian class A case, whose corresponding non-Hermitian 
symmetry classes are class AIII, AIII + ${\cal S}_{-}$, BDI + ${\cal S}_{-+}$ and CI + ${\cal S}_{-+}$. 
A general Hermitian class-A Hamiltonian $\tilde{\cal H}_{\rm A}$ is given in Eq.~(\ref{a}), while
Hamiltonians in these non-Hermitian symmetry classes ${\cal H}$ 
are given by Eqs.~(\ref{nh-aiii},\ref{nh-aiii-s-},\ref{nh-bdi-s-+},\ref{nh-ci-s-+}) respectively. 
The symmetry-conserving energies $E$ in these four non-Hermitian classes 
are $E={\rm i}E_{\rm i}$ with $E_{\rm i}\in \mathbb{R}$, 
$E=0$, $E=0$, and $E=0$ respectively.

\subsubsection{Hermitian class A}
A general Hermitian Hamiltonian in class A can always be written as 
\begin{align}
\tilde{\cal H}_{\rm A} = \left(\begin{array}{cc}
	 \tilde{h}_1 & h_{12} \\
	h_{12}^{\dagger} & \tilde{h}_2
\end{array}\right), \label{a}
\end{align}
with $\tilde{h}_{1}^{\dag} = \tilde{h}_{1},\tilde{h}_{2}^{\dag} = \tilde{h}_{2}$, and 
$ \tilde{h}_1, \tilde{h}_2,h_{12}\neq 0$. Let $n_1$ and $n_2$ be the 
dimensions of the square matrices $\tilde{h}_{1}$ and $\tilde{h}_{2}$, respectively.
In general, $n_1$ and $n_2$ can be different. Let $\tilde{E} \in \mathbb{R}$ 
be an eigenenergy of $\tilde{\cal H}_{\rm A}$ and $\ket{\psi}$ be the corresponding eigenstate. 

\paragraph{From Hermitian class A to non-Hermitian class AIII:}
For the given Hermitian Hamiltonian $\tilde{\cal H}_{\rm A}$ in Eq.~(\ref{a}), we construct  
a non-Hermitian Hamiltonian ${\cal H}$ in class AIII, which takes a generic form of class AIII Hamiltonians 
and whose eigenmode at symmetry-conserving energy shares the same localization properties 
with $|\psi\rangle$ of $\tilde{\cal H}_{\rm A}$. Here the symmetry-conserving energy of 
class AIII is pure imaginary, $E={\rm i}E_{\rm i}$ and $E_{\rm i}\in \mathbb{R}$.  
Then, consider a non-Hermitian Hamiltonian ${\cal H}$ and its chiral symmetry by
\begin{align}
{\rm AIII} \!\ :  
{\cal H} \equiv  \left(\begin{array}{cc}
	 -{\rm i}\,( \tilde{h}_1 - \tilde{E})  + {\rm i} E_{\rm i}
	 & {\rm i} h_{12} \\
	-{\rm i}h_{12}^{\dagger} & 
	{\rm i}\,( \tilde{h}_2 - \tilde{E}) + {\rm i} E_{\rm i} 
\end{array}\right), \!\ \!\ \!\ 
\cal{U_C} \equiv \begin{pmatrix}
1_{n_1\times n_1} & 0 \\
0 & -1_{n_2\times n_2}  \label{nh-aiii}
\end{pmatrix}. 
\end{align}
The non-Hermitian Hamiltonian ${\cal H}$ respects chiral symmetry ${\cal H} = - \cal{U_C} {\cal H}^{\dagger}\cal{U^{\dagger}_C}$ and hence belongs to class AIII.
Moreover, ${\cal H}$ has a right eigenstate at $E={\rm i} E_{\rm i}$, 
that share the same localization properties as $\ket{\psi}$ of $\tilde{\cal H}_{\rm A}$.
In fact, ${\cal U}_{\cal C} \ket{\psi}$ is such an eigenstate,
\begin{align}
{\cal H} \!\ {\cal U}_{\cal C} \ket{\psi} = {\rm i}E_{\rm i} \!\ {\cal U}_{\cal C} \ket{\psi}. \label{nh-aiii-1}
\end{align} 
Since ${\cal U}_{\cal C}$ is a local transformation, ${\cal U}_{\cal C} \ket{\psi}$ 
has the same localization properties as $\ket{\psi}$. Under the chiral 
symmetry defined in Eq.~(\ref{nh-aiii}), non-Hermitian Hamiltonians 
in class AIII take following general form,
\begin{align}
{\cal H}_{{\rm AIII}} =  \left(\begin{array}{cc}
	 {\rm i}\, \tilde{g}_1 
	 & {\rm i} g_{12} \\
	-{\rm i}g_{12}^{\dagger} & 
	{\rm i}\, \tilde{g}_2 
\end{array}\right),
\end{align} 
with $\tilde{g}^{\dagger}_1=\tilde{g}_1$, $\tilde{g}^{\dagger}_2=\tilde{g}_2$. 
Since $\tilde{E}, E_{\rm i}\in \mathbb{R}$, ${\cal H}$ in Eq.~(\ref{nh-aiii}) indeed 
takes this general form.  

\paragraph{From Hermitian class A to non-Hermitian class AIII + ${\cal S}_{-}$:}  
Let us introduce two random Hermitian 
Hamiltonians in class A defined by Eq.~(\ref{a}), 
$\tilde{\cal H}_{\rm A}$ and $\tilde{\cal H}^{\prime}_{\rm A}$, 
which have the same dimension and are independent 
of each other. For example, one can pick up two different disorder 
realizations in the same Hermitian lattice model with the same disorder strength in class A. 
Let $\tilde{E}$ and $\tilde{E}^{\prime}$ be real eigenenergies of $\tilde{\cal H}_{\rm A}$ 
and $\tilde{\cal H}^{\prime}_{\rm A}$, 
and $\ket{\psi}$ and $\ket{\psi'}$ be the corresponding eigenstates, respectively. 
Then, we introduce a non-Hermitian Hamiltonian ${\cal H}$ of class AIII + ${\cal S}_{-}$ 
together with its CS and SLS by 
\begin{align}
{\rm AIII}+{\cal S}_{-} \!\ : 
{\cal H} \equiv \begin{pmatrix}
    0 & \tilde{\cal H}_{\rm A} -\tilde{E} \\
    \tilde{\cal H}^{\prime}_{\rm A} -\tilde{E}' & 0
\end{pmatrix},  \!\ \!\  \!\   
\mathcal{U_{C}} = \tau_y,  \!\ \!\ \!\  \mathcal{U_{S}} = \tau_z.  \label{nh-aiii-s-}
\end{align}
Here, $\tau_y$ and $\tau_z$ are Pauli matrices. 
Thus, ${\cal H}$ respects 
$\mathcal{U_{C}}{\cal H}^{\dagger}\mathcal{U^{\dagger}_{C}} = -{\cal H}$ and $\mathcal{U_{S}}{\cal H}\mathcal{U^{\dagger}_{S}} = -{\cal H}$ with $[{\cal U}_{\cal C},{\cal U}_{\cal S}]_{-}=0$, 
and hence belongs to class AIII + $\cal{S}_-$. 
 The symmetry-conserving energy of class AIII + ${\cal S}_{-}$ is zero, $E=0$. 
${\cal H}$ in Eq.~(\ref{nh-aiii-s-}) has right eigenmodes at the zero energy, $\left( 0~\ket{\psi} \right)^T$ and 
$\left( \ket{\psi'}~0 \right)^T$, which have the same localization properties 
as $\ket{\psi}$ and $\ket{\psi'}$ of $\tilde{\cal H}_{\rm A}$ and 
$\tilde{\cal H}^{\prime}_{\rm A}$. 
Under the chiral and sublattice symmetries in Eq.~(\ref{nh-aiii-s-}), non-Hermitian Hamiltonians 
in class AIII + ${\cal S}_{-}$ take following general form,
\begin{align}
{\cal H}_{{\rm AIII}+{\cal S}_{-}} =  \left(\begin{array}{cc}
	 0  & \tilde{\cal G}_{\rm A} \\
    \tilde{\cal G}^{\prime}_{\rm A} & 0 \\ 
\end{array}\right), \label{nh-aiii-general}
\end{align} 
with two independent Hermitian Hamiltonians $\tilde{\cal G}_{\rm A}$ 
and $\tilde{\cal G}^{\prime}_{\rm A}$. 
Since $\tilde{\cal H}_{\rm A}$ and $\tilde{\cal H}^{\prime}_{\rm A}$ in Eq.~(\ref{nh-aiii-s-})  
are two independent Hermitian class-A Hamiltonians, Eq.~(\ref{nh-aiii-s-}) takes the general form 
of Eq.~(\ref{nh-aiii-general}).

\paragraph{From Hermitian class A to non-Hermitian class BDI + ${\cal S}_{-+}$:}
For a generic Hermitian Hamiltonian $\tilde{\cal H}_{\rm A}$, we introduce a non-Hermitian 
Hamiltonian ${\cal H}$ in class BDI + ${\cal S}_{-+}$ 
together with its SLS, TRS, and PHS by  
\begin{align}
\text{BDI}+{\cal S}_{-+} \!\ : {\cal H} = \begin{pmatrix}
    0 & {\rm i}\,(\tilde{\cal H}_{\rm A} - \tilde{E}) \\
    -{\rm i}\,(\tilde{\cal H}^{*}_{\rm A} - \tilde{E}) & 0
\end{pmatrix},  \!\ \!\ {\cal U}_{\cal S} = \tau_z,  \!\ \!\ 
{\cal U}_{\cal T_+} = \tau_x, \!\ \!\  {\cal U}_{\cal P_-} = \tau_0.   \label{nh-bdi-s-+}
\end{align}
Here, $\tau_x$ and $\tau_z$ are Pauli matrices. The non-Hermitian Hamiltonian
${\cal H}$ respects $\mathcal{U_{S}}{\cal H}\mathcal{U^{\dagger}_{S}} = -{\cal H}$, 
$\mathcal{U_{T_+}}{\cal H}^*\mathcal{U^{\dagger}_{T_+}} = {\cal H}$, and 
$\mathcal{U_{P_-}}{\cal H}^T\mathcal{U^{\dagger}_{P_-}} = -{\cal H}$
with $\mathcal{U_{T_+}} \mathcal{U_{T_+}^{*}} = + 1$,  $\mathcal{U_{P_-}}\mathcal{U_{P_-}^{*}} = -1$, and $[{\cal U}_{\cal S},{\cal U}_{\cal T_{+}}]_{-}=[{\cal U}_{\cal S},{\cal U}_{\cal P_{-}}]_{+}=0$. 
Thus, ${\cal H}$ belongs to class BDI + $\cal{S}_{-+}$. Symmetry-conserving 
energy of class BDI + ${\cal S}_{-+}$ is zero, $E=0$. 
${\cal H}$ in Eq.~(\ref{nh-bdi-s-+}) has right eigenmodes at the zero energy, 
$\left( \ket{\psi}^{*}~0 \right)^T$ and $\left( 0~\ket{\psi} \right)^T$, which have the same localization 
properties as the original eigenstate $\ket{\psi}$ of $\tilde{\cal H}_{\rm A}$. 
Under the symmetries in Eq.~(\ref{nh-bdi-s-+}), non-Hermitian Hamiltonians 
in class BDI + ${\cal S}_{-+}$ take following general form,
\begin{align}
{\cal H}_{{\rm BDI}+{\cal S}_{-+}} =  \left(\begin{array}{cc}
	 0  & {\rm i}\!\ \tilde{\cal G}_{\rm A} \\
    -{\rm i} \!\ \tilde{\cal G}^*_{\rm A} & 0 \\ 
\end{array}\right),
\end{align} 
with general Hermitian Hamiltonian $\tilde{\cal G}_{\rm A}$. 
With $\tilde{\cal H}_{\rm A}$ in Eq.~(\ref{a}), 
Eq.~(\ref{nh-bdi-s-+}) takes this general form.

\paragraph{From Hermitian class A to non-Hermitian class CI + ${\cal S}_{-+}$:}
For a general Hermitian Hamiltonian $\tilde{\cal H}_{\rm A}$ in class A, we 
introduce a non-Hermitian Hamiltonian ${\cal H}$ in 
class CI + ${\cal S}_{-+}$ together with its SLS, TRS, and PHS by 
\begin{align}
\text{CI}+{\cal S}_{-+} \!\ : 
{\cal H} = \begin{pmatrix}
    0 & \tilde{\cal H}_{\rm A}-\tilde{E} \\ 
    - \sigma_y\,(\tilde{\cal H}^{*}_{\rm A} - \tilde{E})\,\sigma_y & 0
\end{pmatrix}, \!\ \!\ \!\ 
\mathcal{U_{S}} = \tau_z, \!\ \!\ 
\mathcal{U_{T_+}} = \tau_y \otimes \sigma_y,
\!\ \!\ \mathcal{U_{P_-}} = \sigma_y. \label{nh-ci-s-+}
\end{align}
Then, ${\cal H}$ respects
$\mathcal{U_{S}}{\cal H}\mathcal{U^{\dagger}_{S}} = -{\cal H}$, 
$\mathcal{U_{T_+}}{\cal H}^*\mathcal{U^{\dagger}_{T_+}} = {\cal H}$, 
and
$\mathcal{U_{P_-}}{\cal H}^T\mathcal{U^{\dagger}_{P_-}} = -{\cal H}$
with $\mathcal{U_{T_+}}\mathcal{U_{T_+}^{*}} = +1$,
$\mathcal{U_{P_-}}\mathcal{U_{P_-}^{*}} = -1$,
and $[{\cal U}_{\cal S},{\cal U}_{\cal T_{+}}]_{-}=[{\cal U}_{\cal S},{\cal U}_{\cal P_{-}}]_{+}=0$. 
Thus, ${\cal H}$ belongs to class CI + $\cal{S}_{-+}$. Here we assume that the dimension of 
${\cal H}_{\rm A}$ is even so that ${\cal H}_{\rm A}$ can be regarded as a Hamiltonian with 
local pseudospin-$\frac{1}{2}$ degree of freedom. 
$\sigma_y$ in Eq.~(\ref{nh-ci-s-+}) flips the spin locally. Symmetry-conserving energy of  
class CI + ${\cal S}_{-+}$ is zero, $E=0$. ${\cal H}$ in Eq.~(\ref{nh-ci-s-+})
 has right eigenmode at the zero energy,  
$\left( \sigma_y \ket{\psi}^{*}~0\right)^T$ and $\left( 0~\ket{\psi}\right)^T$, which have the 
same localization properties as the original eigenstate $\ket{\psi}$ of $\tilde{\cal H}_{\rm A}$.
Under the symmetries in Eq.~(\ref{nh-ci-s-+}), non-Hermitian Hamiltonians 
in class CI + ${\cal S}_{-+}$ take following general form,
\begin{align}
{\cal H}_{{\rm CI}+{\cal S}_{-+}} =  \left(\begin{array}{cc}
	 0  & \tilde{\cal G}_{\rm A} \\
    -\sigma_y\tilde{\cal G}^*_{\rm A} \sigma_y & 0 \\ 
\end{array}\right),
\end{align} 
with general Hermitian Hamiltonian $\tilde{\cal G}_{\rm A}$. 
Eq.~(\ref{nh-ci-s-+}) takes this general form.  

\bigskip
In summary, for 
a generic Hermitian class-A Hamiltonian $\tilde{\cal H}_{\rm A}$ and its eigenstate 
$|\psi\rangle$, we construct a 
non-Hermitian Hamiltonian ${\cal H}$ in its corresponding symmetry classes 
whose right eigenmode $|\phi_r\rangle$ at symmetry-conserving energy $E$ 
has the same localization properties as $|\psi\rangle$ of $\tilde{\cal H}_{\rm A}$. 
The non-Hermitian Hamiltonian thus introduced  
takes a generic Hamiltonian form of respective non-Hermitian symmetry classes.  

\bigskip
In the next subsection, we consider Hermitian class AI case, whose corresponding 
non-Hermitian symmetry classes are class CI, BDI$^{\dagger}$, and CI + ${\cal S}_{+-}$. 
A general Hermitian class-AI Hamiltonian $\tilde{\cal H}_{\rm AI}$ is given in Eq.~(\ref{ai}), while
Hamiltonians in these non-Hermitian symmetry classes ${\cal H}$ 
are given by Eqs.~(\ref{nh-ci},\ref{nh-bdi^dagger},\ref{nh-ci-s+-}) respectively. The 
symmetry-conserving energies $E$ in these non-Hermitian classes 
are $E=0$, $E={\rm i}E_{\rm i}$ with $E_{\rm i}\in \mathbb{R}$, 
and $E=0$ respectively.   

\subsubsection{Hermitian class AI}
A generic Hermitian Hamiltonian $\tilde{\cal H}_{\rm AI}$ in class AI respects 
TRS ${\cal V_{T}}\tilde{\cal H}_{\rm AI} {\cal V_{T}^{\dagger}} = 
\tilde{\cal H}^*_{\rm AI}$ with a unitary matrix ${\cal V_{T}}$ satisfying ${\cal V_{T}}{\cal V_{T}^* } = 1$. 
Without loss of generality, 
$\tilde{\cal H}_{\rm AI}$ and ${\cal V}_{\cal T}$ can be put into the following form, 
\begin{align}
\tilde{\cal H}_{\rm AI} \equiv \left(\begin{array}{cc}
	 \tilde{h}_1 & h_{12} \\
	h_{12}^{\dagger} & \tilde{h}_2
\end{array}\right), \!\ \!\ {\cal V}_{\cal T} = \left(\begin{array}{cc}
	 1_{n_1\times n_1} & 0 \\
	0 & -1_{n_2\times n_2} \end{array}\right),  \label{ai}
\end{align}
with $\tilde{h}_{1}^{\dag} = \tilde{h}_{1}=\tilde{h}_{1}^*$, 
$\tilde{h}_{2}^{\dag} = \tilde{h}_{2}=\tilde{h}_{2}^*$, and $h^*_{12} = -h_{12}$. 
Let $n_1$ and $n_2$ be the dimensions of the square matrices $\tilde{h}_1$ and $\tilde{h}_2$, respectively.

\paragraph{From Hermitian class AI to non-Hermitian class CI:}
For a generic Hermitian Hamiltonian $\tilde{\cal H}_{\rm AI}$ in class AI with even rows and columns, 
we  construct a non-Hermitian Hamiltonian ${\cal H}$ in class CI whose $|\phi_r\rangle$ shares 
the same localization properties with $|\psi\rangle$ of $\tilde{\cal H}_{\rm AI}$. 
Since the rows and columns are assumed to be even, we can take $n_1 = n_2 \equiv n$ 
without loss of generality. Then, under an appropriate unitary transformation, we can choose 
$\tilde{\cal H}_{\rm AI}$ and ${\cal V}_{\cal T}$ to be
\begin{align}
\tilde{\cal H}_{\rm AI} \equiv \left(\begin{array}{cc}
	 \tilde{h}^{\prime}_1 & h^{\prime}_{12} \\
	h^{\prime \dagger}_{12} & \tilde{h}^{\prime}_2
\end{array}\right), \!\ \!\ {\cal V}_{\cal T} = \left(\begin{array}{cc}
	0 & 1_{n\times n} \\
	1_{n\times n} & 0 \end{array}\right) \equiv \sigma_x, \label{ai2}
\end{align}  
with  $\tilde{h}_{1}^{\prime \dag} = \tilde{h}^{\prime}_{1}=\tilde{h}_{2}^{\prime *}$, 
$\tilde{h}_{2}^{\prime \dag} = \tilde{h}^{\prime}_{2}=\tilde{h}_{1}^{\prime *}$, and 
$h^{\prime {\rm T}}_{12} = h^{\prime}_{12}$. 
Let $\tilde{E} \in \mathbb{R}$ be an eigenenergy of $\tilde{\cal H}_{\rm AI}$ in Eq.~(\ref{ai2}) 
and $\ket{\psi}$ be the corresponding eigenstate. We introduce a non-Hermitian Hamiltonian ${\cal H}$ in 
class CI and its symmetry operations, 
\begin{align}
{\rm CI} \!\ : \!\ {\cal H} = \left(\begin{array}{cc}
	 -{\rm i}\,( \tilde{h}^{\prime}_1 - \tilde{E} ) & {\rm i} h^{\prime}_{12} \\
	-{\rm i}h_{12}^{\prime \dagger} & 
	{\rm i}\,( \tilde{h}^{\prime}_2 - \tilde{E} )
\end{array}\right), \!\ \!\ {\cal U_{T_+}} = \sigma_x, \!\ \!\ {\cal U_{P_-}} = \sigma_y,  
\label{nh-ci}
\end{align}
with ${\cal U_{T_+}}{\cal H}^*{\cal U_{T_+}^{\dagger}}={\cal H}$, 
${\cal U_{P_-}}{\cal H}^{\rm T}{\cal U_{P_-}^{\dagger}}= - {\cal H}$, 
${\cal U_{T_+}}{\cal U^*_{T_+}} = 1$, and ${\cal U_{P_-}}{\cal U^*_{P_-}} = -1$. 
Symmetry-conserving energy of class CI is zero, $E=0$.  
${\cal H}$ in Eq.~(\ref{nh-ci}) has a zero-energy right eigenmode $\sigma_z \ket{\psi}$, which has the same 
localization properties as the original eigenstate $\ket{\psi}$ of $\tilde{\cal H}_{\rm AI}$. Under 
the symmetries in Eq.~(\ref{nh-ci}), a generic non-Hermitian Hamiltonian in class CI 
takes a following Hamiltonian form, 
\begin{align}
{\cal H}_{{\rm CI}} =  \left(\begin{array}{cc}
	 {\rm i}\!\ \tilde{g}  & f  \\
     {f}^{\dagger} & - {\rm i} \!\ \tilde{g}^* \\ 
\end{array}\right).
\end{align}
Here $g$ and $f$ are $n$ by $n$ square matrices satisfying 
$\tilde{g}^{\dagger}=\tilde{g}$ and $f^T=f$. With Eq.~(\ref{ai2}), one can see 
that Eq.~(\ref{nh-ci}) indeed takes this general form. 

\paragraph{From Hermitian class AI to non-Hermitian class BDI$^{\dagger}$:} 
For the given Hermitian Hamiltonian $\tilde{\cal H}_{\rm AI}$ in Eq.~(\ref{ai}), 
we construct a non-Hermitian Hamiltonian ${\cal H}$ in class BDI$^{\dagger}$ 
whose $|\phi_r\rangle$ shares the same localization properties with $|\psi\rangle$ of 
$\tilde{\cal H}_{\rm AI}$. The symmetry-conserving 
energy of class BDI$^{\dagger}$ takes pure imaginary values, 
$E={\rm i} E_{\rm i}$. Let 
$\tilde{E} \in \mathbb{R}$ be an eigenenergy of $\tilde{\cal H}_{\rm AI}$ in Eq.~(\ref{ai}) 
and $\ket{\psi}$ be the corresponding eigenstate. We introduce a non-Hermitian 
Hamiltonian ${\cal H}$ in class BDI$^{\dagger}$ and its symmetry 
operations by
\begin{align}
{\rm BDI}^{\dagger} \!\ : \!\ {\cal H} = 
\left(\begin{array}{cc}
	 -{\rm i}\,( \tilde{h}_1 - \tilde{E})  + {\rm i} E_{\rm i}
	 & {\rm i}h_{12} \\
	 -{\rm i}h_{12}^{\dagger} & 
	 {\rm i}\,( \tilde{h}_2 - \tilde{E}) + {\rm i} E_{\rm i}
\end{array}\right), \!\ \!\ \!\ {\cal U}_{\cal P_{+}} = 1, 
\!\ \!\ {\cal U}_{\cal T_{-}} = \left(\begin{array}{cc}
	 1_{n_1\times n_1} & 0 \\
	0 & -1_{n_2\times n_2} \end{array}\right), \label{nh-bdi^dagger}
\end{align}
with ${\cal U_{P_+}}{\cal H}^{\rm T}{\cal U_{P_+}^{\dagger}}={\cal H}$, 
${\cal U_{T_-}}{\cal H}^*{\cal U_{T_-}^{\dagger}}= - {\cal H}$, and 
${\cal U_{P_+}}{\cal U^*_{P_+}} = {\cal U_{T_-}}{\cal U^*_{T_-}} = 1$. 
${\cal H}$ has a right eigenmode at $E={\rm i} E_{\rm i}$,  
which has the same localization properties as the original 
eigenstate $\ket{\psi}$ of $\tilde{\cal H}_{\rm AI}$. In fact, $\mathcal{U}_{\cal T_{-}} \ket{\psi}$ 
is such an eigenmode, 
\begin{align}
{\cal H} \!\ \mathcal{U}_{\cal T_{-}} \ket{\psi} = {\rm i} E_{\rm i} \!\ 
\mathcal{U}_{\cal T_{-}} \ket{\psi}. 
\end{align} 
Since $\mathcal{U}_{\cal T_{-}}$ is a local transformation, $\mathcal{U}_{\cal T_{-}} \ket{\psi}$ 
and $\ket{\psi}$ share the same localization properties. Under the symmetries 
in Eq.~(\ref{nh-bdi^dagger}), non-Hermitian Hamiltonians 
in class BDI$^{\dagger}$ take the following general form,
\begin{align}
{\cal H}_{{\rm BDI}^{\dagger}} =  \left(\begin{array}{cc}
	 {\rm i}\!\ \tilde{g}_1  & {g}_{12} \\
     {g}^{\dagger}_{12} & {\rm i} \!\ \tilde{g}_2 \\ 
\end{array}\right). 
\end{align}
Here $\tilde{g}_1$ and $\tilde{g}_2$ are $n_1 \times n_1$ and 
$n_2 \times n_2$ general real symmetric Hamiltonians respectively, 
and $g_{12}$ is $n_1 \times n_2$ general real Hamiltonian $g^*_{12}=g_{12}$. 
Given $\tilde{h}_1$, $\tilde{h}_2$ and $h_{12}$ in Eq.~(\ref{ai}) and 
real $\tilde{E}$ and $E_{\rm i}$, one can see that 
Eq.~(\ref{nh-bdi^dagger}) takes this general form. 
 
\paragraph{From Hermitian class AI to non-Hermitian class CI + ${\cal S}_{+-}$:}
For generic Hermitian Hamiltonians $\tilde{\cal H}_{\rm AI}$ in Eqs.~(\ref{ai},\ref{ai2}), we construct 
a non-Hermitian Hamiltonian in class CI + ${\cal S}_{+-}$ whose $|\phi_r\rangle$ 
shares the same 
localization properties with $|\psi\rangle$ of $\tilde{\cal H}_{\rm AI}$. 
Let us have two independent  
random Hermitian Hamiltonians in class AI, $\tilde{\cal H}_{\rm AI}$ and $\tilde{\cal H}^{\prime}_{\rm AI}$. 
They have the same dimensions and respect the same symmetry as in Eq.~(\ref{ai}) or Eq.~(\ref{ai2}). 
Let $\tilde{E}\in\mathbb{R}$ and $\tilde{E}' \in \mathbb{R}$ be eigenenergies of 
$\tilde{\cal H}_{\rm AI}$ and $\tilde{\cal H}^{\prime}_{\rm AI}$, and $\ket{\psi}$ and $\ket{\psi'}$ be 
the corresponding eigenstates, respectively. Out of the two, we
introduce a non-Hermitian Hamiltonian in 
class CI + ${\cal S}_{+-}$ together with 
SLS, TRS, and PHS operations by
\begin{align}
& {\rm CI} + {\cal S}_{+-} \!\ : \!\ \nonumber \\
& \ {\cal H} = \begin{pmatrix}
    0 &\tilde{\cal H}_{\rm AI} -\tilde{E} \\
    \tilde{\cal H}^{\prime}_{\rm AI} -\tilde{E}' & 0
\end{pmatrix}, \!\ \!\ \mathcal{U_{S}} = \tau_z \otimes 1_{N \times N}, \!\ \!\ 
\mathcal{U_{T_+}} = \tau_0 \otimes {\cal V}_{\cal T}, \!\ \!\ 
\mathcal{U_{P_-}} = \tau_y \otimes {\cal V}_{\cal T}, \label{nh-ci-s+-}
\end{align}
with $\mathcal{U_{S}}{\cal H}\mathcal{U^{\dagger}_{S}} = -{\cal H}$, 
$\mathcal{U_{T_+}}{\cal H}^*\mathcal{U^{\dagger}_{T_+}} = {\cal H}$, 
$\mathcal{U_{P_-}}{\cal H}^T\mathcal{U^{\dagger}_{P_-}} = -{\cal H}$, 
$[{\cal U}_{\cal S},{\cal U}_{\cal T_{+}}]_{+}=[{\cal U}_{\cal S},{\cal U}_{\cal P_{-}}]_{-}=0$ 
and $N \equiv n_1+n_2$. ${\cal V}_{\cal T}$ is defined in Eqs.~(\ref{ai},\ref{ai2}). 
Here $\tau_0$, $\tau_y$, and $\tau_z$ act in 
the enlarged space, and $\tau_y$ exchanges 
$\tilde{\cal H}_{\rm AI}$ and $\tilde{\cal H}^{\prime}_{\rm AI}$ in ${\cal H}$. 
Symmetry-conserving energy of class CI + ${\cal S}_{+-}$ is zero, $E=0$. Then, 
${\cal H}$ in Eq.~(\ref{nh-ci-s+-}) 
has right eigenmodes at the zero energy, $( 0~\ket{\psi} )^T$ and $( \ket{\psi'}~0 )^T$, 
which have the same localization properties as the original eigenstates $\ket{\psi}$ and $\ket{\psi'}$ of 
$\tilde{\cal H}_{\rm AI}$ 
and $\tilde{\cal H}^{\prime}_{\rm AI}$. Under the symmetries in Eq.~(\ref{nh-ci-s+-}), 
non-Hermitian Hamiltonians in class CI + ${\cal S}_{+-}$ take the following general form,
\begin{align}
{\cal H}_{{\rm CI} + {\cal S}_{+-}} =  \left(\begin{array}{cc}
	 0  & \tilde{\cal G}_{\rm AI} \\
     \tilde{\cal G}^{\prime}_{\rm AI} & 0 \\ 
\end{array}\right).
\end{align}
Here two independent Hermitian Hamiltonians $\tilde{\cal G}_{\rm AI}$ 
and $\tilde{\cal G}^{\prime}_{\rm AI}$ 
respect the class AI symmetry; 
${\cal V}_{\cal T}\tilde{\cal G}_{\rm AI} {\cal V}^{\dagger}_{\cal T}=\tilde{\cal G}^*_{\rm AI}$ 
and ${\cal V}_{\cal T}\tilde{\cal G}^{\prime}_{\rm AI} {\cal V}^{\dagger}_{\cal T}=
\tilde{\cal G}^{\prime *}_{\rm AI}$. 
Since $\tilde{\cal H}_{\rm AI}$ and $\tilde{\cal H}^{\prime}_{\rm AI}$ 
are two independent class-AI Hamiltonians,
Eq.~(\ref{nh-ci-s+-})  takes this general form.

\bigskip
In summary, for a generic Hermitian class-AI Hamiltonian $\tilde{\cal H}_{\rm AI}$ and its 
eigenstate $|\psi\rangle$, we 
constructed a non-Hermitian Hamiltonian ${\cal H}$ in its  
corresponding symmetry classes whose right eigenmode $|\phi_r\rangle$ at the 
symmetry-conserving energy $E$ has the 
same localization properties as $|\psi\rangle$ of $\tilde{\cal H}_{\rm AI}$. 
The non-Hermitian Hamiltonian thus constructed 
also takes a generic Hamiltonian form of respective symmetry classes. 

\bigskip
In the next subsection, we consider Hermitian chiral unitary case, whose corresponding non-Hermitian symmetry 
classes are class A, AIII$^{\dagger}$, 
AIII + ${\cal S}_{+}$, D + ${\cal S}_{+}$, C + ${\cal S}_{+}$, and AI + ${\cal S}_{-}$. 
A generic Hamiltonian of the chiral unitary class is given in Eq.~(\ref{aiii}). 
Hamiltonians in the last five non-Hermitian classes are given by 
Eqs.~(\ref{nh-aiii^dagger},\ref{nh-aiii-s+},\ref{nh-d-s+},\ref{nh-c-s+},\ref{nh-ai-s+}). 
Symmetry-conserving energy of the chiral unitary class is zero, $\tilde{E}=0$. Symmetry-conserving 
energy of non-Hermitian class A takes a complex value, while symmetry-conserving energies of 
the other non-Hermitian classes are all zero, $E=0$. 

\subsubsection{Hermitian class AIII}
A generic Hermitian Hamiltonian $\tilde{\cal H}_{\rm AIII}$ in class AIII respects 
CS ${\cal V_{S}}\tilde{\cal H}_{\rm AIII} {\cal V_{S}^{\dagger}} = -\tilde{\cal H}_{\rm AIII}$. 
Without loss of generality, it takes the 
following form with ${\cal V_{S}} = \sigma_z$,
\begin{align}
\tilde{\cal H}_{\rm AIII} = \begin{pmatrix}
    0 & h \\
    h^{\dagger} & 0
\end{pmatrix}. \label{aiii}
\end{align}
$h$ is a general non-Hermitian Hamiltonian (i.e., $h^{\dag} \neq h$)
and has no symmetries. The symmetry-conserving energy of class AIII is zero, $\tilde{E}=0$.  
We assume the presence of zero modes in $\tilde{\cal H}_{\rm AIII}$. Because of CS, a 
zero mode with positive (negative) chirality can be written as 
$( \ket{\psi_{+}}~0 )^T$ [$( 0~\ket{\psi_{-}} )^T$], satisfying 
$h^{\dag} \ket{\psi_+} = 0$ ($h \ket{\psi_{-}} = 0$).

There clearly exists a non-Hermitian Hamiltonian in class A whose right 
eigenstate at complex energy $E$ shares the same localization properties with the zero mode   
of $\tilde{\cal H}_{\rm AIII}$; $h^{\prime} \equiv h+E$, 
$h^{\prime} \ket{\psi_{-}} = E \ket{\psi_{-}}$.  
Since $h$ is general non-Hermitian, $h^{\prime}$ also 
takes the general Hamiltonian form of the non-Hermitian symmetry class A.   

\paragraph{From Hermitian class AIII to non-Hermitian class AIII$^{\dagger}$:}
For generic Hermitian Hamiltonians in class AIII, we construct 
a non-Hermitian Hamiltonian ${\cal H}$ in class AIII$^{\dag}$ that shares the same localization 
properties of zero modes. To this end, we consider two independent Hermitian random 
Hamiltonians in class AIII,   
\begin{align}
\tilde{\cal H}_{\rm AIII} = \begin{pmatrix}
    0 & h \\
    h^{\dagger} & 0
\end{pmatrix}, \!\ \!\  \!\ \tilde{\cal H}^{\prime}_{\rm AIII} = \begin{pmatrix}
    0 & h^{\prime} \\
    h^{\prime \dagger} & 0
\end{pmatrix},  \label{aiii-b}
\end{align}
and assume the presence of zero modes in 
$\tilde{\cal H}_{\rm AIII}$ and/or $\tilde{\cal H}^{\prime}_{\rm AIII}$. 
let $( 0~\ket{\psi_-} )^T$ [$( 0~\ket{\psi_{-}^{'}} )^T$] be a zero mode of $\tilde{\cal H}_{\rm AIII}$ 
($\tilde{\cal H}^{\prime}_{\rm AIII}$) with negative chirality. Out of these two, 
we introduce a non-Hermitian Hamiltonian 
${\cal H}$ that belongs to class AIII$^{\dagger}$,  
\begin{align}
{\rm AIII}^{\dagger} \!\ : \!\ {\cal H} = \begin{pmatrix}
    0 & h \\
    h^{\prime} & 0
\end{pmatrix}.  \label{nh-aiii^dagger}
\end{align}
By construction, 
${\cal H}$ has SLS ${\cal U_S} {\cal H} {\cal U^{\dagger}_S}= -{\cal H}$ with ${\cal U_S} = \tau_z$  
and belongs to class AIII$^{\dagger}$. Symmetry-conserving energy 
of class AIII$^{\dagger}$ is zero, $E=0$. ${\cal H}$ in Eq.~(\ref{nh-aiii^dagger}) has right 
eigenmodes at the zero energy,  
$( 0~\ket{{\psi}_{-}} )^T$ and/or $( \ket{{\psi}^{\prime}_{-}}~0 )^T$, which have the 
same localization properties as the zero modes of ${\cal H}_{\rm AIII}$ and  
${\cal H}^{\prime}_{\rm AIII}$. Since $h$ and $h^{\prime}$ are two independent 
non-Hermitian Hamiltonians,  Eq.~(\ref{nh-aiii^dagger}) takes a general Hamiltonian form of 
class AIII$^{\dagger}$ under the SLS with ${\cal U_S} = \tau_z$.

\paragraph{From Hermitian class AIII to non-Hermitian class AIII + ${\cal S}_{+}$:}
For the given Hermitian Hamiltonian $\tilde{\cal H}_{\rm AIII}$ in Eq.~(\ref{aiii}), we 
consider the following non-Hermitian Hamiltonian ${\cal H}$, 
\begin{align}
{\rm AIII} + {\cal S}_{+} \!\ : \!\ 
{\cal H} = \begin{pmatrix}
    0 & h \\
    -h^{\dagger} & 0
\end{pmatrix}, \!\ \!\ \!\ \!\ 
{\cal U_C} = \tau_0, \!\ {\cal U_S} = \tau_z. \label{nh-aiii-s+}
\end{align}
${\cal H}$ satisfies CS 
${\cal U_C} {\cal H}^{\dagger} {\cal U^{\dagger}_C}=-{\cal H}$ 
and SLS ${\cal U_S} {\cal H} {\cal U^{\dagger}_S}=-{\cal H}$ and hence 
belongs to class AIII + ${\cal S}_+$. Symmetry-conserving energy of class 
AIII + ${\cal S}_+$ is zero, $E=0$. ${\cal H}$ in Eq.~(\ref{nh-aiii-s+})  
has right eigenmodes at the zero energy, $(\ket{\psi_{+}}~0)^T$ and $(0~\ket{\psi_{-}})^T$, 
which are identical to the original zero modes of $\tilde{\cal H}_{\rm AIII}$. Since $h$ 
is a general non-Hermitian Hamiltonian from Eq.~(\ref{aiii}), ${\cal H}$ in 
Eq.~(\ref{nh-aiii-s+}) takes a general form of class AIII + ${\cal S}_{+}$ 
under the CS and SLS symmetries.

\paragraph{From Hermitian class AIII to non-Hermitian class D + ${\cal S}_{+}$:}
For the given Hermitian Hamiltonian $\tilde{\cal H}_{\rm AIII}$ in Eq.~(\ref{aiii}), 
we introduce the following non-Hermitian Hamiltonian ${\cal H}$ in class D + ${\cal S}_+$ 
and its symmetry operations, 
\begin{align}
\text{D}+{\cal S}_+ \!\ : \!\ {\cal H} = \begin{pmatrix}
    0 & h \\
    h^{\rm T}& 0
\end{pmatrix}, \!\ \!\  {\cal U_{P_-}} = {\cal U_{S}} = \tau_z. \label{nh-d-s+}
\end{align}
${\cal H}$ has PHS ${\cal U_{P_-}} {\cal H}^{\rm T}{\cal U^{\dagger}_{P_-}} 
= - {\cal H}$, TRS$^{\dagger}$ $ {\cal H}^{\rm T} 
= {\cal H}$, and SLS ${\cal U_{S}} {\cal H}{\cal U^{\dagger}_{S}} = - {\cal H}$, 
and hence belongs to class D + ${\cal S}_+$. Symmetry-conserving energy of 
class D + ${\cal S}_+$ is zero, $E=0$. ${\cal H}$ in Eq.~(\ref{nh-d-s+}) 
has right eigenmodes at the zero energy, 
$(\ket{\psi_{+}}^{*}~0)^T$ and $(0~\ket{\psi_-})^T$, which have the same localization 
properties as the original zero modes $(\ket{\psi_{+}}~0)^T$ and $(0~\ket{\psi_-})^T$ 
of $\tilde{\cal H}_{\rm AIII}$. Since $h$ 
is a general Hamiltonian from Eq.~(\ref{aiii}),
${\cal H}$ in Eq.~(\ref{nh-d-s+}) takes a 
general form of class D + ${\cal S}_{+}$ 
under the PHS, TRS$^{\dagger}$ and SLS symmetries.

\paragraph{From Hermitian class AIII to non-Hermitian class C + ${\cal S}_{+}$:}
For the given Hermitian Hamiltonian $\tilde{\cal H}_{\rm AIII}$ in Eq.~(\ref{aiii}), 
we define the following non-Hermitian Hamiltonian ${\cal H}$ in 
class C + ${\cal S}_+$ and its symmetry operations,
\begin{align}
\text{C}+{\cal S}_+ \!\ : \!\ {\cal H}  = \begin{pmatrix}
    0 & h \\
    \sigma_y h^{\rm T} \sigma_y
    & 0
\end{pmatrix}, \!\ \!\  
{\cal U_{P_-}} = \tau_z \otimes \sigma_y,
\!\ \!\ 
{\cal U_{P_+}} = \tau_0 \otimes \sigma_y,
\!\ \!\ {\cal U_{S}} = \tau_z \otimes 1.  \label{nh-c-s+}
\end{align}
By construction, the non-Hermitian Hamiltonian ${\cal H}$ 
has PHS ${\cal U_{P_-}} {\cal H}^{\rm T}{\cal U^{\dagger}_{P_-}} 
= - {\cal H}$, TRS$^{\dagger}$ ${\cal U_{P_+}} {\cal H}^{\rm T} {\cal U^{\dagger}_{P_+}} 
= {\cal H}$, and SLS ${\cal U_{S}} {\cal H} {\cal U^{\dagger}_{S}} = - {\cal H}$, and hence
belongs to class C + ${\cal S}_+$. Here we assume that the dimension of 
$h$ is even so that $h$ can be regarded as a Hamiltonian with 
local pseudospin-$\frac{1}{2}$ degree of freedom. 
$\sigma_y$ in Eq.~(\ref{nh-c-s+}) flips the spin locally. Symmetry-conserving 
energy of class C+${\cal S}_+$ 
is zero, $E=0$. Then ${\cal H}$ in Eq.~(\ref{nh-c-s+}) has right eigenmodes at the zero energy,  
$(\sigma_y \ket{\psi_{+}}^{*}~0)^T$ and $(0~\ket{\psi_-})^T$, which have the same 
localization properties as the original zero modes $(\ket{\psi_{+}}~0)^T$ and $(0~\ket{\psi_-})^T$ 
of the Hermitian Hamiltonian $\tilde{\cal H}_{\rm AIII}$. Since $h$ 
is a general Hamiltonian from Eq.~(\ref{aiii}), ${\cal H}$ in 
Eq.~(\ref{nh-c-s+}) takes a general Hamiltonian form of class C + ${\cal S}_{+}$ 
under the PHS, TRS$^{\dagger}$ and SLS symmetries.  

\paragraph{From Hermitian class AIII to non-Hermitian class AI + ${\cal S}_{-}$:}
For the given Hermitian Hamiltonian $\tilde{\cal H}_{\rm AIII}$ in Eq.~(\ref{aiii}), 
we introduce the following non-Hermitian Hamiltonian in class AI + ${\cal S}_-$ 
and its symmetry operations,
\begin{align}
\text{AI}+{\cal S}_- \!\ : \!\ {\cal H} = 
\begin{pmatrix}
    0 & h \\
    h^* & 0
\end{pmatrix}, \!\ \!\ {\cal U_{T_+}} = \tau_x, \!\ \!\  {\cal U_{T_-}} = \tau_y, \!\ \!\
{\cal U_{S}} = \tau_z. \label{nh-ai-s-}
\end{align}
${\cal H}$ belongs to class AI + ${\cal S}_{-}$ because of TRS 
${\cal U_{T_+}}{\cal H}^*{\cal U^{\dagger}_{T_+}}={\cal H}$, 
PHS$^{\dagger}$ ${\cal U_{T_-}}{\cal H}^*{\cal U^{\dagger}_{T_-}}=-{\cal H}$,  
SLS (${\cal U_{S}}{\cal H}{\cal U^{\dagger}_{S}}=-{\cal H}$), and the 
anti-commutation relation between ${\cal U_{T_+}}$ and ${\cal U}_{\cal S}$. 
Symmetry-conserving energy of class AI + ${\cal S}_{-}$ is zero, $E=0$.  
${\cal H}$ in Eq.~(\ref{nh-ai-s-}) has  
right eigenmodes at the zero energy, $(\ket{\psi_{-}}^{*}~0)^T$ and $(0~\ket{\psi_-})^T$, 
which have the same localization properties as the original zero mode 
$(0~\ket{\psi_-})^T$ of the Hermitian Hamiltonian $\tilde{\cal H}_{\rm AIII}$. ${\cal H}$ in 
Eq.~(\ref{nh-ai-s-}) takes a general Hamiltonian form of class AI + ${\cal S}_{-}$ 
under the TRS, PHS$^{\dagger}$ and SLS symmetries.  

\bigskip
In the next subsection, we consider Hermitian chiral orthogonal case, whose corresponding 
non-Hermitian symmetry classes are class AI, 
AI + ${\cal S}_{+}$, BDI + ${\cal S}_{++}$, and CI + ${\cal S}_{++}$. 
A generic Hamiltonian of chiral orthogonal class is given in Eq.~(\ref{bdi}). 
Hamiltonians in the last three non-Hermitian classes are given by 
Eqs.~(\ref{nh-ai-s+},\ref{nh-bdi-s++},\ref{nh-ci-s++}). 
Symmetry-conserving energy of the chiral orthogonal class is zero, $\tilde{E}=0$. 
Symmetry-conserving energy of non-Hermitian 
class AI takes a real value ($E= E^*$), while symmetry-conserving energies  
of the other non-Hermitian classes are all zero, $E=0$. 

\subsubsection{Hermitian class BDI}
A Hermitian Hamiltonian $\tilde{\cal H}_{\rm BDI}$ in class BDI 
respects TRS, PHS, and CS, where the signs of TRS and PHS are $+1$.
Under an appropriate unitary transformation, $\tilde{\cal H}$ can be put into the 
following form together with the three symmetry operations, 
\begin{align}
\tilde{\cal H}_{\rm BDI}=\begin{pmatrix}
    0 & h \\
    h^{\dagger} & 0
\end{pmatrix},  \!\ \!\ {\cal V}_{\cal T}=1, \!\ \!\ {\cal V}_{\cal P} = {\cal V}_{\cal S} = \sigma_z.   
\label{bdi}
\end{align}
$h$ is real non-Hermitian satisfying $h = h^*$. 
$\tilde{\cal H}_{\rm BDI}$ respects 
$\tilde{\cal H}^*_{\rm BDI} = \tilde{\cal H}_{\rm BDI}$ and ${\cal V}_{\cal P}\tilde{\cal H}^T_{\rm BDI} 
{\cal V}_{\cal P} = {\cal V}_{\cal S}\tilde{\cal H}_{\rm BDI} {\cal V}_{\cal S} = - \tilde{\cal H}_{\rm BDI}$. 
Symmetry-conserving energy of BDI class is zero, $\tilde{E}=0$. Let us assume the 
presence of zero modes in $\tilde{\cal H}_{\rm BDI}$. 
A zero mode with positive (negative) chirality can be written as 
$( \ket{\psi_{+}}~0 )^T$ [$( 0~\ket{\psi_{-}} )^T$], satisfying $h^{\dag} \ket{\psi_+} = 0$ 
($h \ket{\psi_{-}} = 0$).

Then, there exists a non-Hermitian Hamiltonian in class AI, whose 
right eigenmode at real energy $E$ shares the same 
localization properties with the zero mode of $\tilde{\cal H}_{\rm BDI}$; 
$h^{\prime}\equiv h+E$, $h^{\prime}\ket{\psi_{-}}=E \ket{\psi_{-}}$. Since 
$h$ is general real non-Hermitian 
Hamiltonian and the symmetry-conserving energy $E$ is real in class AI, 
$h^{\prime}\equiv h+E$ also takes the general form of non-Hermitian 
Hamiltonian in the symmetry class AI.

\paragraph{From Hermitian class BDI to non-Hermitian class AI + ${\cal S}_{+}$:}
Symmetry-conserving energy of class AI + ${\cal S_+}$ 
is zero, $E=0$. From generic Hermitian Hamiltonians in class BDI, we construct 
a non-Hermitian Hamiltonian ${\cal H}$ in class AI + ${\cal S}_{+}$ 
whose right eigenmode at the zero energy 
shares the same localization properties with the zero modes of the BDI Hamiltonians. 
To this end, we consider 
two independent Hermitian Hamiltonians in class BDI, 
\begin{align}
\tilde{\cal H}_{\rm BDI}=\begin{pmatrix}
    0 & h \\
    h^{\dagger} & 0
\end{pmatrix}, \!\ \!\ \tilde{\cal H}^{\prime}_{\rm BDI}=\begin{pmatrix}
    0 & h^{\prime} \\
    h^{\prime \dagger} & 0
\end{pmatrix}, \label{two-bdi}
\end{align} 
with $h= h^*$, $h^{\prime} = h^{\prime *}$, 
$h\ne h^{\dagger}$, and $h^{\prime} \ne h^{\prime \dagger}$ and assume 
the presence of their zero modes. Let $(0~\ket{\psi_{-}})^T$ and 
$(0~\ket{\psi'_{-}})^T$ be zero modes of $\tilde{\cal H}_{\rm BDI}$ 
and $\tilde{\cal H}^{\prime}_{\rm BDI}$ with 
negative chirality, respectively. From Eq.~(\ref{two-bdi}), we construct  
the following  non-Hermitian Hamiltonian: 
\begin{align} 
\text{AI}+{\cal S}_+ \!\ : {\cal H} = \begin{pmatrix}
    0 & h \\
    h^{\prime} & 0
\end{pmatrix}. \label{nh-ai-s+}
\end{align}
${\cal H}$ belongs to class AI + ${\cal S_+}$ because of TRS ${\cal H}^*={\cal H}$ and SLS 
$\tau_z {\cal H} \tau_z = -{\cal H}$. ${\cal H}$ in Eq.~(\ref{nh-ai-s+}) has right eigenmodes at the zero energy,  
$(0~\ket{\psi_-})^T$ and $(\ket{\psi'_-}~0)^T$, which share the same localization properties as 
the original zero modes of $\tilde{\cal H}_{\rm BDI}$ and $\tilde{\cal H}^{\prime}_{\rm BDI}$.
Since real $h$ and $h^{\prime}$  
are independent of each other, ${\cal H}$ in Eq.~(\ref{nh-ai-s+}) takes a general 
Hamiltonian form of class AI + ${\cal S}_{+}$ under the TRS and SLS symmetries.

\paragraph{From Hermitian class BDI to non-Hermitian class BDI + ${\cal S}_{++}$:}
From the given Hermitian Hamiltonian $\tilde{\cal H}_{\rm BDI}$ in Eq.~(\ref{bdi}), 
we introduce the following non-Hermitian Hamiltonian,  
\begin{align} 
\text{BDI}+{\cal S}_{++} \!\ : \!\ 
{\cal H}= \begin{pmatrix}
    0 & h \\
    - h^{\dag} & 0
\end{pmatrix}. \label{nh-bdi-s++}
\end{align}
${\cal H}$ has TRS 
with ${\cal U_{T_+}} = 1$ (${\cal H}^* = {\cal H}$),
PHS with ${\cal U_{P_-}} = \tau_0$ 
(${\cal U_{P_-}} {\cal H}^{\rm T} {\cal U^{\dagger}_{P_-}} = -{\cal H}$), and 
SLS with ${\cal U_{S}} = \tau_z$ 
(${\cal U_{S}} {\cal H} {\cal U^{\dagger}_{S}} = -{\cal H}$), and  
hence belongs to class BDI + ${\cal S}_{++}$. Symmetry-conserving energy of class 
BDI + ${\cal S}_{++}$ is zero, $E=0$. ${\cal H}$ in Eq.~(\ref{nh-bdi-s++}) has right 
eigenmodes at the zero energy, $(\ket{\psi_{+}}~0)^T$ and $(0~\ket{\psi_-})^T$, which are identical to the 
original zero modes of $\tilde{\cal H}_{\rm BDI}$. Since $h$ is real 
non-Hermitian without any other symmetries, ${\cal H}$ in Eq.~(\ref{nh-bdi-s++}) takes 
a generic Hamiltonian form of class BDI + ${\cal S}_{++}$ under the TRS, PHS and 
SLS symmetries. 

\paragraph{From Hermitian class BDI to non-Hermitian class CI + ${\cal S}_{++}$:}
From the given Hermitian Hamiltonian $\tilde{\cal H}_{\rm BDI}$ in Eq.~(\ref{bdi}), 
we consider the following non-Hermitian Hamiltonian, 
\begin{align}
\text{CI}+{\cal S}_{++} \!\ : \!\ {\cal H} 
= \begin{pmatrix}
    0 & h \\
    -\sigma_y h^{\dag} \sigma_y
    & 0
\end{pmatrix}. \label{nh-ci-s++}
\end{align}
${\cal H}$ respects TRS with ${\cal U_{T_+}} = 1$ (${\cal H}^* = {\cal H}$),  PHS with 
${\cal U_{P_-}} = \sigma_y$ (${\cal U_{P_-}} {\cal H}^{\rm T} {\cal U^{\dagger}_{P_-}} 
= -{\cal H}$), and SLS with ${\cal U_{S}} = \tau_z$ 
(${\cal U_{S}} {\cal H} {\cal U^{\dagger}_{S}} = -{\cal H}$),  and hence 
belongs to class CI + ${\cal S}_{++}$. 
Here we assume that the dimension of 
$h$ is even so that ${\cal H}_{\rm A}$ can be regarded as a Hamiltonian with 
local pseudospin-$\frac{1}{2}$ degree of freedom. 
$\sigma_y$ in Eq.~(\ref{nh-ci-s++}) flips the spin locally.
Symmetry-conserving energy of class CI + ${\cal S}_{++}$ 
is zero, $E=0$. ${\cal H}$ in Eq.~(\ref{nh-ci-s++}) has right eigenmodes at the 
zero energy, $(\sigma_y\ket{\psi_{+}}~0)^T$ 
and $(0~\ket{\psi_-})^T$, which have the same localization properties as the original zero modes 
$(\ket{\psi_{+}}~0)^T$ and $(0~\ket{\psi_-})^T$ of $\tilde{\cal H}_{\rm BDI}$. 
With real non-Hermitian $h$, ${\cal H}$ in Eq.~(\ref{nh-ci-s++}) 
takes a generic Hamiltonian form of class  CI + ${\cal S}_{++}$ under the TRS, PHS and 
SLS symmetries. 

\subsubsection{Other Hermitian symmetry classes}
For general Hermitian Hamiltonians in classes AII, C, and D, 
where only one anti-unitary symmetry is relevant,
we can construct the corresponding non-Hermitian Hamiltonians in a 
similar manner to class AI. For general Hermitian Hamiltonians
in classes CI, CII, and DIII, where two anti-unitary symmetries are relevant,
we can construct the corresponding non-Hermitian 
Hamiltonians in a similar manner to class BDI.

\section{Model, localization length, and polynomial fitting}
\begin{table*}[tb]
	\centering
	\caption{Polynomial fitting results for the normalized localization lengths around 
		the Anderson transition points for three-dimensional 
		classes AI, AII, and AII$^{\dagger}$
		at different 
		complex energies $E$. 
		The goodness of fit (GOF), critical disorder $W_{c}$, critical 
		exponents $\nu$, scaling dimensions $-y$ of the least irrelevant scaling variable, 
		and critical normalized localization length $\Lambda_c$ are 
		shown for the different system sizes and for the different 
		orders of the 
		expansion $(m_1,n_1,m_2,n_2)$. 
		The square bracket is the 95\% confidence interval.}
	\begin{tabular}{c|c|cccccccccc}
		\hline\hline
		~Class~ & ~$E$~	& ~$L$~ & ~$m_1$ & $n_1$ & $m_2$ & $n_2$~ & ~GOF~ & ~$W_c$~ & ~$\nu$~ & ~$y$~ & ~$\Lambda_c$~   \\
		\hline
		\multirow{4}*{AI}&	\multirow{2}*{$0.5\text{i}$}
		&8-20&3&3&0&1&0.12&12.842[12.834, 12.852]&0.988[0.965, 1.008]&0.94[0.78, 1.10]&0.584[0.571, 0.593]\\
		&	&10-20&3&3&0&1&0.24&12.841[12.835, 12.847]&0.980[0.959, 0.999]&1.31[1.15, 1.48]&0.593[0.588, 0.598]\\
		\cline{2-12}
		&	\multirow{2}*{0}
		&10-20&2&3&0&1&0.21&21.540[21.471, 21.564]&0.933[0.799, 1.041]&0.512[0.468, 0.668]&0.269[0.259, 0.293]\\
		&&10-20&3&3&0&1&0.22&21.576[21.503, 21.616]&0.943[0.816, 1.068]&0.439[0.372, 0.588]&0.253[0.234, 0.282]\\
		\hline
		\multirow{8}*{AII}&	\multirow{4}*{$\text{i}$}
		&4-18&1&3&0&1&0.69&8.068[8.063, 8.072]&1.021[0.997, 1.042]&0.48[0.42, 0.54]&0.528[0.505, 0.548]\\
		&	&4-18&2&3&0&1&0.70&8.067[8.062, 8.072]&1.021[0.999, 1.041]&0.49[0.43, 0.55]&0.532[0.510, 0.551]\\
		&	&8-18&1&3&0&1&0.30&8.066[8.050, 8.082]&0.996[0.879, 1.058]&0.50[0.22, 0.97]&0.537[0.397, 0.602]\\
		&	&8-18&2&3&0&1&0.25&8.055[8.042, 8.074]&1.005[0.910, 1.044]&0.79[0.32, 1.36]&0.585[0.473, 0.623]\\
		\cline{2-12}
		&	\multirow{4}*{0}
		&	6-18&2&3&0&0&0.10&6.3200[6.3194, 6.3205]&0.8741[0.8719, 0.8763]&-&0.9363[0.9355, 0.9371]\\
		&	&	6-18&3&3&0&0&0.15&6.3193[6.3187, 6.3199]&0.8791[0.8761, 0.8817]&-&0.9371[0.9362, 0.9379]\\
		&	&	8-18&2&3&0&0&0.10&6.3201[6.3193, 6.3208]&0.8745[0.8710, 0.8783]&-&0.9358[0.9346, 0.9371]\\
		&	&	8-18&3&3&0&0&0.12&6.3196[6.3187, 6.3206]&0.8762[0.8720, 0.8803]&-&0.9365[0.9350, 0.9380]\\
		\hline
		\multirow{4}*{AII$^{\dagger}$}& \multirow{4}*{0}
		&	10-18&2&3&0&1&0.11&7.706[7.703, 7.708]&0.903[0.896, 0.908]&2.65[2.25, 3.15]&0.581[0.576, 0.586]\\
		&	&	10-18&3&3&0&1&0.13&7.712[7.706, 7.720]&0.908[0.898, 0.922]&1.75[1.16, 2.44]&0.565[0.541, 0.579]\\
		&	&	12-18&2&3&0&1&0.19&7.712[7.708, 7.718]&0.899[0.876, 0.914]&1.90[1.09, 2.82]&0.566[0.542, 0.577]\\
		&	&	12-18&3&3&0&1&0.18&7.711[7.702, 7.719]&0.899[0.880, 0.914]&2.06[1.06, 4.02]&0.569[0.537, 0.589]\\
		\hline
		\hline
	\end{tabular}
	\label{table_3D}
\end{table*}

\begin{table*}[tb]
	\centering
	\caption{Polynomial fitting results for the normalized localization lengths around 
		the Anderson transition points for two-dimensional 
		classes AII$^{\dagger}$, AII, CII$^{\dagger}$, and DIII
		at different 
		complex energies $E$. 
		The goodness of fit (GOF), critical disorder $W_{c}$, critical 
		exponents $\nu$, scaling dimensions $-y$ of the least irrelevant scaling variable, 
		and critical normalized localization lengths $\Lambda_c$ are 
		shown for the different system sizes and for the different 
		orders of the expansion $(m_1,n_1,m_2,n_2)$. 
		The square bracket is the 95\% confidence interval.}
	\begin{tabular}{c|c|cccccccccc}
		\hline\hline
		~Class~ & ~$E$~	& ~$L$~ &	~$m_1$ & $n_1$ & $m_2$ & $n_2$~ & ~GOF~ & ~$W_c$~ & ~$\nu$~ & ~$y$~ & ~$\Lambda_c$~   \\
		\hline
		\multirow{4}*{AII$^{\dagger}$} & \multirow{4}*{0}
		&60-250&1&3&0&1&0.20&4.312[4.307, 4.316]&1.377[1.331, 1.439]&0.31[0.16, 0.54]&0.48[0.29, 0.61]\\
		&	&60-250&2&3&0&1&0.11&4.312[4.310, 4.315]&1.372[1.346, 1.413]&0.33[0.25, 0.41]&0.49[0.41, 0.55]\\
		&	&100-250&1&3&0&1&0.16&4.306[4.301, 4.316]&1.375[1.328, 1.567]&0.73[0.13, 1.72]&0.65[0.24, 0.73]\\
		&	&100-250&2&3&0&1&0.17&4.311[4.307, 4.317]&1.348[1.286, 1.456]&0.38[0.16, 0.66]&0.52[0.30, 0.64]\\
		\hline
		\multirow{2}*{AII}	& \multirow{2}*{$0.01\text{i}$}
		&	80-150&2&3&0&1&0.12&2.622[2.619, 2.626]&1.562[1.524, 1.609]&2.49[1.97, 3.09]&1.290[1.276, 1.303]\\
		&&	80-150&3&3&0&1&0.11&2.622[2.618, 2.626]&1.607[1.557, 1.656]&2.00[1.66, 2.47]&1.286[1.269, 1.303]\\
		\hline
		\multirow{2}*{CII$^{\dagger}$}&\multirow{2}*{0}
		&16-144 & 1 & 3 &0&0&0.30& 6.192[6.189, 6.196]&2.740[2.706, 2.773]&-& 1.852[1.848, 1.855]\\
		&&24-144&1&3&0&0&0.23&6.195[6.190, 6.199]&2.738[2.694, 2.783]&-&1.849[1.844, 1.854]\\
		\hline
		\multirow{2}*{DIII}&\multirow{2}*{0}
		&8-96 & 1   &  2 &  0 & 0  & 0.13  &  6.194[6.191, 6.197]  & 2.751[2.731, 2.772] & - &  1.850[1.846, 1.852]   \\
		&&16-96 & 1   &  2  &  0 & 0  & 0.56  &   6.193[6.189, 6.197] &2.757[2.726, 2.788] & -& 1.852[1.847, 1.855]  \\
		\hline
		\hline
	\end{tabular}
	\label{table_2D}
\end{table*}

To study the AT in class AI, we introduce 
the following O(1) tight-binding model on 2D square and 3D cubic lattices:
\begin{align}
	{H}=\sum_i \varepsilon_i c^{\dagger}_ic_i+\sum_{\langle i,j\rangle}
	V_{i,j}c^{\dagger}_ic_j,
\end{align}
where $\varepsilon_i$ is the random potential with the uniform distribution
in $[-W/2, W/2]$ with the disorder strength $W$. 
Here, $\langle i , j \rangle$ stands for 
nearest neighbor lattice sites. 
$V_{i,j}$ is set 
to $-1$ or $+1$ randomly with the equal probability.  
$V_{i,j}$ and $V_{j,i}$ are treated as independent random variables, 
so that 
Hermiticity will be broken by $V_{i,j}^* \ne V_{j,i}$. 
According to the 
symmetry classification, $H$ 
belongs to class AI with $H=H^*$.

To study the ATs in 
classes AII, AII$^{\dagger}$, CII$^{\dagger}$, and DIII, 
we introduce the following 
non-Hermitian extension of the 
SU(2) model~\cite{Asada02,Asada04,Asada05} on 2D square and 3D cubic lattices,
\begin{align}
	{H}=\sum_{i,\sigma} \varepsilon_{i,\sigma} c_{i,\sigma}^{\dagger}c_{i,\sigma}
	+\sum_{\langle i, j\rangle,\sigma,\sigma'}R(i,j)_{\sigma,\sigma'}c_{i,\sigma}^{\dagger}c_{j,\sigma'}.
\end{align}
We parametrize the matrix $R(i, j)$ as
\begin{align}
	R(i,j)=
	\begin{pmatrix}
		e^{\text{i}\alpha_{i,j}}\cos(\beta_{i,j}) & e^{\text{i}\gamma_{i,j}}\sin(\beta_{i,j}) \\
		-e^{-\text{i}\gamma_{i,j}}\sin(\beta_{i,j}) & e^{-\text{i}\alpha_{i,j}}\cos(\beta_{i,j}) \\
	\end{pmatrix}
\end{align}
with the imaginary unit i. 
Here, we distribute $\alpha_{i,j}$ and $\gamma_{i, j}$ with the uniform probability in the range $[0, 2\pi)$, and $\beta_{i, j}$ according to the probability density
$P(\beta)d\beta=\sin(2\beta)d\beta$
in the range $[0, \pi/2]$. 
The parameters in the hopping terms satisfy $\alpha_{i,j}=-\alpha_{j,i}$, $\beta_{i,j}=\beta_{j,i}$, and $\gamma_{i,j}=\gamma_{j,i}+\pi$ 
for classes AII, AII$^{\dagger}$, and CII$^{\dagger}$, 
leading to $R^{\dag}(i,j)=R(j,i)$. 
For class DIII, on the other hand, we have $\alpha_{i,j}=-\alpha_{j,i}+\pi$ and $\gamma_{i,j}=\gamma_{j,i}+\pi$, leading to $\sigma_z R^{\dag}(i,j) \sigma_z = - R(j, i)$.
We set the on-site potential as 
$\varepsilon_{j,\sigma}=\omega_{j,\sigma}^r+\text{i}\omega_{j,\sigma}^i$, 
where $\omega_{j,\sigma}^r$ and $\omega_{j,\sigma}^i$ are 
independently and uniformly distributed random numbers in 
$[-W_r/2,W_r/2]$ and $[-W_i/2,W_i/2]$, respectively.

With $\varepsilon_{j,\uparrow}=\varepsilon_{j,\downarrow}$,
$H$ satisfies $\sigma_y H^{\rm T} \sigma_y=H$.
If we set $W_r\ne0$ and $W_i\ne 0$, $H$ belongs to class 
AII$^{\dagger}$. If we set $W_r=0$ and $W_i\ne 0$, $H$ 
on the bipartite lattice also has CS, 
$\mu_z \sigma_y H^* \sigma_y 
\mu_z = - H$,  where $\mu_z$ is diagonal in 
the sublattice degrees of 
freedom, taking different signs on the 
different sublattices. Thus $H$ 
belongs to class  CII$^{\dagger}$. 
With $\varepsilon_{j,\uparrow}=\varepsilon_{j,\downarrow}^*$, 
$H$ satisfies $\sigma_y H^* \sigma_y=H$. 
If we set $W_r\ne0$ and $W_i\ne 0$, $H$ belongs to 
class AII.  If we set $W_r=0$, $W_i\ne 0$, $\varepsilon_{j,\uparrow}=-\varepsilon_{j,\downarrow}$, 
and require $\sigma_z R^{\dag}(i,j) \sigma_z = - R(j, i)$, $H$ belongs to class DIII, 
$\sigma_y{\cal H}^{*} \sigma_y={\cal H}$ and 
$\sigma_x{\cal H}^{\rm T} \sigma_x=-{\cal H}$.

In order to extract the critical exponents,  localization lengths $\xi(W,L)$
are calculated by the transfer matrix method.
We note that the transfer matrix along the transmission direction $z$ can be put into the unit matrix by proper gauge transformations. In classes AII and DIII, however, the onsite disorder is different for spin up and spin down, and the spin-dependent hopping along the transmission direction cannot be gauged away.

For class AI, the localization lengths at $E=0$ 
for different disorder strength are calculated with the transmission length $L_z=10^8$ for $L=6,8,10,12,16,18,20$ [Fig.~\ref{AI}(a)].
Similarly, the localization lengths at $E=0.5\text{i}$ are calculated with the transmission length
$L_z=10^7$ for $L=6,8,10,12,16,18,20$ [Fig.~\ref{AI}(b)].

We set $W_r=W_i=W$ for the following 
calculations of 
the SU(2) models in classes AII$^{\dagger}$ and AII.
For 2D class AII$^{\dagger}$ at $E=0$, the localization lengths are calculated with $L_z=10^8$ for $L=60,100$, $L_z=4\times10^7$ for $L=150$,
$L_z=2\times10^7$ for $L=200$, and $L_z=10^7$ for  $L=250$ [Fig.~\ref{AII_dagger_lambda}(a)].
For 3D class AII$^{\dagger}$ at $E=0$, the localization lengths are calculated with $L_z=10^7$ for $L=4,6,8,10,12,14,16$, and $L_z=6\times10^6$ for $L=18$ [Fig.~\ref{AII_dagger_lambda}(b)].   
For 2D class AII at $E=0.01\text{i}$, the localization lengths are calculated with $L_z=5\times10^7$ for $L=20, 60, 80, 100, 120, 150$ [Fig.~\ref{AII_3D_lambda}(a)].
For 3D class AII at $E=0$, the localization lengths are calculated with $L_z=10^6$ for $L=6,8,10,12,16,18$ [Fig.~\ref{AII_3D_lambda}(b)].
For 3D class AII at $E=\text{i}$, the localization length 
for $L=4,6,8,10,12,14,16,18$ has been calculated with $L_z=5\times 10^6$ [Fig.~\ref{AII_3D_lambda}(c)].

We set $W_r=0$ and $W_i=W$ for the calculations of 
the SU(2) models in classes CII$^{\dagger}$ and DIII.
For 2D class CII$^{\dag}$ at $E=0$, the localization lengths are calculated with $L_z=10^7$ for $L=16, 24, 32, 48, 64, 96,144$ [Fig.~\ref{CII^dDIII}(a)].
For 2D class DIII at $E=0$, the localization lengths are calculated with $L_z=10^7$ for $L=8,16, 24, 32, 48, 64, 96$ [Fig.~\ref{CII^dDIII}(b)].

According to the finite-size scaling, the dimensionless normalized localization length $\Lambda=\xi(W,L)/L$ follows a 
scaling function that depends on the relevant scaling variable and possibly many other 
irrelevant scaling variables. 
Empirically, we do not need to consider most of the irrelevant scaling variables, 
and numerical data of $\Lambda$ for the larger system size $L$ depend only on 
the relevant scaling variable $\phi_1$ and the least irrelevant scaling variable 
$\phi_2$:
\begin{equation}\label{key0}
	\Lambda(W,L)=f(\phi_1,\phi_2).
\end{equation} 
The scaling argument leads to 
\begin{align}
	\phi_1\equiv u_1(w) L^{1/\nu},\quad 
	\phi_2\equiv u_2(w) L^{-y} 
\end{align}
with 
$w \equiv (W-W_c)/W_c$ and some functions $u_1, u_2$. Here, $\nu$ is the 
critical exponent for the correlation length, and $-y$ is the scaling dimension 
of the least irrelevant scaling variable. Taking account of the non-linearity 
of $u_{i}(w)$ in $w$, these functions can be expanded in small $w$:
\begin{equation}\label{key1}
	u_{i}(w) \equiv \sum^{m_i}_{j=0} b_{i,j} w^{j}
\end{equation} 
with $b_{1,0}=0$.  
For sufficiently small $w$ and large $L$, the scaling function can 
be also expanded by 
\begin{equation}\label{key2}
	f(\phi_1,\phi_2) = \sum^{n_1}_{j_1=0} \sum^{n_2}_{j_2=0} a_{j_1,j_2} \phi^{j_1}_1 \phi^{j_2}_2.
\end{equation}
To fix the ambiguity, we set $a_{1,0}=a_{0,1}=1$. Then, the numerical data of $\Lambda$ 
for different $L$ and $W$ are fitted by the polynomial  
with the fitting parameters $W_c$, $\nu$, $y$, $a_{i,j}$, and $b_{i,j}$.
We minimize
\begin{equation}\label{key3}
	\chi^2 \equiv \sum^{N_D}_{k=1} 
	\frac{(\Lambda_{k}-f_{k})^2}{\sigma^2_{\Lambda_{k}}}
\end{equation}
in terms of the fitting parameters,    
where $N_D$ is the number of total data points, $\Lambda_{k}$ the 
$k$-th data point calculated with precision $\sigma_{\Lambda_{k}}$, and 
$f_{k}$ the $k$-th fitted data point by the polynomial. 
Goodness of fit~\cite{Slevin14} is
calculated to quantify the quality of the fits. 
The critical parameters, 
such as $W_c$, $\nu$, $y$, and $\Lambda_c$, 
are
estimated with 95\% confidence interval~\cite{Slevin14}.
The results are shown in Table~\ref{table_3D} for 3D and Table~\ref{table_2D} for 2D.

\section{Phase diagrams for 
	classes AI and AII}
The ATs 
at real energy and non-real energy for 
classes AI and AII belong to the different
symmetry and universality classes. In this section, we show 
the
phase diagrams for 
the O(1) model in class AI and the SU(2) model in class AII
in terms of the disorder strength $W$ and the 
imaginary part ${\rm Im}\,E$ of 
energy. 
We set ${\rm Re}\,E=0$ and take 
several values of ${\rm Im}\,E$. For each value of ${\rm Im}\,E$, we 
calculate the localization lengths with different system sizes and 
disorder strength, to determine the critical disorder for the ATs. 
This gives the phase diagrams 
shown in Fig.~\ref{phaseDiagram_AI_AII}.


As shown in Fig.~\ref{phaseDiagram_AI_AII}\,(b) and (c),
when 
the disorder strength $W$ increases, the 2D and 3D 
SU(2) models in class AII 
exhibit reentrant behavior
of the localization-delocalization-localization
transition for small ${\rm Im}\,E$. This is because 
the 
models at $W_r=W_i=W=0$ reduce to Hermitian models that have no 
eigenstates at 
nonzero ${\rm Im}\, E$. 
When weak complex-valued 
disorder 
$W$ is 
introduced in these Hermitian models, 
eigenstates acquire nonzero but small ${\rm Im}\,E$ and 
tend to be localized.
When increasing the 
disorder strength, 
the density of states (DoS) at ${\rm Im}\,E \ne 0$ increases, 
and states in this 
energy region may undergo a localization-delocalization transition. 
Further increase of the 
disorder leads to 
localization. 
Consequently, the 
reentrant localization-delocalization-localization transition 
is observed in the SU(2) models in class AII.
This situation is similar to 
reentrant behavior 
of the localization-delocalization-localization 
transition near band edges of 
Hermitian band insulators~\cite{Kramer93,Bulka87}. 

On the other hand, the 3D O(1) model in class AI 
without 
the on-site potential disorder $W$ 
is a non-Hermitian system with the random hopping $V_{i,j}^* \ne V_{j,i}$. 
This model shows no reentrant behavior but
a delocalization-localization transition 
at ${\rm Im}\,E \neq 0$ and $W=0$ 
[Fig.~\ref{phaseDiagram_AI_AII}\,(a)]. 
A delocalized phase with the larger DoS appears 
for small ${\rm Im}\,E$, and a localized phase with 
the smaller DoS appears for large ${\rm Im}\,E$ in 
the outer region. 
The mobility edge 
decreases as a function of ${\rm Im}\,E$
on introducing the on-site potential disorder $W$. 

\section{Distribution of eigenenergies, and inverse participation ratio}

The distribution of eigenenergies provides complementary information about 
criticality 
of disorder-driven quantum phase transitions 
in Hermitian systems. 
A prime example is non-Anderson transitions in 
disordered semimetals~\cite{Syzranov18}, where the DoS plays a role of 
the order parameter of a semimetal-metal quantum phase 
transition~\cite{Fradkin86}, and 
the 
scaling property of 
the DoS is characterized by a dynamical critical exponent~\cite{Kobayashi14}. The DoS in the complex energy plane 
may also provide important information about quantum 
criticality of the ATs in non-Hermitian system.
Motivated by this anticipation, we study 
the distribution of eigenenergies 
in the complex energy plane
for the O(1) model in class AI and SU(2) model in classes AII and AII$^{\dagger}$.
We numerically diagonalize the 
2D and 3D Hamiltonians with smaller system size 
under 
periodic 
boundary conditions. We take an average of the distribution over 
many different disorder realizations. To characterize eigenstates $\psi({\bm r})$, 
we also calculate the inverse participation ratio (IPR):
\begin{equation}\label{key}
	I=\frac{\sum_{\bf r}|\psi({\bf r})|^4}{\Big( \sum_{\bf r} |\psi({\bf r})|^2 \Big)^2}.
\end{equation}
For extended states, $1/I$ scales with the system's volume $L^d$, 
where $L$ is the system's length and $d$ is the spatial dimension. 
For localized states, it remains to be around $\xi^d$ 
with a localization length $\xi$.
In the following, we summarize the results 
for 3D class AI, 2D and 3D class AII, and 2D and 3D class AII$^{\dagger}$. 

\subsection{3D class AI}
In Fig.~\ref{DoS_AI}\,(a), (c), and (e),
we show the distributions of 
eigenenergies for the 3D O(1) model in class AI 
for different values of 
the disorder strength $W$. For $W=0$, 
the distribution is statistically symmetric with respect to 
the exchange between ${\rm Re} \, E$ and ${\rm Im}\, E$. 
The symmetry comes from a gauge transformation that assigns 
$+{\rm i}$ ($+1$) to one (the other) of the sublattice sites in the cubic lattice. On increasing the real-valued on-site potential disorder $W$, eigenenergies collapse 
into the real axis. 
In Fig.~\ref{DoS_AI}(b), (d), and (f), we show 
the DoS as a function of the 
imaginary part of 
eigenenergies. The DoS shows a singular peak 
on the real axis. The peak becomes sharper for larger $W$. The IPR 
shows that eigenmodes near $E=0$ 
are more delocalized than the other eigenmodes. 

\subsection{2D and 3D class AII}
Figure~\ref{AII_2D3D} shows that in the presence of 
nonzero complex-valued 
potential disorder $W$, the DoS shows a soft gap around the 
real axis 
and gets largest away from the real axis.
The soft gap in the DoS 
is observed both in 2D and 3D.
Note that the 2D 
model does not 
show any ATs at $E=0$, 
but undergoes the AT for $\mathrm{Im}\, E\ne 0$ 
[Fig.~\ref{phaseDiagram_AI_AII}(c)]; the disorder-driven localization-delocalization 
and delocalization-localization transition points are located at $W_{c,1} \simeq 0.46$ 
and at $W_{c,2} \simeq 2.62$ for $E = 0.01\text{i}$, respectively [Fig.~\ref{phaseDiagram_AI_AII}(d)]. 
On the other hand, the 3D model shows the AT at 
$E=0$ as well as ${\rm Im}\, E \ne 0$ [Fig.~\ref{phaseDiagram_AI_AII}(b)]. 
We emphasize that 
the soft gap of the DoS around the real axis appears both in 2D and 3D although the phase diagrams are qualitatively different.

\subsection{2D and 3D class AII$^{\dagger}$}
Figure~\ref{AII_dagger_2D3D} shows that 
eigenenergies in 
the 2D and 3D SU(2) models in class AII$^{\dagger}$ 
have no singular structure 
around the real axis 
and 
are quite equally distributed in 
the complex plane. 
The IPR 
shows that both in 2D and 3D 
the AT occurs at $E=0$ as well 
as ${\rm Im}\, E \ne 0$. 
We also confirm 
that the critical 
behavior on the real axis is similar to the critical behavior 
far away from the real axis.

\section{Density of states around the real axis}

In this section, We heuristically discuss the 
DoS around the real axis (i.e., $\mathrm{Im} E=0$), 
assuming that the non-Hermitian disorder is weak and 
treating it as a perturbation.

\subsection{Class AII} 
For the non-Hermitian SU(2) model in class AII, 
we observe the soft gap of DoS around the real axis. 
To explain it heuristically, we begin with a disordered 
Hermitian Hamiltonian $H$ in class AII, 
where we have a Kramers pair on the real axis, 
$\psi$ and $\psi'={\cal T}\psi$, for 
eigenenergy $\varepsilon$. 
Here, ${\cal T}$ 
is the time-reversal anti-unitary operator.
In the real space basis, they are expressed explicitly as
$\psi = [a_1, b_1, a_2, b_2, \cdots, a_N, b_N]^\mathrm{T}$ and ${\psi'}=[b_1^*, -a_1^*, b_2^*, -a_2^*,\cdots, b_N^*, -a_N^*]^\mathrm{T}$,
where $a$ and $b$ refer to the amplitudes on $\uparrow$-spin and 
$\downarrow$-down, and $N=L^d$ 
with the spatial dimension $d$.
We then introduce a non-Hermitian on-site potential 
that respects the time-reversal symmetry,
\begin{equation}
	V=\text{i} \!\ \mathrm{diag}[w_1, w_2, \cdots, w_N]\otimes \sigma_z\,,
\end{equation}
with real $w_i$. 
To study how the Kramers pair on the real axis 
is split by the non-Hermitian on-site 
potential $V$, let us consider a 2$\times$2 effective Hamiltonian $h$, 
\begin{equation}
	h = [\psi,\psi']^\dag ({H}+V)[\psi, \psi']=\left(
	\begin{array}{cc}
		\varepsilon+\langle\psi |V|\psi\rangle & \langle \psi |V|\psi'\rangle \\
		\langle \psi' |V|\psi\rangle & \varepsilon+\langle \psi' |V|\psi' \rangle
	\end{array}
	\right) \equiv \left(\begin{array}{cc}
		\varepsilon + \text{i}\!\ h_{1,1} & \text{i} \!\ h_{1,2} \\
		\text{i} \!\ h_{2,1} & \varepsilon + \text{i} \!\ h_{2,2} 
	\end{array}\right)\,.
\end{equation}
The matrix elements are calculated as
\begin{align}
	h_{1,1}&=\sum_i^N w_i (|a_i|^2-|b_i|^2)=:\Delta_z\, ,\\\nonumber
	h_{1,2}&=2\sum_i^N w_i a_i^* b_i^*=:\Delta_x - {\rm i} \Delta_y\, , \\\nonumber
	h_{2,1}&=2\sum_i^N w_i a_i b_i=\Delta_x + {\rm i} \Delta_y\, , \\\nonumber
	h_{2,2}&=\sum_i^N w_i (-|a_i|^2+|b_i|^2)=-\Delta_z\, , \\\nonumber 
\end{align}
which are expressed as
\begin{eqnarray}
	h=\varepsilon+\text{i}\,(\sigma_x\Delta_x+\sigma_y\Delta_y+\sigma_z\Delta_z)\, .
\end{eqnarray}
Here, $\Delta_i \, (i=x,y, z)$ are all real and random numbers.
Consistently, $h$ respects time-reversal symmetry $\sigma_y h^{*} \sigma_y = h$.

The eigenenergies of $h$ are $\varepsilon\pm \text{i}\sqrt{\Delta_x^2+\Delta_y^2+\Delta_z^2}$,
and the energy splitting is along the imaginary axis and equals $2\sqrt{\Delta_x^2+\Delta_y^2+\Delta_z^2}$.
Assuming that each of the three $\Delta_i$ obeys the independent 
Gaussian distributions due to the central limit theorem, the 
probability distribution of the energy splitting from the real axis 
is estimated as
\begin{align}
	P(s)&=\frac{1}{N}\int^{\infty}_{-\infty} d \Delta_x \int^{\infty}_{-\infty}d \Delta_y \int^{\infty}_{-\infty} d \Delta_z \!\  \delta\Big(s-\sqrt{\Delta_x^2+\Delta_y^2+\Delta_z^2}\Big) \!\ 
	\exp\left(
	-A(\Delta_x^2+\Delta_y^2+\Delta_z^2)
	\right)\,,  \nonumber \\
	N & \equiv \int_{-\infty}^{\infty} d \Delta_x \int_{-\infty}^{\infty} d \Delta_y \int_{-\infty}^{\infty} d \Delta_z \!\  
	\exp\left(
	-A(\Delta_x^2+\Delta_y^2+\Delta_z^2)\right) = \bigg(\sqrt{\frac{\pi}{A}}\bigg)^{3}, \nonumber 
\end{align}
where $A>0$ is a normalization constant.
This leads to the Wigner distribution for the unitary ensemble:
\begin{equation}
	P(s)= \frac{4A^{\frac{3}{2}}}{\sqrt{\pi}} s^2 \exp\left(-As^2\right).  
\end{equation}
Notably, $P (s)$ vanishes for small $s$, i.e., 
$P (s) \propto s^2 \to 0$ for $s \to 0$. 
This means that the probability of the energy levels having a small 
imaginary part is significantly suppressed, hence the small density
of states around the real axis.

\subsection{Class AI} 
For the non-Hermitian O(1) model in class AI,
we observe the peak of DoS around the real axis, 
$\mathrm{Im}\, E=0$. 
To explain it, 
we begin with a disordered 
Hermitian Hamiltonian $H$ in class AI, 
introduce a non-Hermitian 
on-site potential $V$ that respects time-reversal symmetry, and study 
two nearest-neighbor eigenmodes. 
The non-Hermitian on-site potential 
reads
\begin{eqnarray}
	V = \text{i}\!\ \mathrm{diag}[w_1, w_2, \cdots, w_N]\otimes \sigma_z 
\end{eqnarray}
with real random number $w_i$ ($i=1,2,\cdots, N$).
The system respects time-reversal symmetry: 
\begin{eqnarray}
	\sigma_x (H+V)^{*} \sigma_x = H +V,   
\end{eqnarray}
which imposes a constraint on each of the 
two eigenmodes of the disordered Hermitian Hamiltonian $H$ as 
$\psi_1=[a_1, a^*_1, a_2, a^*_2, \cdots, a_N, a^*_N]^\mathrm{T}$ 
and $\psi_2=[b_1, b_1^*, b_2, b_2^*,\cdots, b_N, b_N^*]^\mathrm{T}$. 
Let $\varepsilon_1$ and $\varepsilon_2$ be the corresponding eigenenergies of $\psi_1$ and $\psi_2$, respectively.
The two by two effective Hamiltonian $h$ reads
\begin{equation}
	h=[\psi_1,\psi_2]^\dag (H+V)[\psi_1, \psi_2]=\left(
	\begin{array}{cc}
		\varepsilon_1+\langle\psi_1 |V|\psi_1 \rangle & \langle \psi_1 |V|\psi_2\rangle \\
		\langle \psi_2 |V|\psi_1\rangle & \varepsilon_2+\langle \psi_2 |V|\psi_2 \rangle
	\end{array}
	\right)  \equiv \left(\begin{array}{cc}
		\varepsilon_1 + \text{i}\!\ h_{1,1} & \text{i} \!\ h_{1,2} \\
		\text{i} \!\ h_{2,1} & \varepsilon_2 + \text{i} \!\ h_{2,2} 
	\end{array}\right)\,.
\end{equation}
The matrix elements are calculated as
\begin{align}
	h_{1,1}&=\sum_i^N w_i (a^*_i a_i - a_i a^*_i)=0\, ,\\\nonumber 
	h_{1,2}&=2\sum_i^N w_i (a_i^* b_i - a_i b^*_i) =:{\rm i} \Delta_0\, ,\\\nonumber
	h_{2,1}&=2\sum_i^N w_i (b^*_i a_i - b_i a^*_i) = - {\rm i}\Delta_0\, ,\\\nonumber
	h_{2,2}&=\sum_i^N w_i (b^*_i b_i - b_i b^*_i)=0, \nonumber
\end{align}
which are expressed as
\begin{eqnarray}
	h= \left(
	\begin{array}{cc}
		\varepsilon_1  & -\Delta_0 \\
		\Delta_0 & \varepsilon_2 
	\end{array}
	\right).
\end{eqnarray}
Here, $\Delta_0$ is a real random number. 
Then, $h$ is a real matrix and indeed respects time-reversal symmetry $h^{*} = h$.


The two eigenenergies of $h$ are 
\begin{equation}
	\frac{\varepsilon_1 + \varepsilon_2}{2}
	\pm \sqrt{\left( \frac{\varepsilon_1 - \varepsilon_2}{2}\right)^2 - \Delta_{0}^2}\,,
\end{equation}
which are real for $\left| \varepsilon_1 - \varepsilon_2 \right|/2 \geq \left| \Delta_0 \right|$. In class AII, by contrast, the eigenenergies cannot be real unless the stronger constraint $\Delta_x = \Delta_y = \Delta_z = 0$ is satisfied. This means that real eigenenergies are more stable against non-Hermitian perturbations in class AI than in class AII. Consequently, the eigenenergies remain real more easily in class AI, which corresponds to the sharp peak of DoS on the real axis.

\section{Density of states for the Ginibre orthogonal and symplectic ensembles}

The Ginibre ensembles are 
ensembles of non-Hermitian random matrices~\cite{Ginibre65}. 
They are useful for understanding the energy level statistics and the ATs in non-Hermitian 
disordered systems. We have three kinds of the Ginibre ensembles: 
Ginibre unitary ensemble (GinUE) (no restriction on $H$; class A), 
Ginibre orthogonal ensemble (GinOE) ($H^*=H$; class AI), 
and Ginibre symplectic ensemble (GinSE) ($\sigma_y H^* \sigma_y=H$; class AII).
Real and imaginary parts of each element of  
non-Hermitian random matrices are independent and produced by 
the same Gaussian distribution. The GinOE is the ensemble of 
non-Hermitian but real random matrices. Each element of 
real random matrices
is independent and produced by the same Gaussian distribution. 
For the GinSE, non-Hermitian random matrices are defined to 
satisfy $\sigma_y H^* \sigma_y=H$ with 
\begin{align}
	\sigma_y=
	\begin{pmatrix}
		0& -\text{i}\\
		\text{i}& 0\\
	\end{pmatrix},
\end{align}
so that $H$ has the following symplectic structure: 
\begin{align}
	H=\begin{pmatrix}
		X & Y \\
		-Y^* & X^* \\
	\end{pmatrix}, \label{symplectic}
\end{align} 
where $X$, $Y$ are 
generic non-Hermitian random matrices with no constraint.

As shown in Figs.~\ref{DoS_AI} and \ref{AII_2D3D}, 
the DoS of non-Hermitian disordered systems in class AI has a sharp peak  
on the real axis, while the DoS in class AII has a soft gap  
on the real axis. 
In order to understand these behavior, we study the DoS of the GinOE and GinSE. 
The eigenenergy distribution and the DoS along the imaginary axis 
are shown in Fig.~\ref{GinibreOSE}. The DoS around the real axis for 
non-Hermitian disordered Hamiltonians in classes AI and AII 
is qualitatively consistent with the DoS of the GinOE and GinSE, respectively. 


\clearpage

\begin{figure}[bt]
	\centering
	\subfigure[$E=0$, (2,3,0,1)]{
		\begin{minipage}[t]{0.5\linewidth}
			\centering
			\includegraphics[width=1\linewidth]{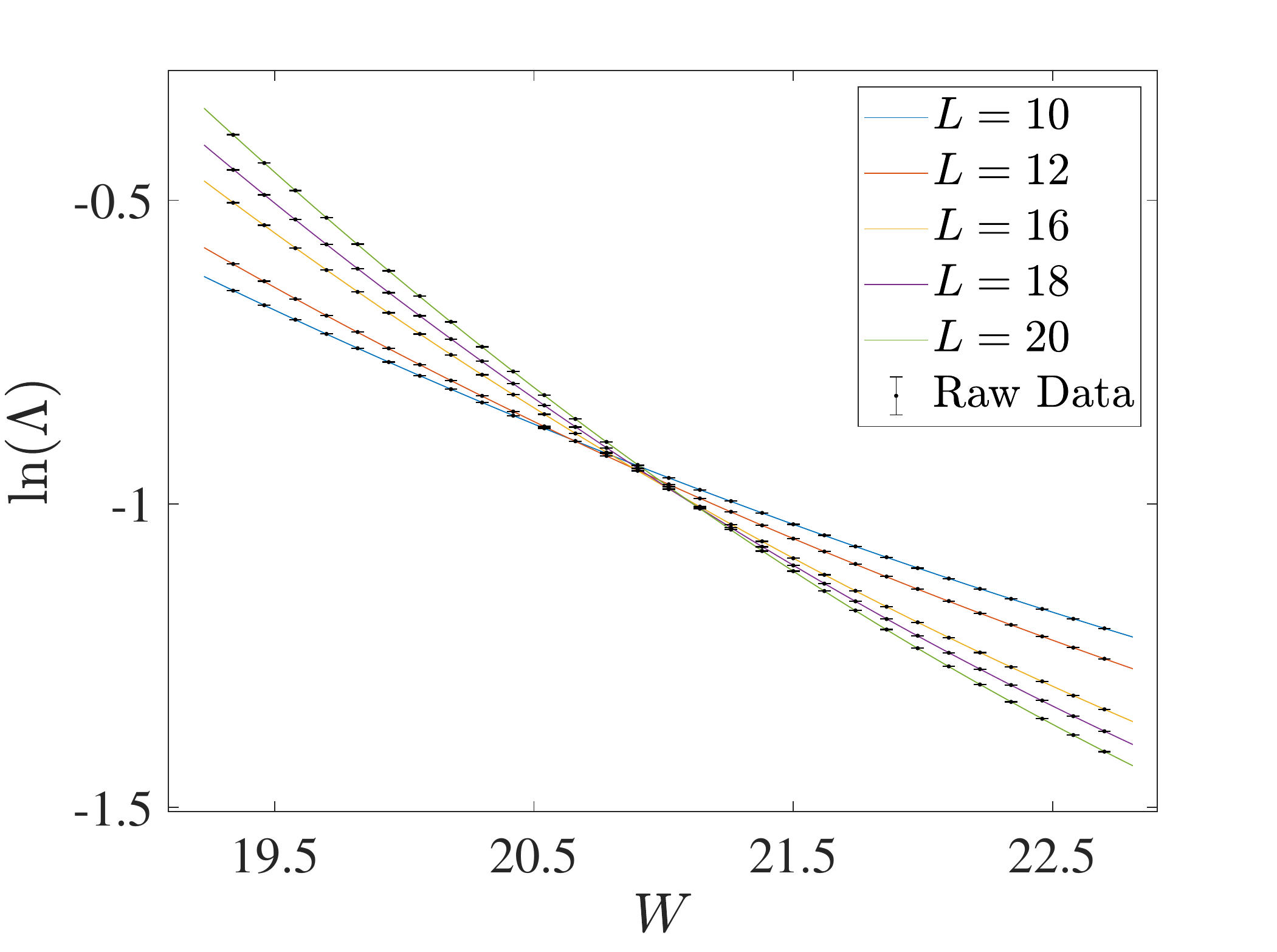}
		\end{minipage}%
	}%
	\subfigure[$E=0.5{\rm i}$, (3,3,0,1)]{
		\begin{minipage}[t]{0.5\linewidth}
			\centering
			\includegraphics[width=1\linewidth]{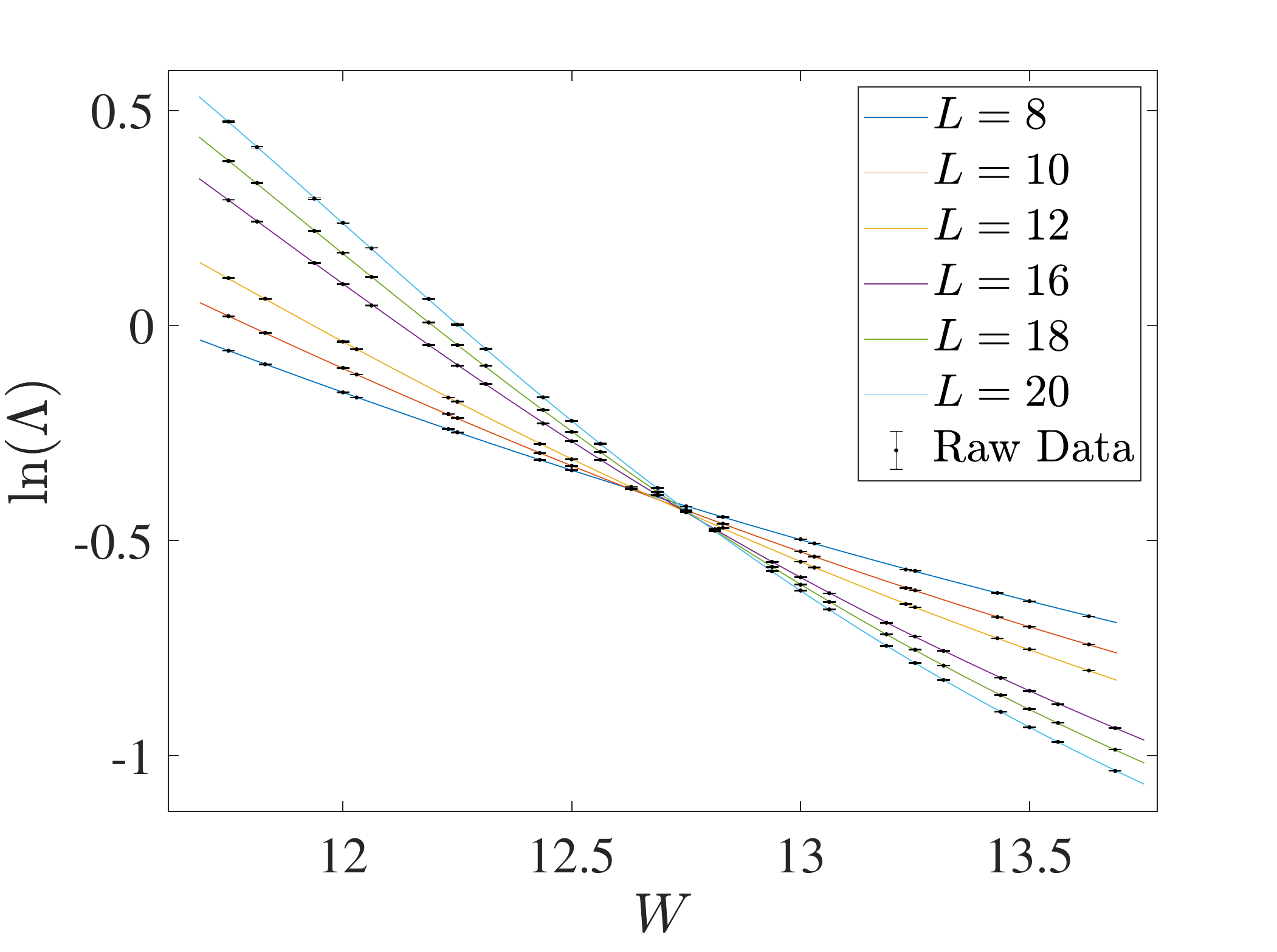}
		\end{minipage}%
	}%
	\caption{
		Normalized localization lengths $\Lambda$ as a function of the disorder strength $W$ for 
		the 3D O(1) model in class AI
		at (a)~$E=0$ and (b)~$E=0.5\text{i}$. 
		The points with the error bars are the 
		numerical data with the different 
		system sizes $L$. 
		The colored curves are the fitted curves with
		the expansion order $(m_1, n_1, m_2, n_2)$. }
	\label{AI}
\end{figure}

\begin{figure}[bt]
	\centering
	\subfigure[2D, (1,3,0,1)]{
		\begin{minipage}[t]{0.5\linewidth}
			\centering
			\includegraphics[width=1\linewidth]{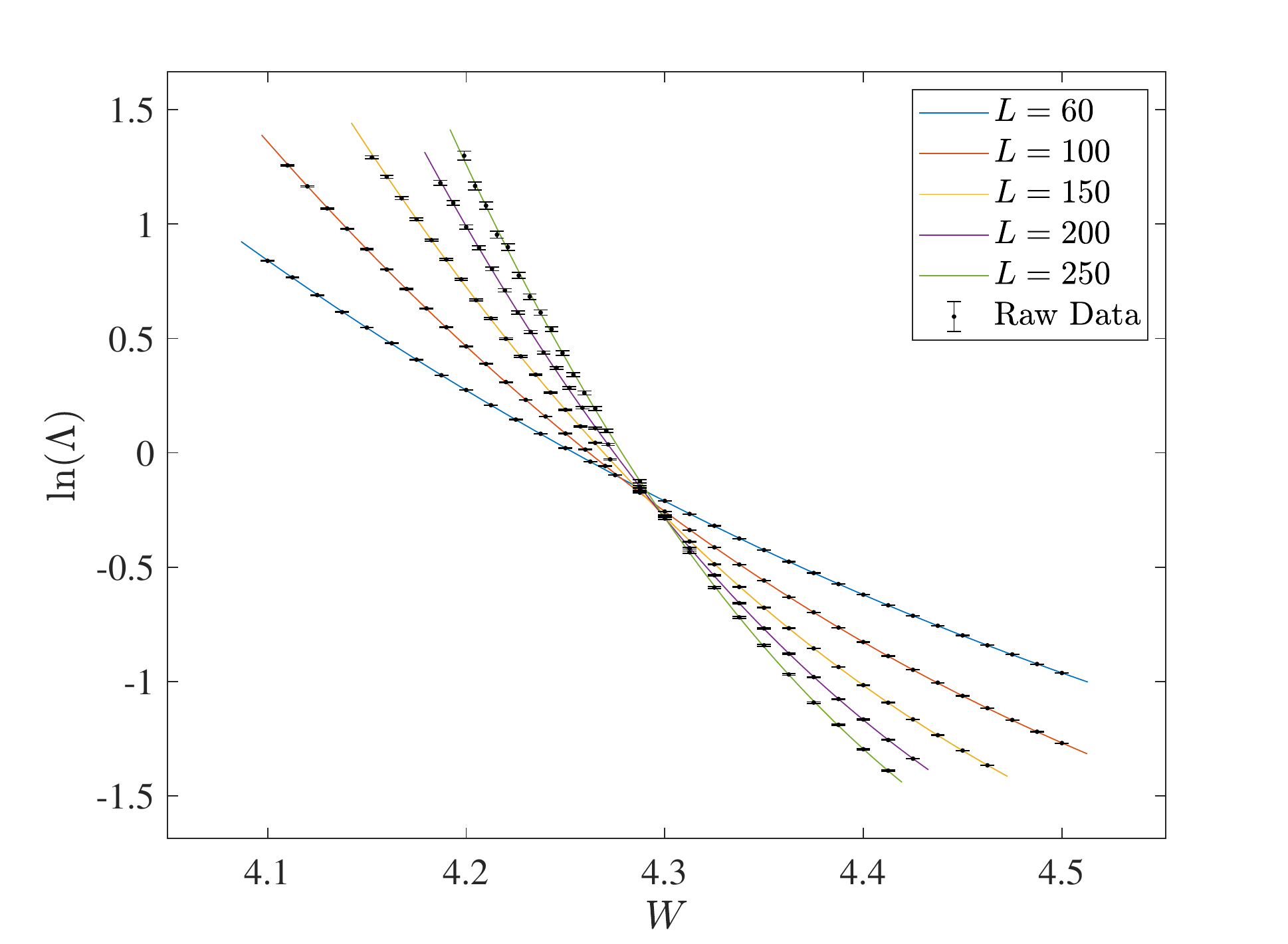}
		\end{minipage}%
	}%
	\subfigure[3D, (2,3,0,1) ]{
		\begin{minipage}[t]{0.5\linewidth}
			\centering
			\includegraphics[width=1\linewidth]{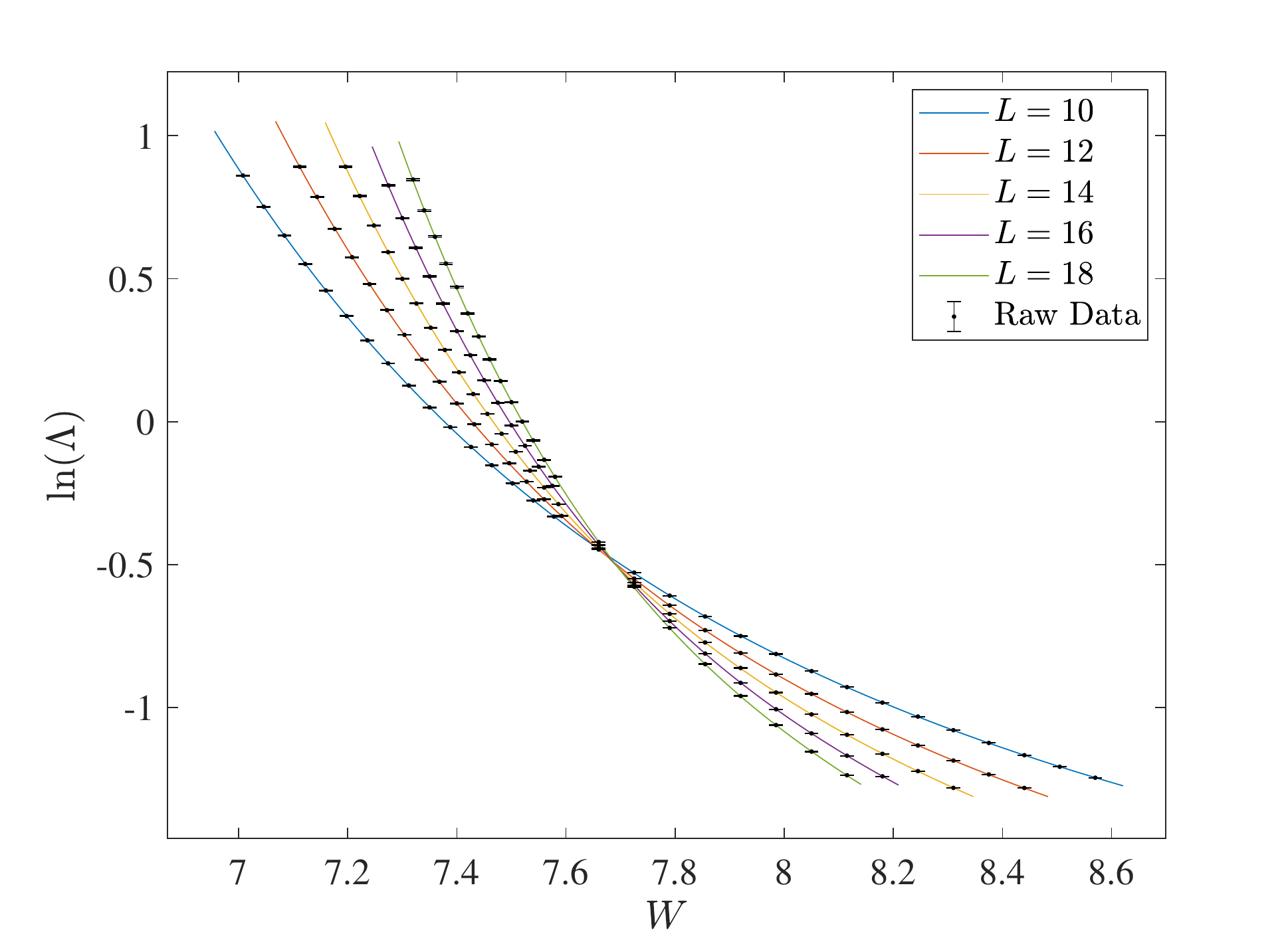}
		\end{minipage}%
	}%
	\caption{
		Normalized localization lengths $\Lambda$ as a function of the disorder strength $W = W_r=W_i$ for the (a)~2D and (b)~3D SU(2) model in class AII$^{\dagger}$ with $E=0$.
		The points with the error bars are the numerical data with the different system sizes $L$. 
		The colored curves are the fitted curves with
		the expansion order $(m_1, n_1, m_2, n_2)$.}
	\label{AII_dagger_lambda}
\end{figure}

\begin{figure}[bt]
	\centering
	\subfigure[2D, $E=0.01{\rm i}$, (3,3,0,1)]{
		\begin{minipage}[t]{0.5\linewidth}
			\centering
			\includegraphics[width=1\linewidth]{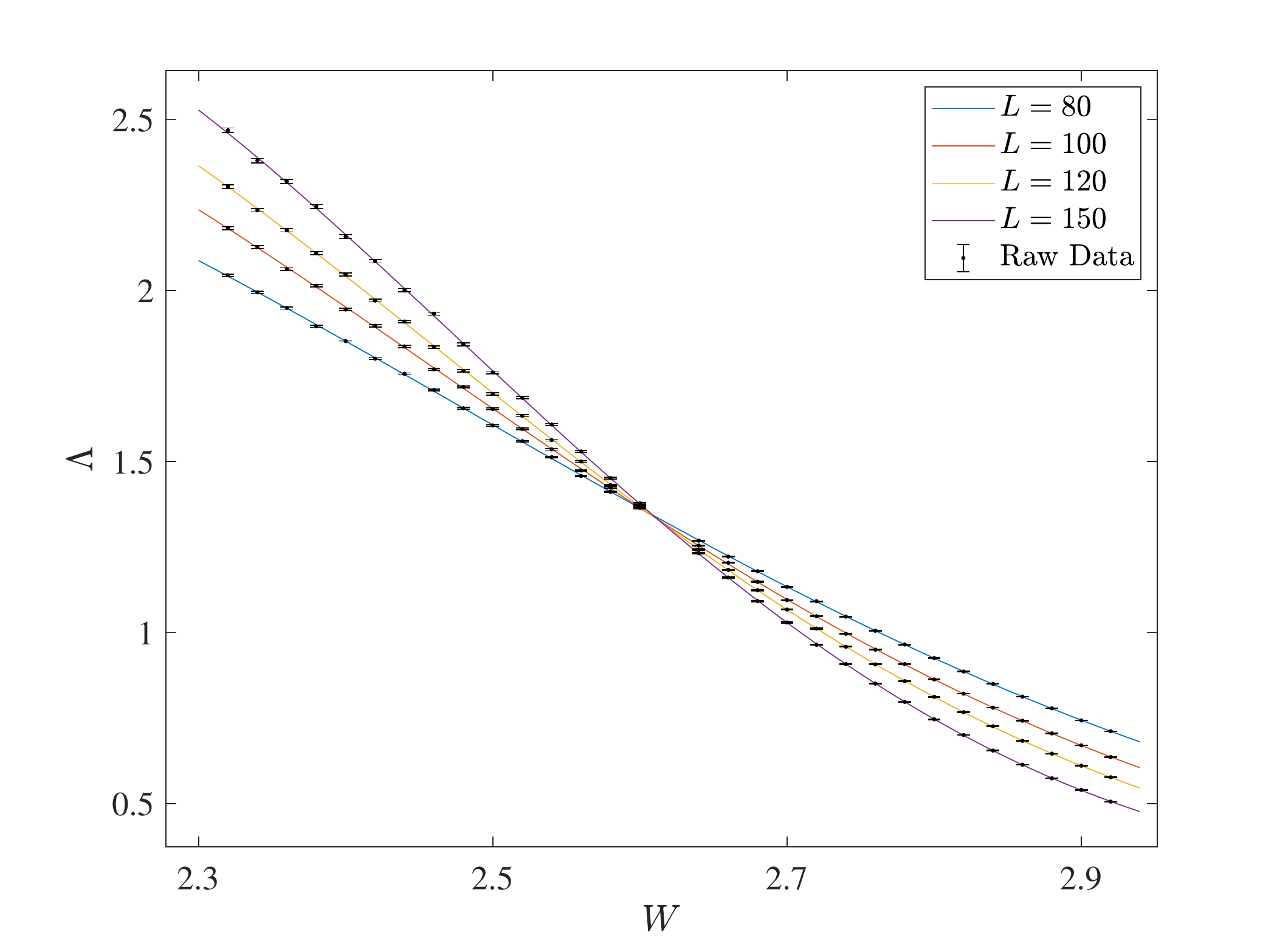}
		\end{minipage}%
	}%
	
	\subfigure[3D, $E=0$, (2,3,0,0)]{
		\begin{minipage}[t]{0.5\linewidth}
			\centering
			\includegraphics[width=1\linewidth]{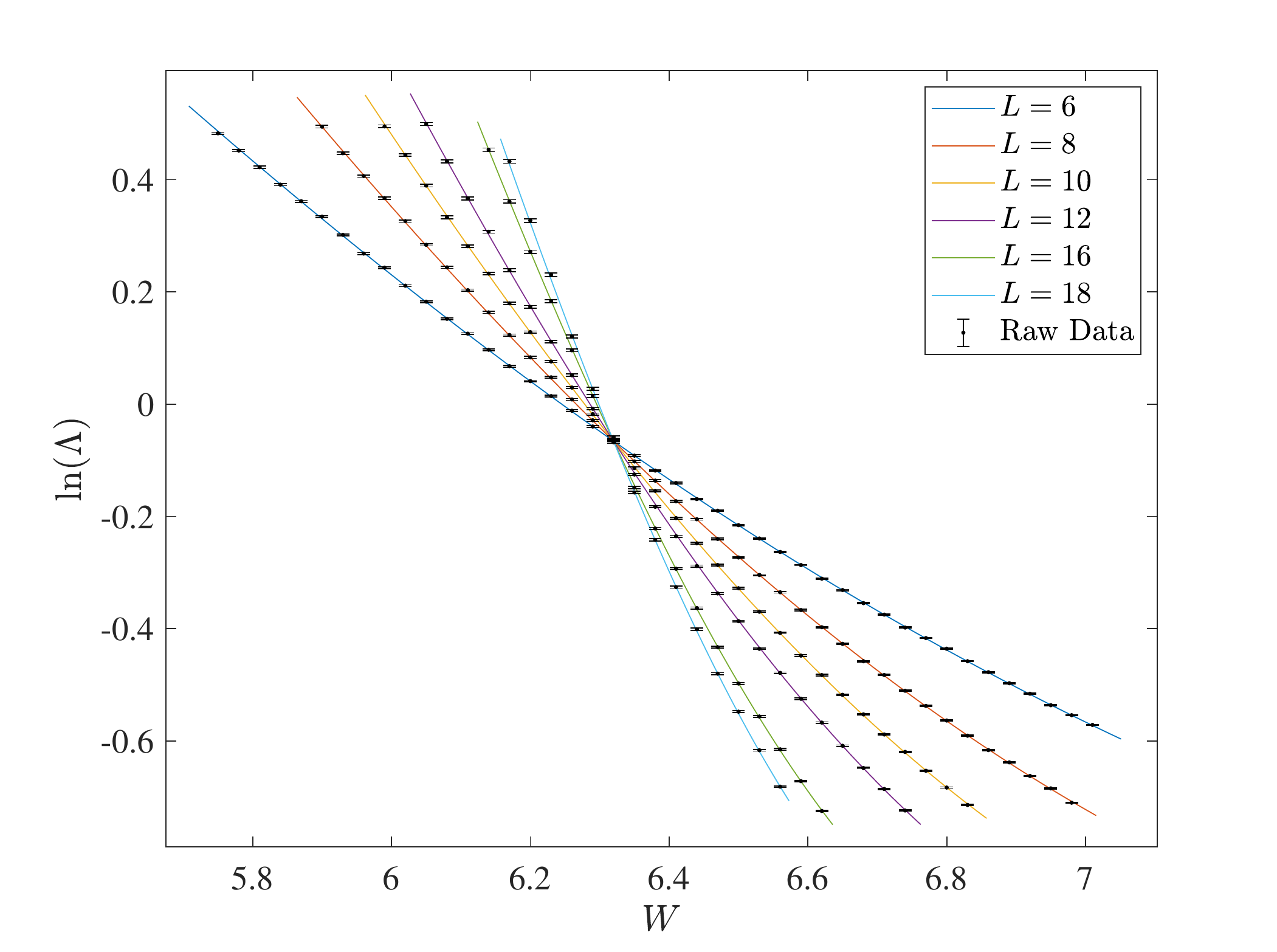}
		\end{minipage}%
	}%
	\subfigure[3D, $E={\rm i}$, (1,3,0,1) ]{
		\begin{minipage}[t]{0.5\linewidth}
			\centering
			\includegraphics[width=1\linewidth]{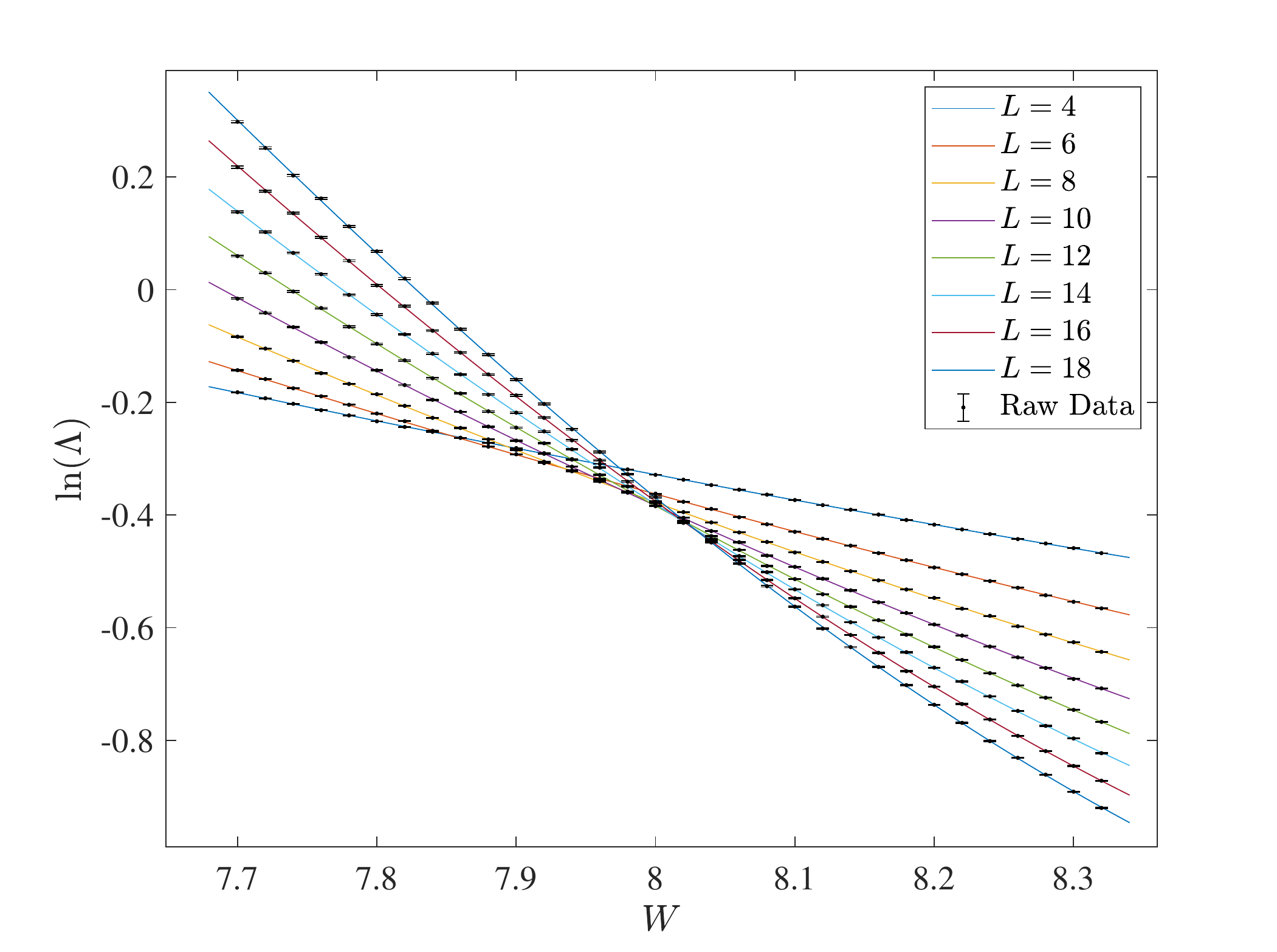}
		\end{minipage}%
	}%
	\caption{
		Normalized localization lengths $\Lambda$ as a function of the disorder strength $W = W_r=W_i$ for 
		(a)~the 2D SU(2) model in class AII with $E = 0.01\text{i}$, 
		(b)~the 3D SU(2) model in class AII with $E = 0$, 
		and (c)~the 3D SU(2) model in class AII with $E = \text{i}$.
		The points with the error bars are the numerical data with the different system sizes $L$. The colored curves are the fitted curves with
		the expansion order $(m_1, n_1, m_2, n_2)$.}
	\label{AII_3D_lambda}
\end{figure}

\begin{figure}[bt]
	\centering
	\subfigure[$E=0$, 2D, class CII$^{\dag}$, (1,3,0,0)]{
		\begin{minipage}[t]{0.5\linewidth}
			\centering
			\includegraphics[width=1\linewidth]{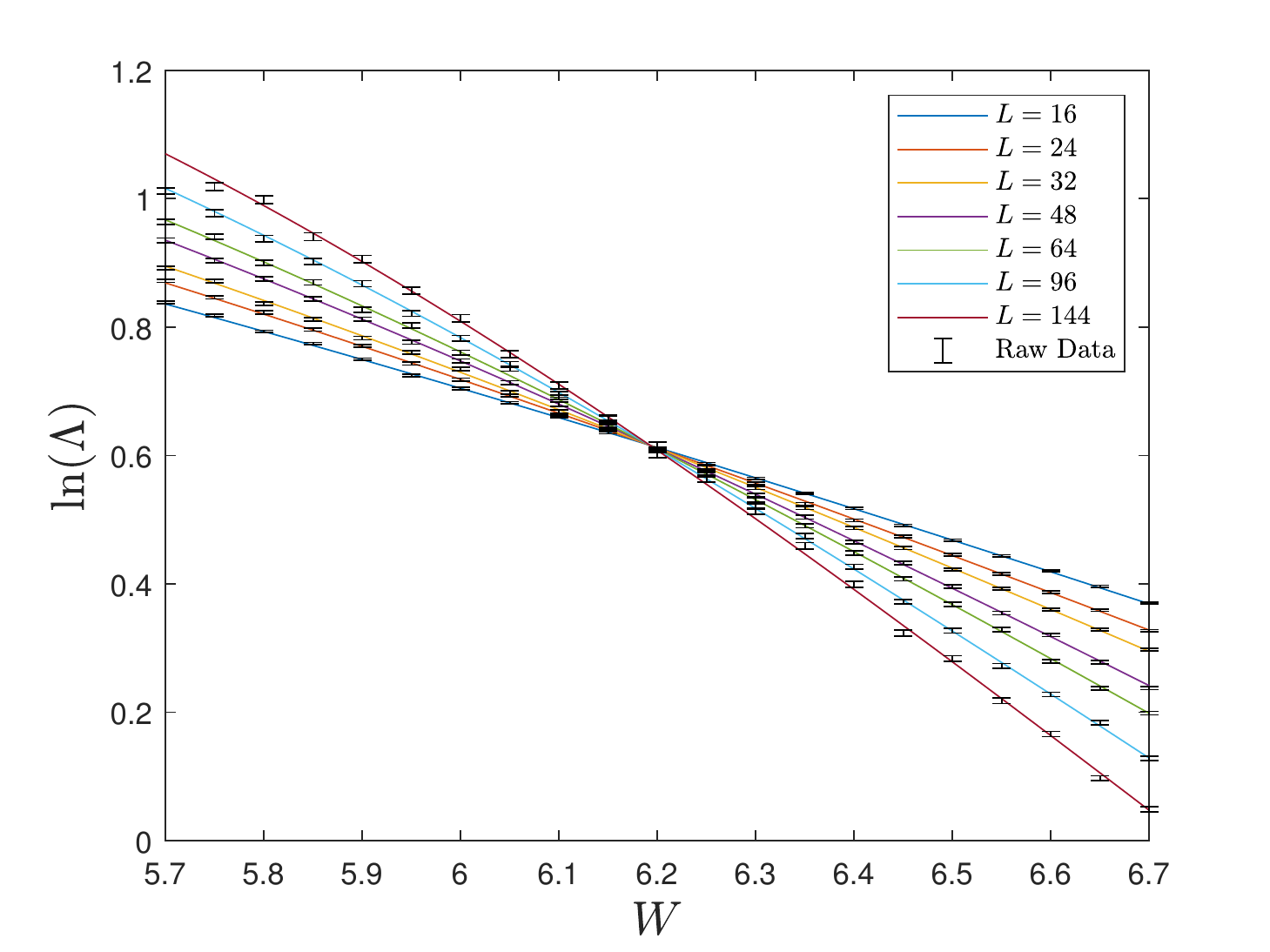}
		\end{minipage}%
	}%
	\subfigure[$E=0$, 2D, class DIII, (1,2,0,0)]{
		\begin{minipage}[t]{0.5\linewidth}
			\centering
			\includegraphics[width=1\linewidth]{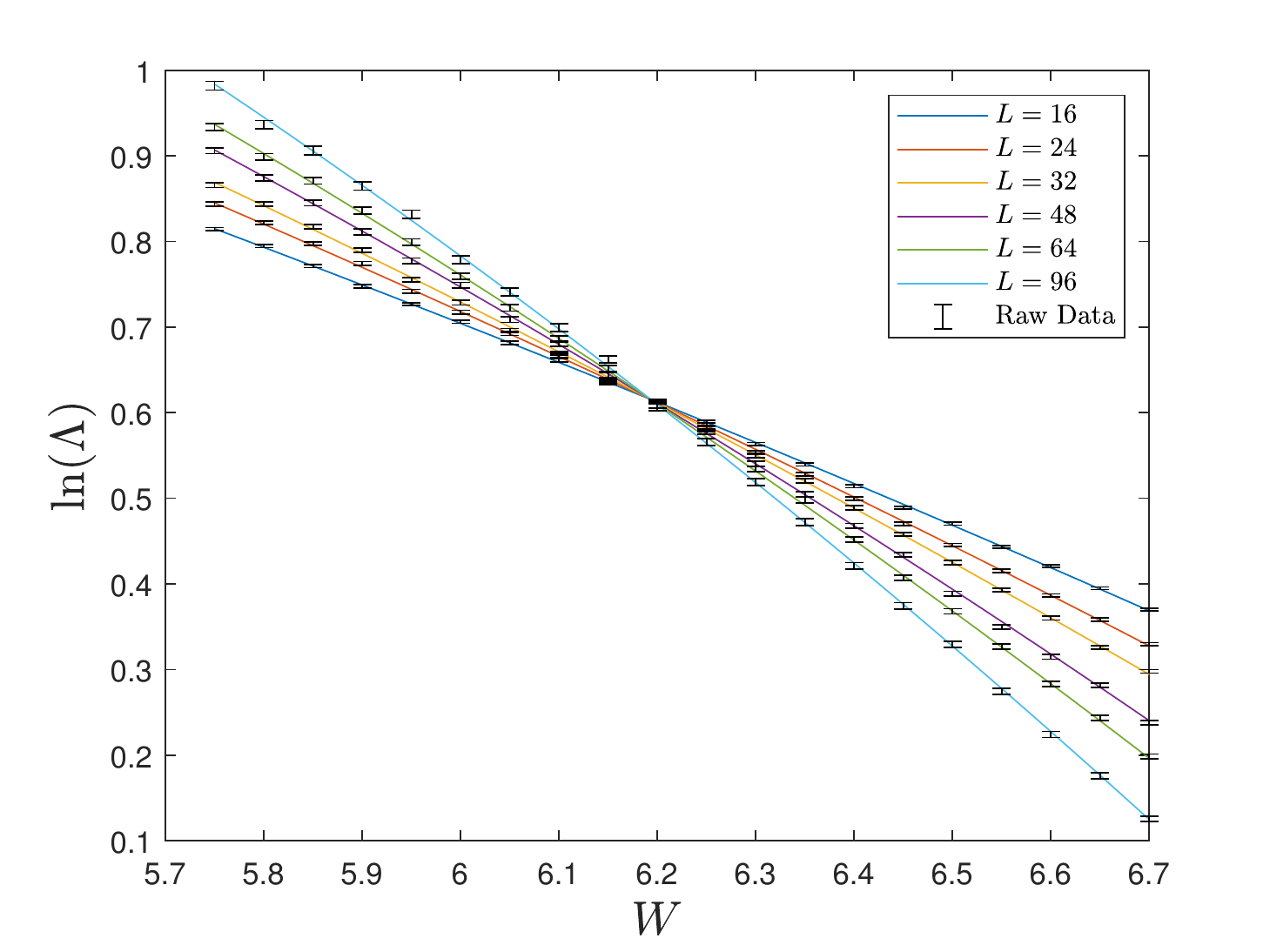}
		\end{minipage}%
	}%
	\caption{Normalized localization lengths $\Lambda$ as a function of the disorder strength $W \equiv W_i$ for the 2D SU(2) model 
                     in (a)~class CII$^{\dag}$ and (b)~class DIII. 
		The points with the error bars are the numerical data with the different system sizes $L$. 
		The colored curves are the fitted curves with the expansion order $(m_1, n_1, m_2, n_2)$. 
	}
	\label{CII^dDIII}
\end{figure}


\begin{figure}[bt]
	\centering
	\subfigure[3D class AI]{
		\begin{minipage}[t]{0.5\linewidth}
			\centering
			\includegraphics[width=1\linewidth]{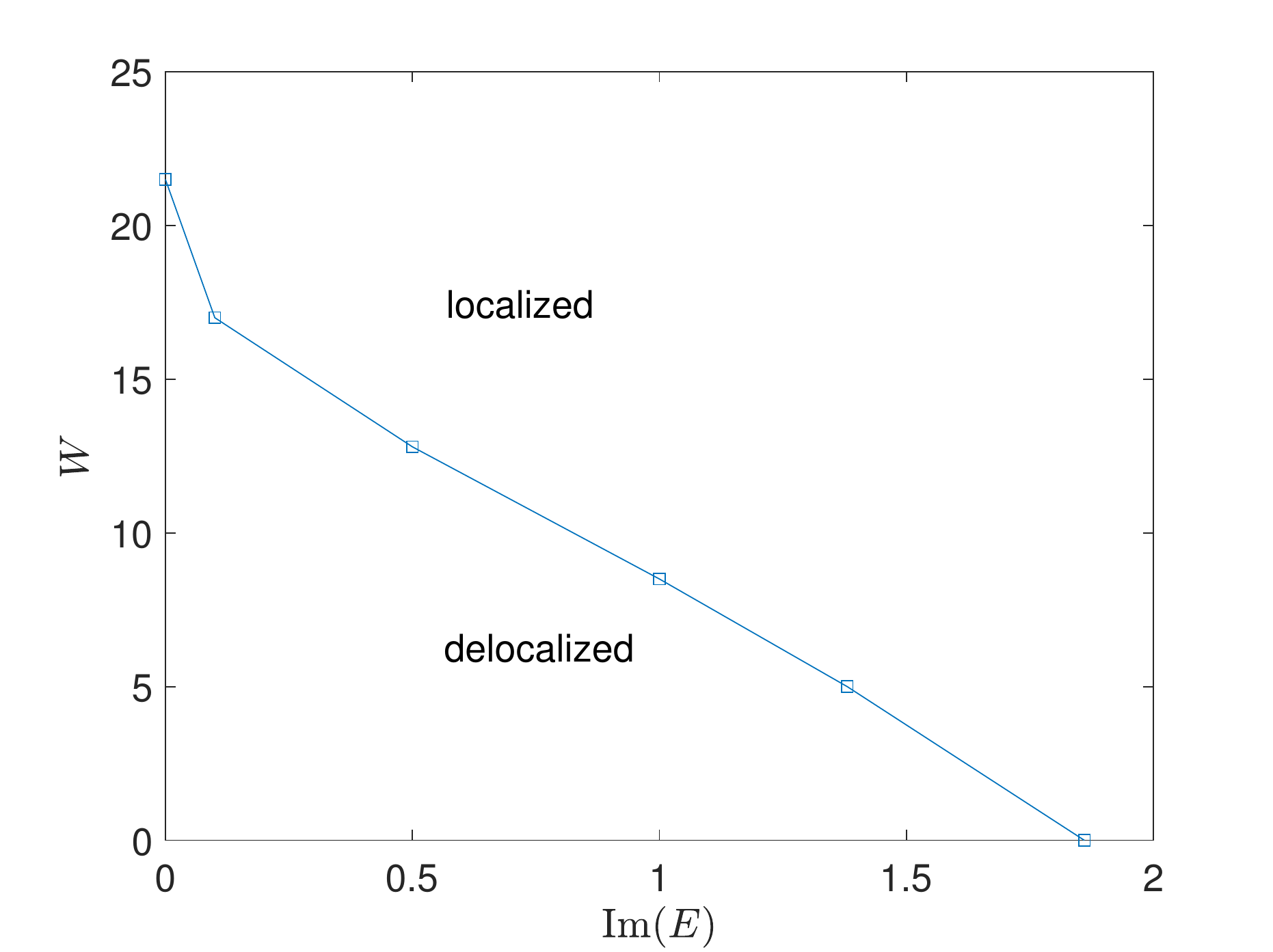}
		\end{minipage}%
	}%
	\subfigure[3D class AII]{
		\begin{minipage}[t]{0.5\linewidth}
			\centering
			\includegraphics[width=1\linewidth]{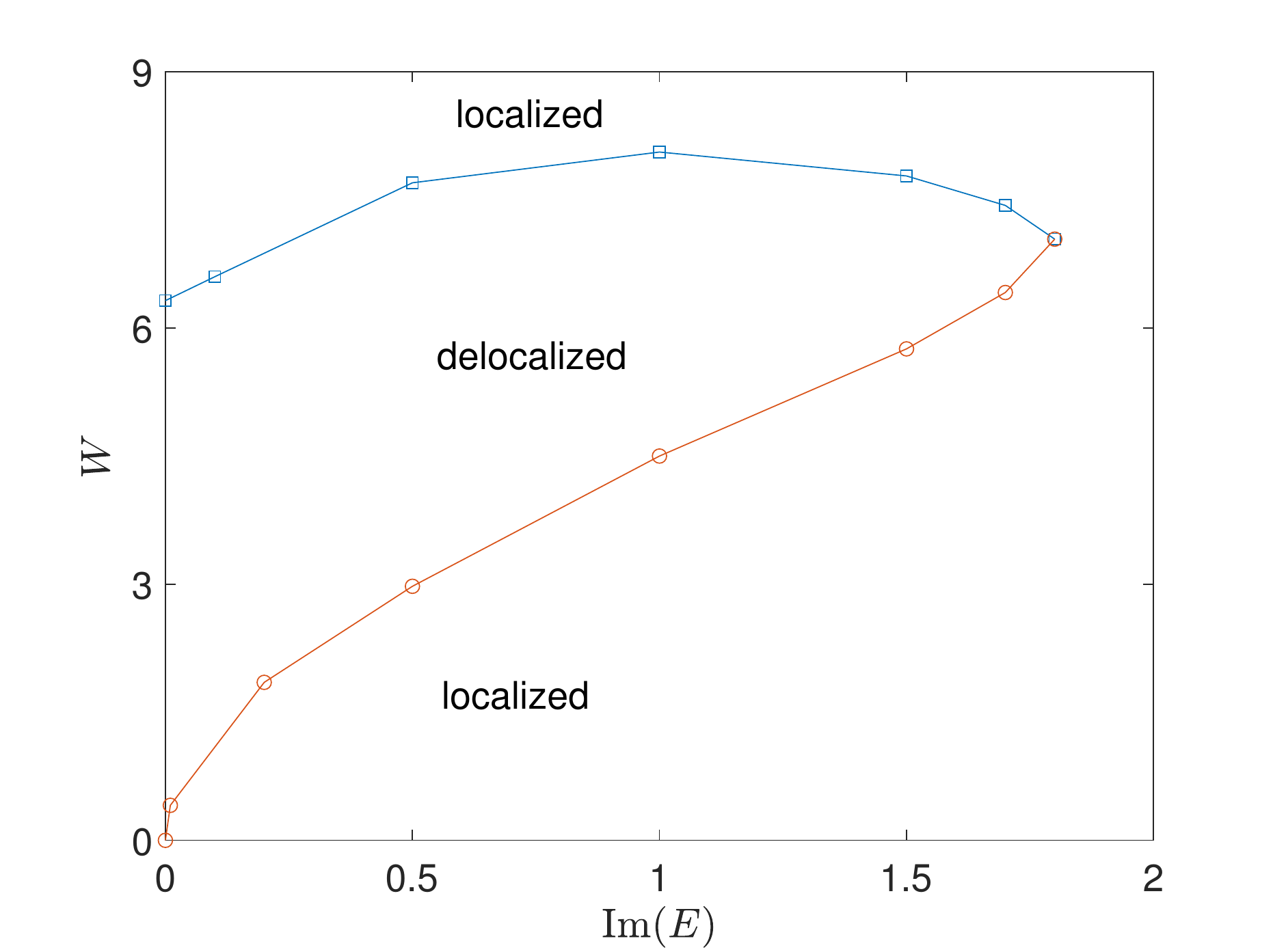}
		\end{minipage}%
	}%

	\subfigure[2D class AII]{
		\begin{minipage}[t]{0.5\linewidth}
			\centering
			\includegraphics[width=1\linewidth]{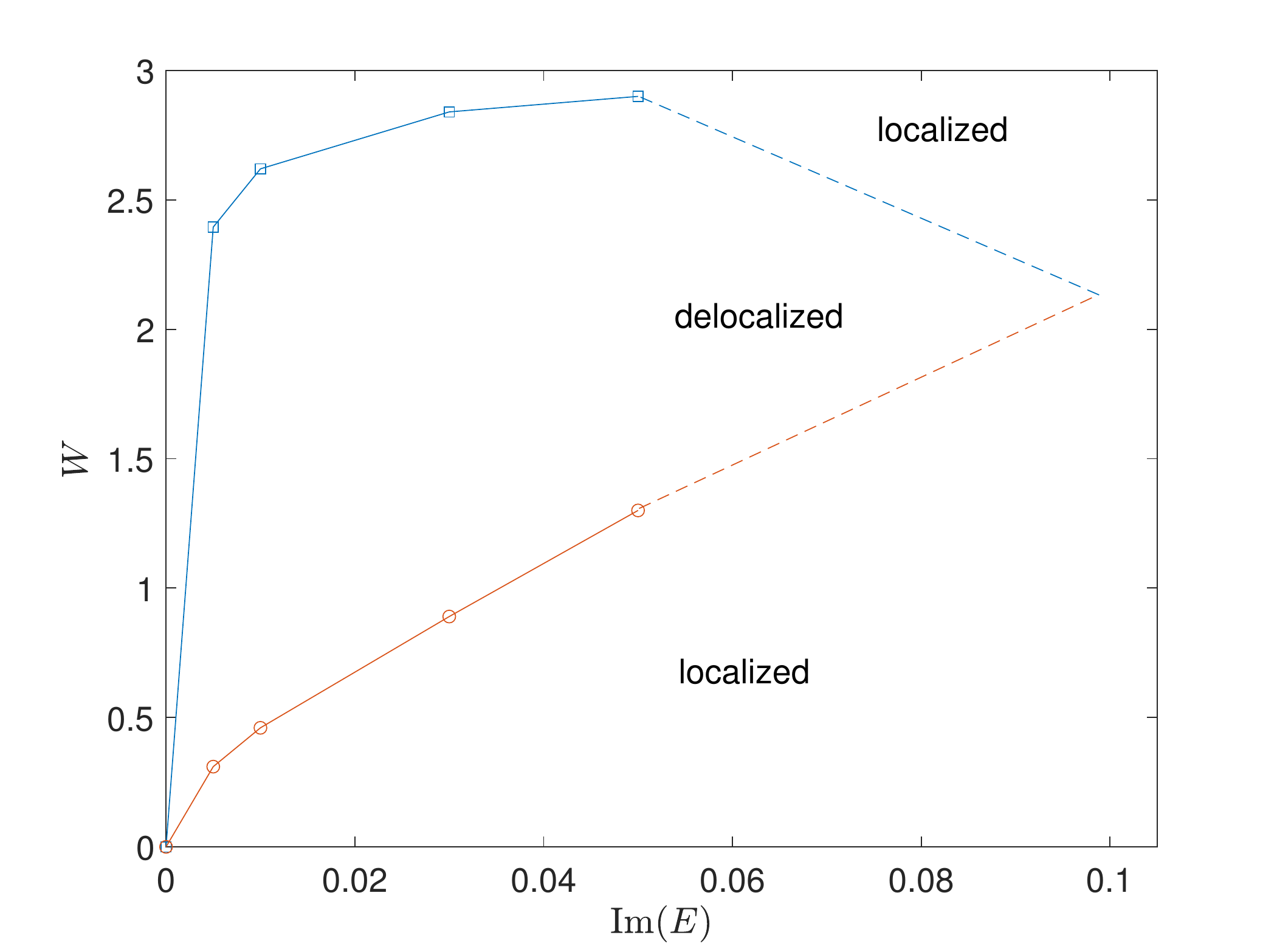}
		\end{minipage}%
	}%
	\subfigure[2D class AII, $E=0.01{\rm i}$]{
		\begin{minipage}[t]{0.5\linewidth}
			\centering
			\includegraphics[width=1\linewidth]{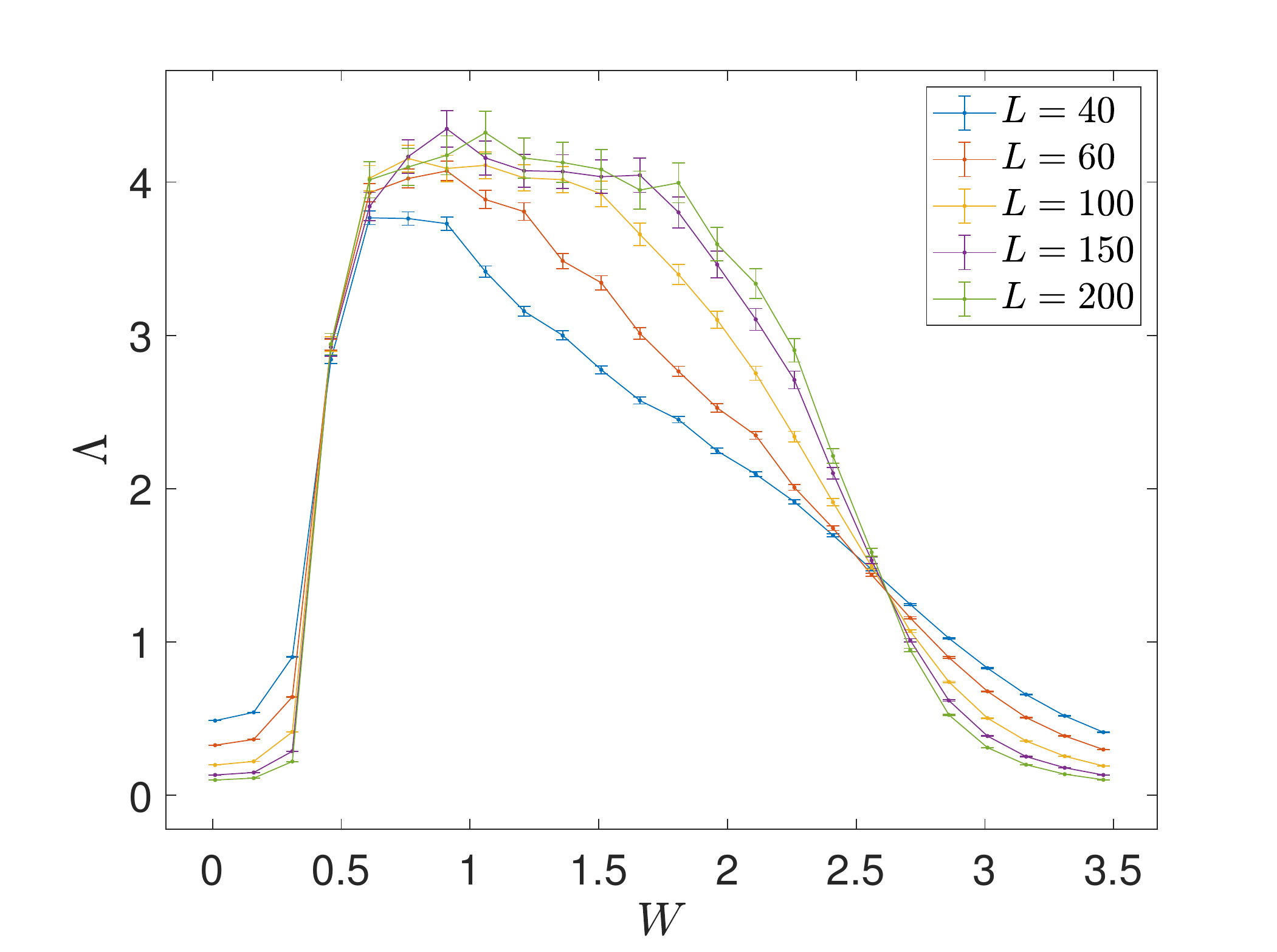}
		\end{minipage}%
	}%
	\caption{Phase diagrams of the O(1) and SU(2) models for (a)~3D class AI, (b)~3D class AII, and (c)~2D class AII in terms of the disorder strength $W$ and the imaginary part of eigenenergies $E$.
		The real part of $E$ is set to $0$.
		The blue squares and red circles are the phase boundaries for the Anderson transitions. The phase boundaries are determined by the localization lengths.
		(d)~Normalized localization length  as a function of $W\equiv W_r=W_i$ for the 2D class AII model at $E=0.01\text{i}$.}
	\label{phaseDiagram_AI_AII}
\end{figure}

\begin{figure}[bt]
	\centering
	\subfigure[$W=0$]{
		\begin{minipage}[t]{0.5\linewidth}
			\centering
			\includegraphics[width=1\linewidth]{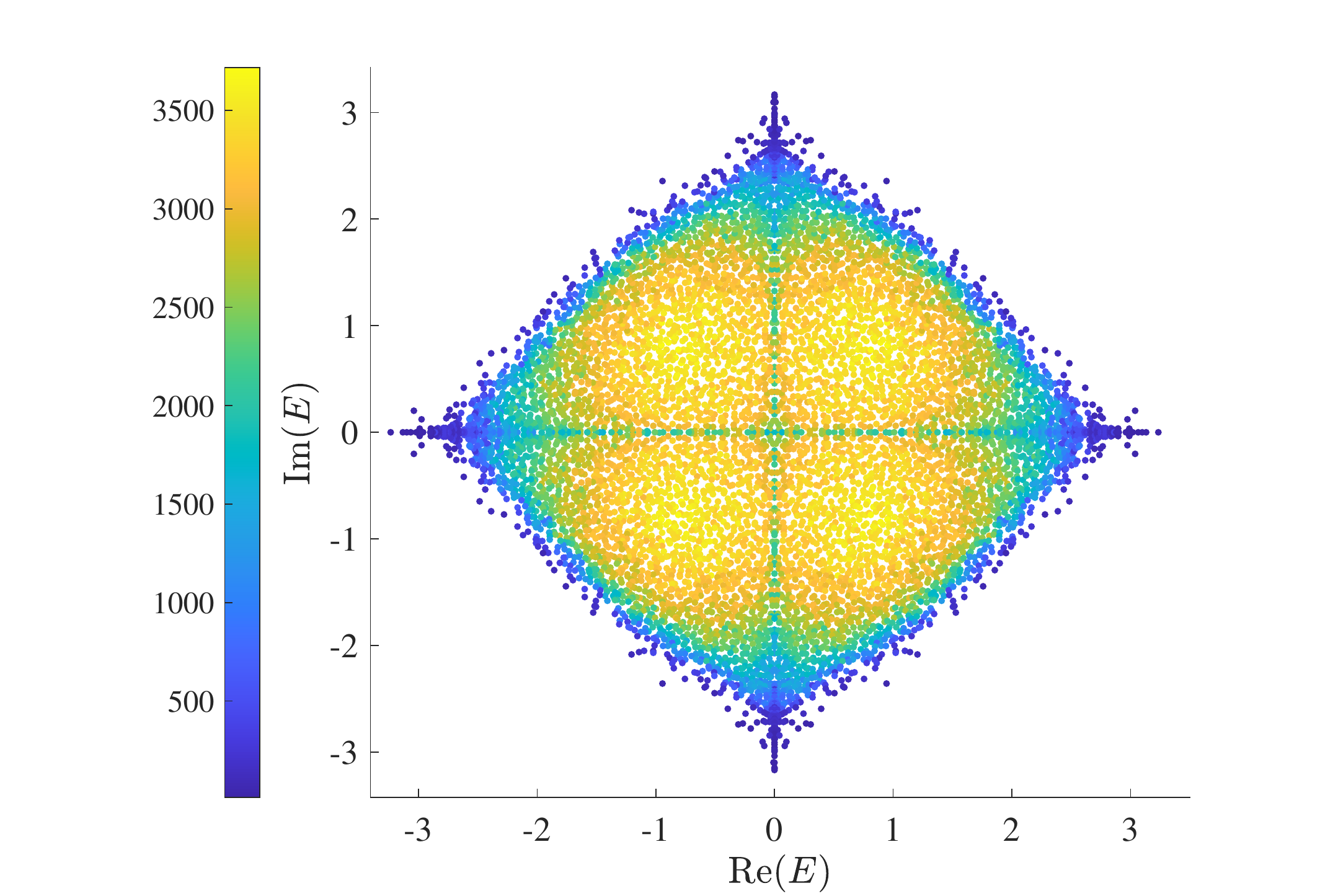}
		\end{minipage}%
	}%
	\subfigure[$W=0$ ]{
		\begin{minipage}[t]{0.5\linewidth}
			\centering
			\includegraphics[width=1\linewidth]{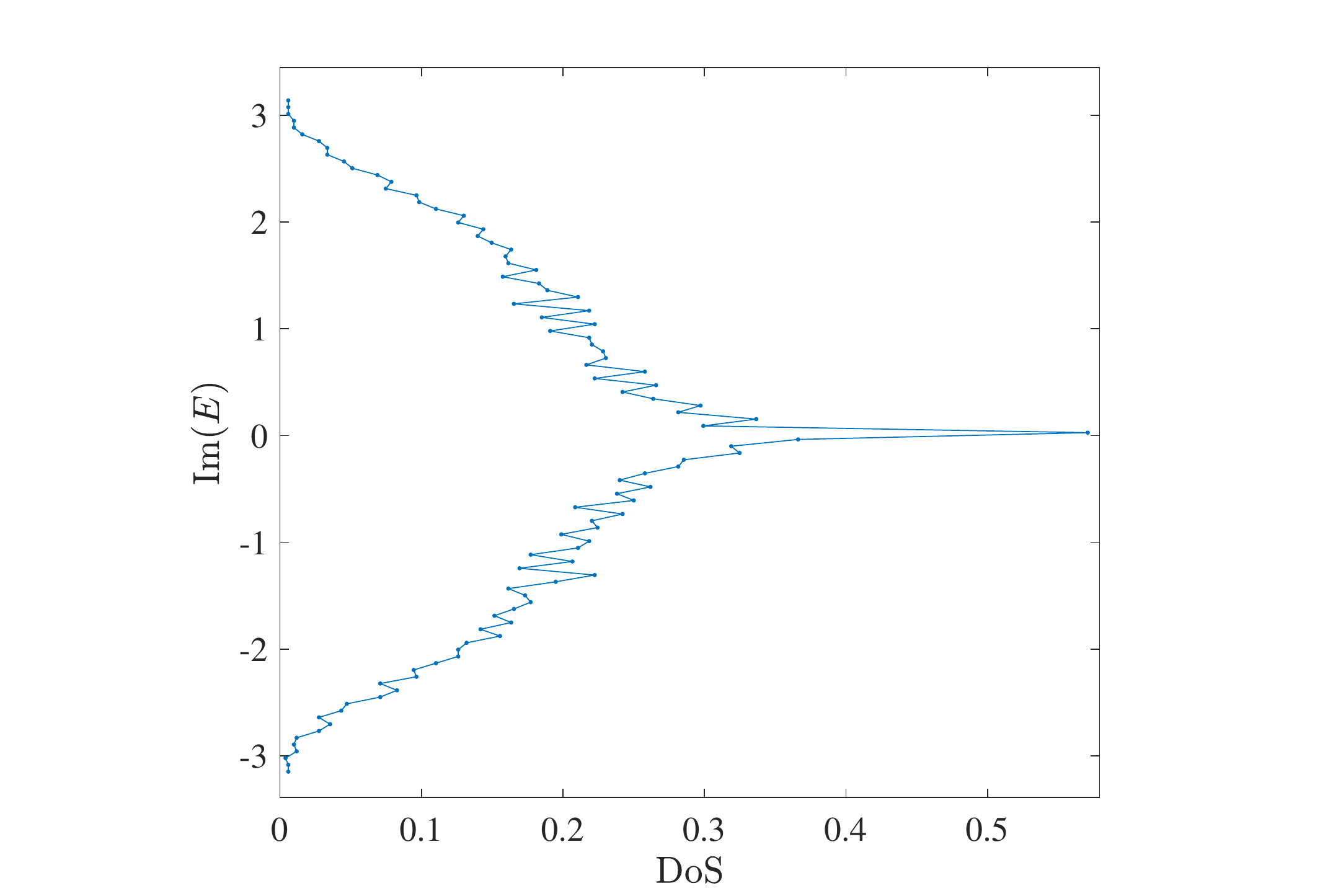}
		\end{minipage}%
	}%

	\subfigure[$W=10$]{
		\begin{minipage}[t]{0.5\linewidth}
			\centering
			\includegraphics[width=1\linewidth]{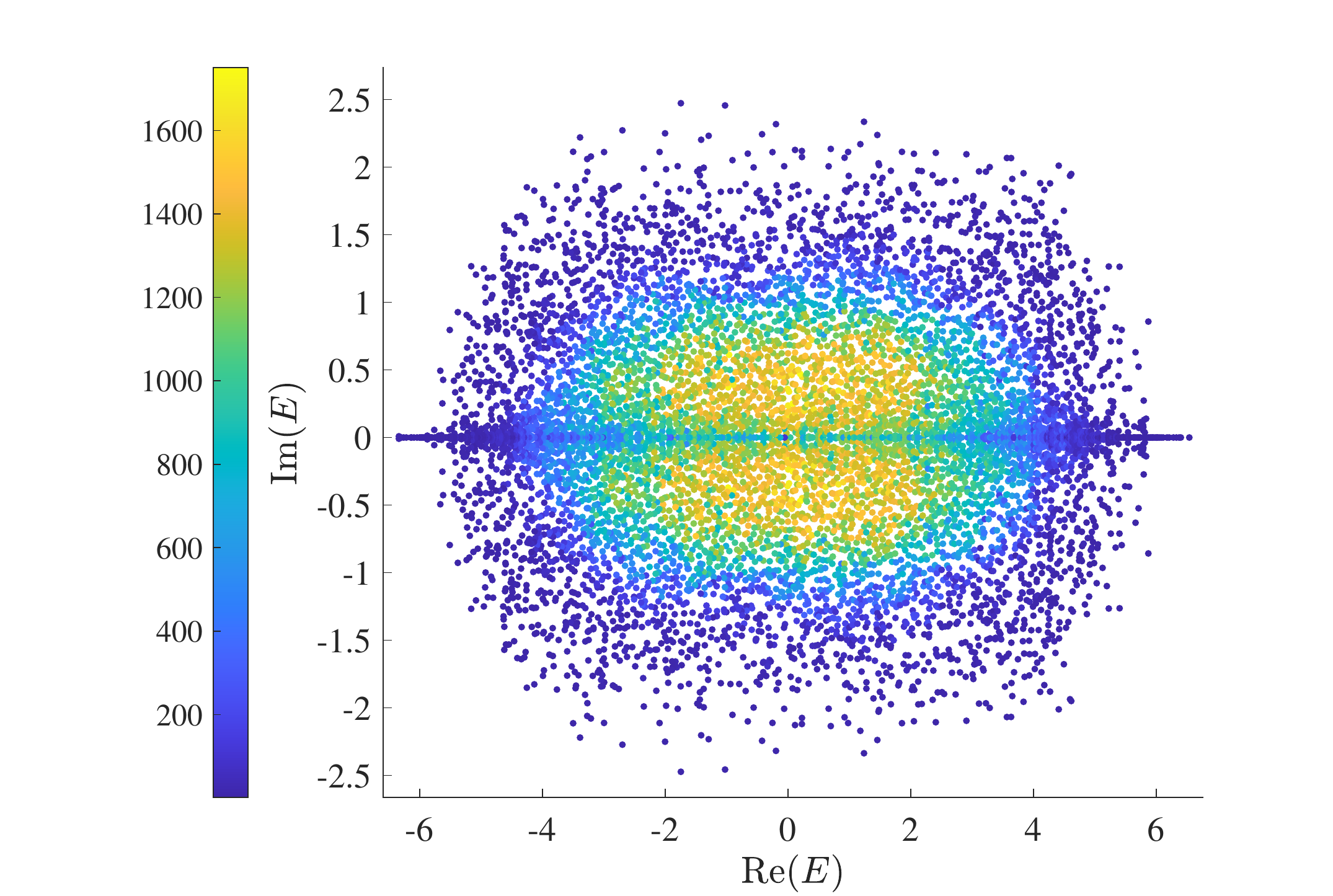}
		\end{minipage}%
	}%
	\subfigure[$W=10$ ]{
		\begin{minipage}[t]{0.5\linewidth}
			\centering
			\includegraphics[width=1\linewidth]{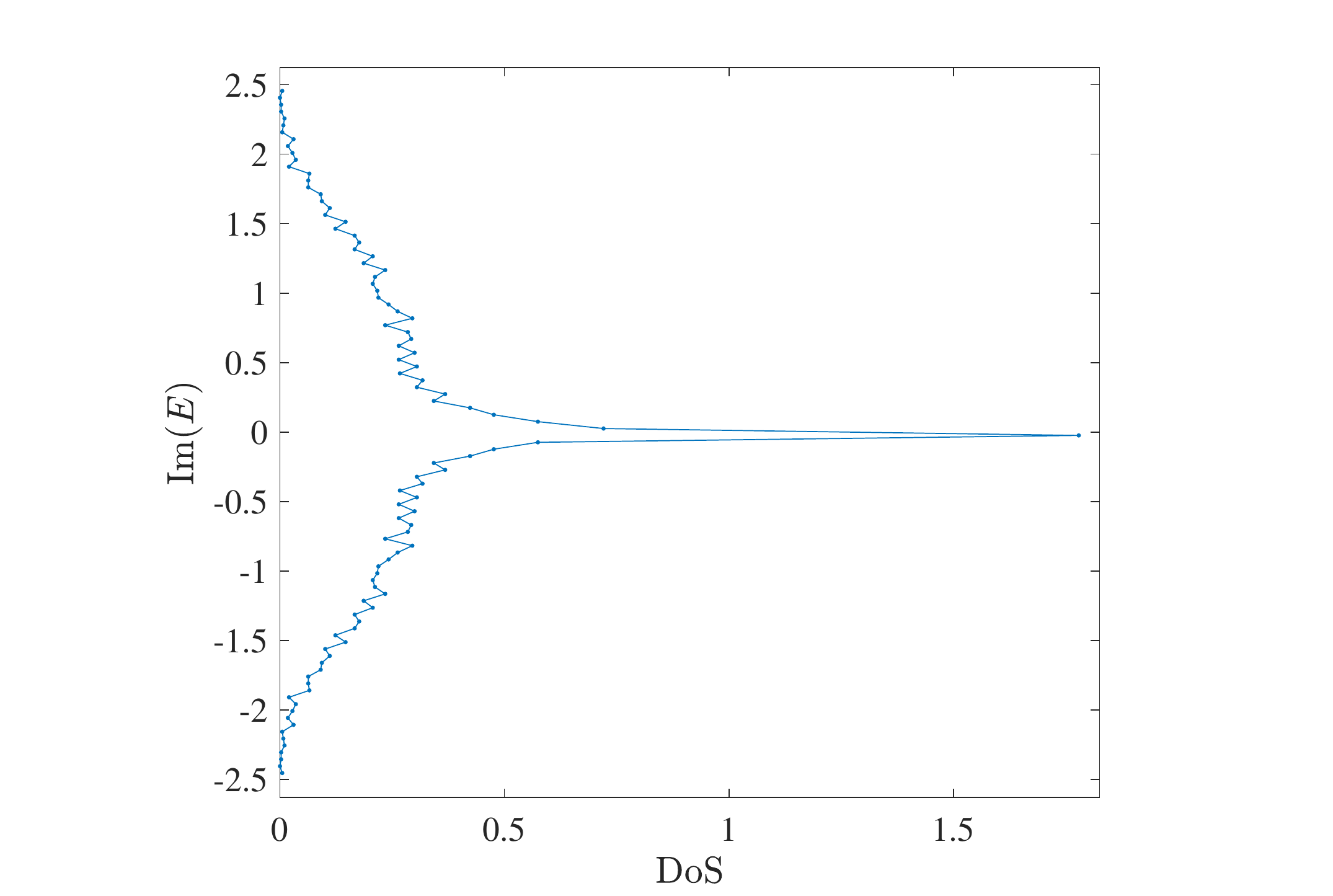}
		\end{minipage}%
	}%

	\subfigure[$W=20$]{
		\begin{minipage}[t]{0.5\linewidth}
			\centering
			\includegraphics[width=1\linewidth]{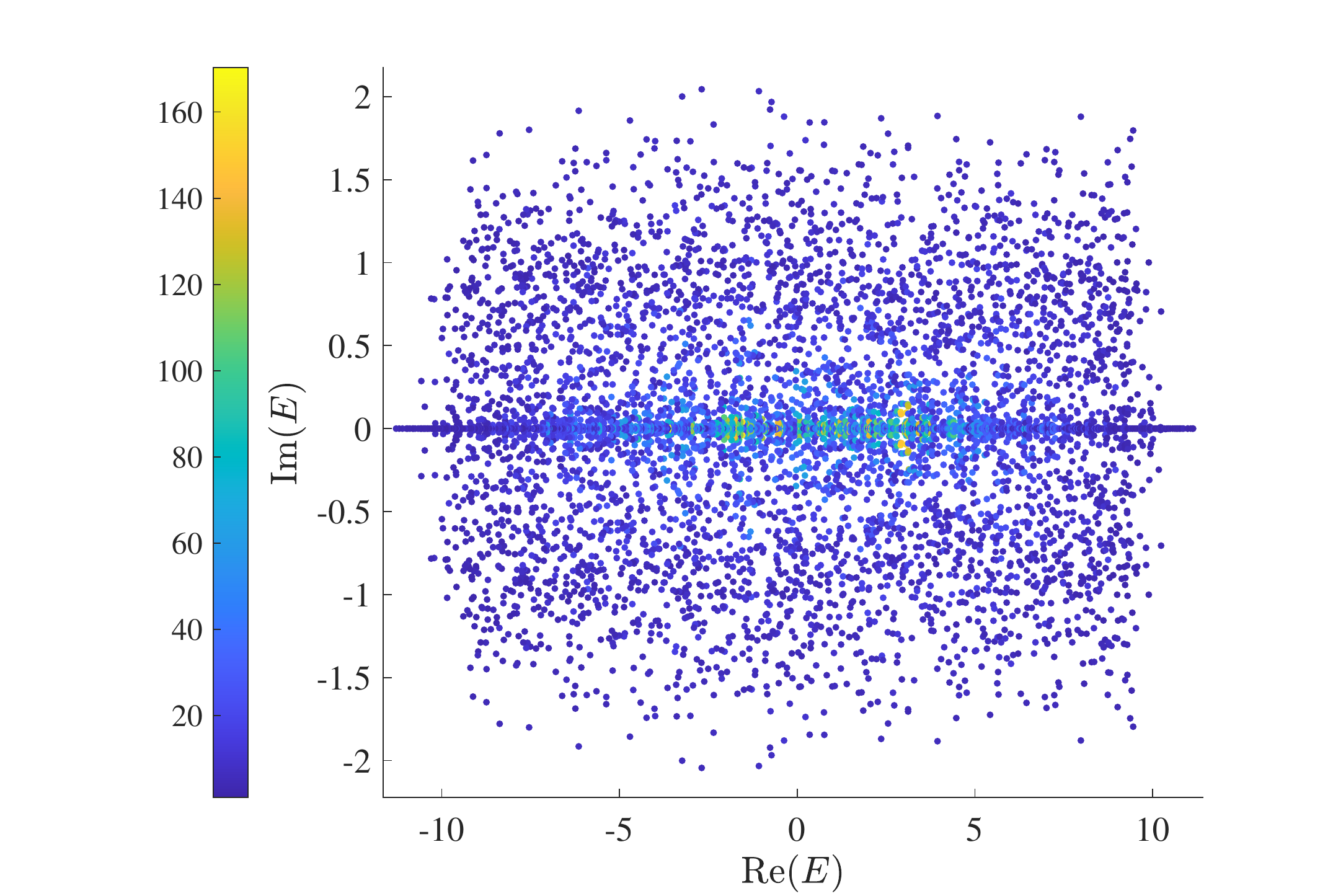}
		\end{minipage}%
	}%
	\subfigure[$W=20$ ]{
		\begin{minipage}[t]{0.5\linewidth}
			\centering
			\includegraphics[width=1\linewidth]{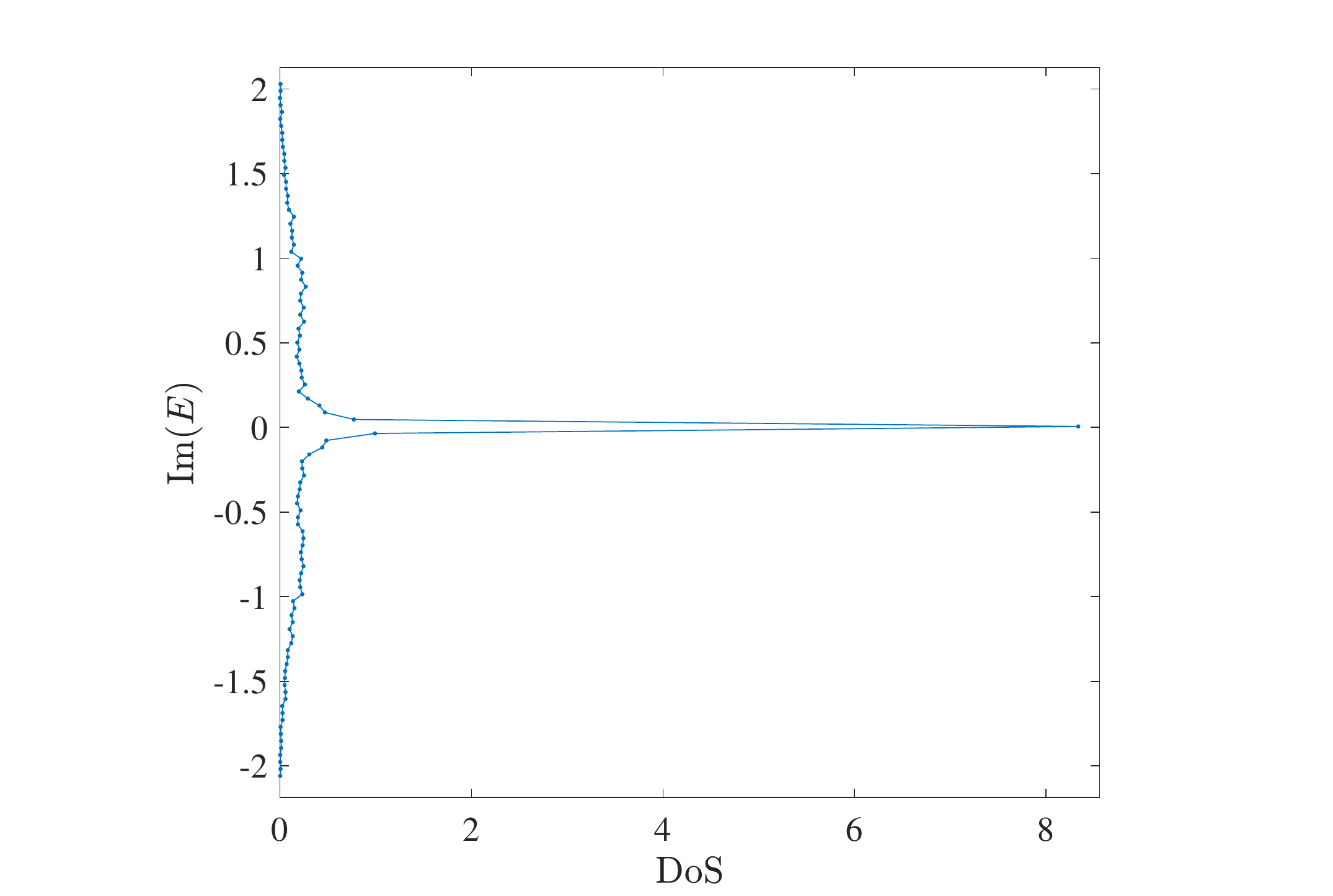}
		\end{minipage}%
	}%
	\caption{(a), (c), (e)~Eigenenergy distributions in the complex plane. 
		Each point corresponds to each eigenenergy in one sample. 
		The color of the points describes $1/I$ with the inverse participation ratios $I$ for the corresponding eigenmodes. 
		(b), (d), (f)~Density of states (DoS) for the imaginary part of eigenenergies. 
		The numerical calculations are performed 
		for the 3D O(1) model in class AI with the cubic system 
		size $L=20$, under the periodic boundary conditions, and 
		with the disorder strength $W = 0, 10, 20$.}
	\label{DoS_AI}
\end{figure}

\begin{figure}[tb]
	\centering
	\subfigure[2D]{
		\begin{minipage}[t]{0.5\linewidth}
			\centering
			\includegraphics[width=1\linewidth]{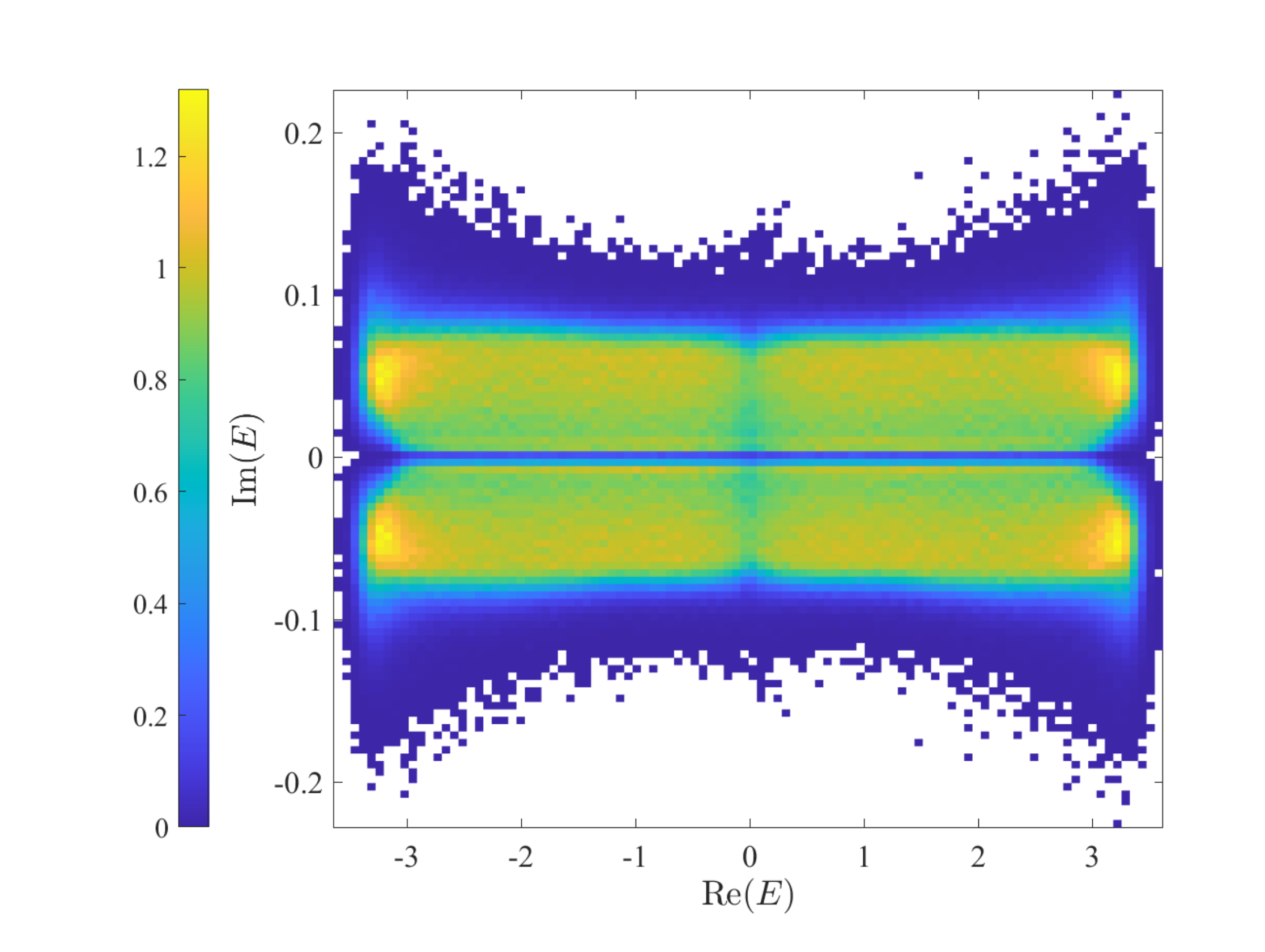}
		\end{minipage}%
	}%
	\subfigure[2D ]{
		\begin{minipage}[t]{0.5\linewidth}
			\centering
			\includegraphics[width=1\linewidth]{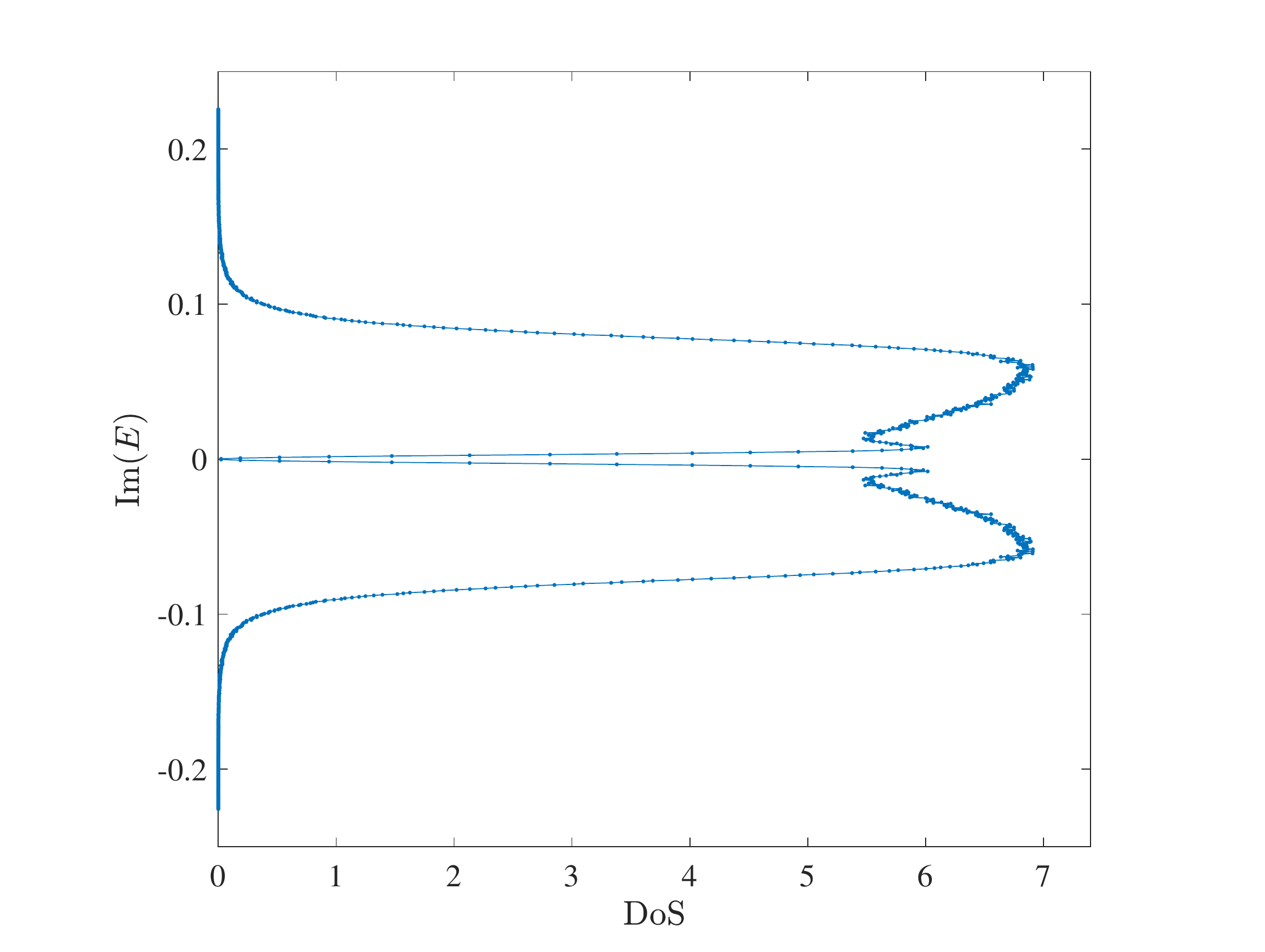}
		\end{minipage}%
	}%
	
	\subfigure[3D]{
		\begin{minipage}[t]{0.5\linewidth}
			\centering
			\includegraphics[width=1\linewidth]{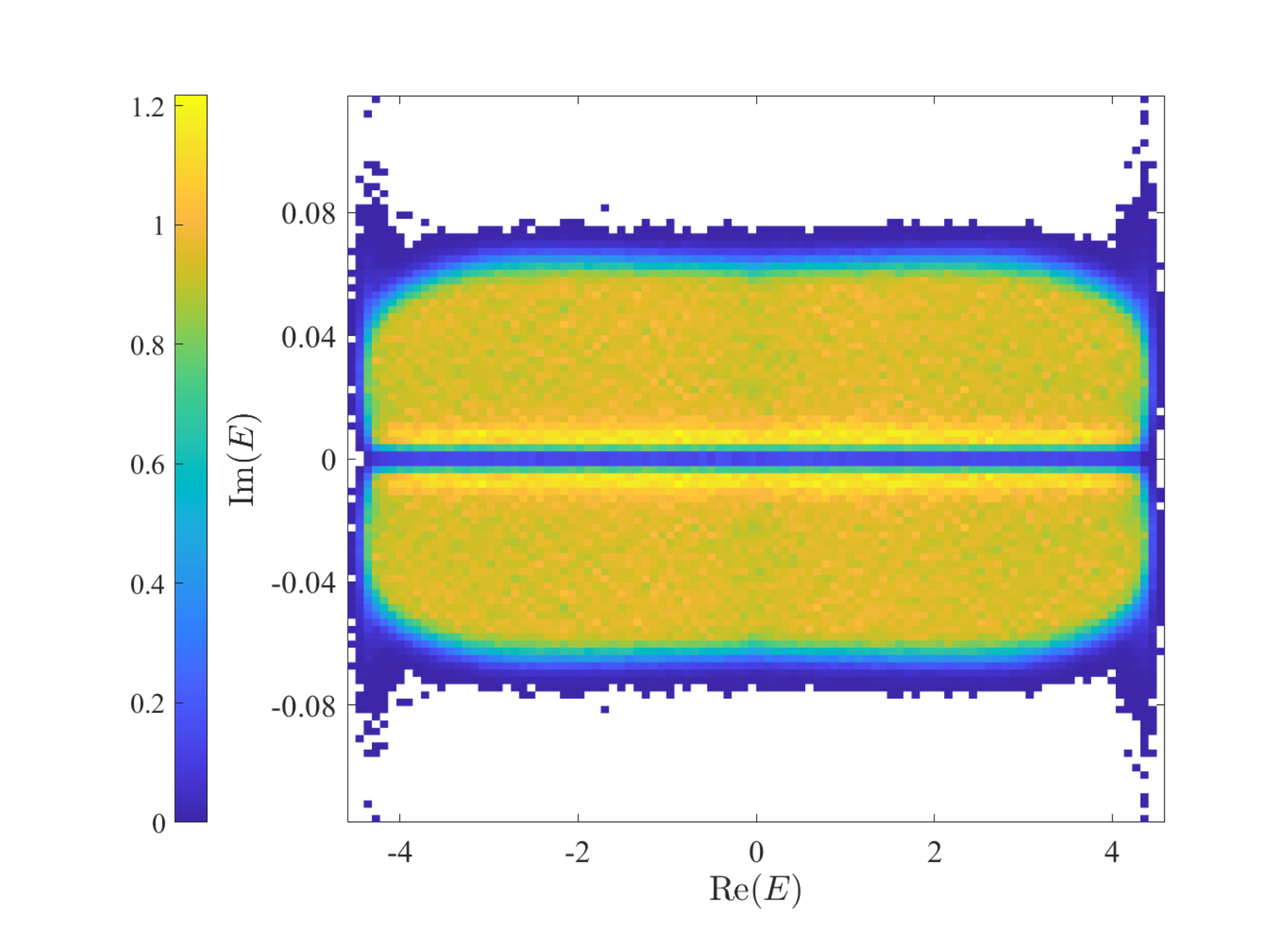}
		\end{minipage}%
	}%
	\subfigure[3D ]{
		\begin{minipage}[t]{0.5\linewidth}
			\centering
			\includegraphics[width=1\linewidth]{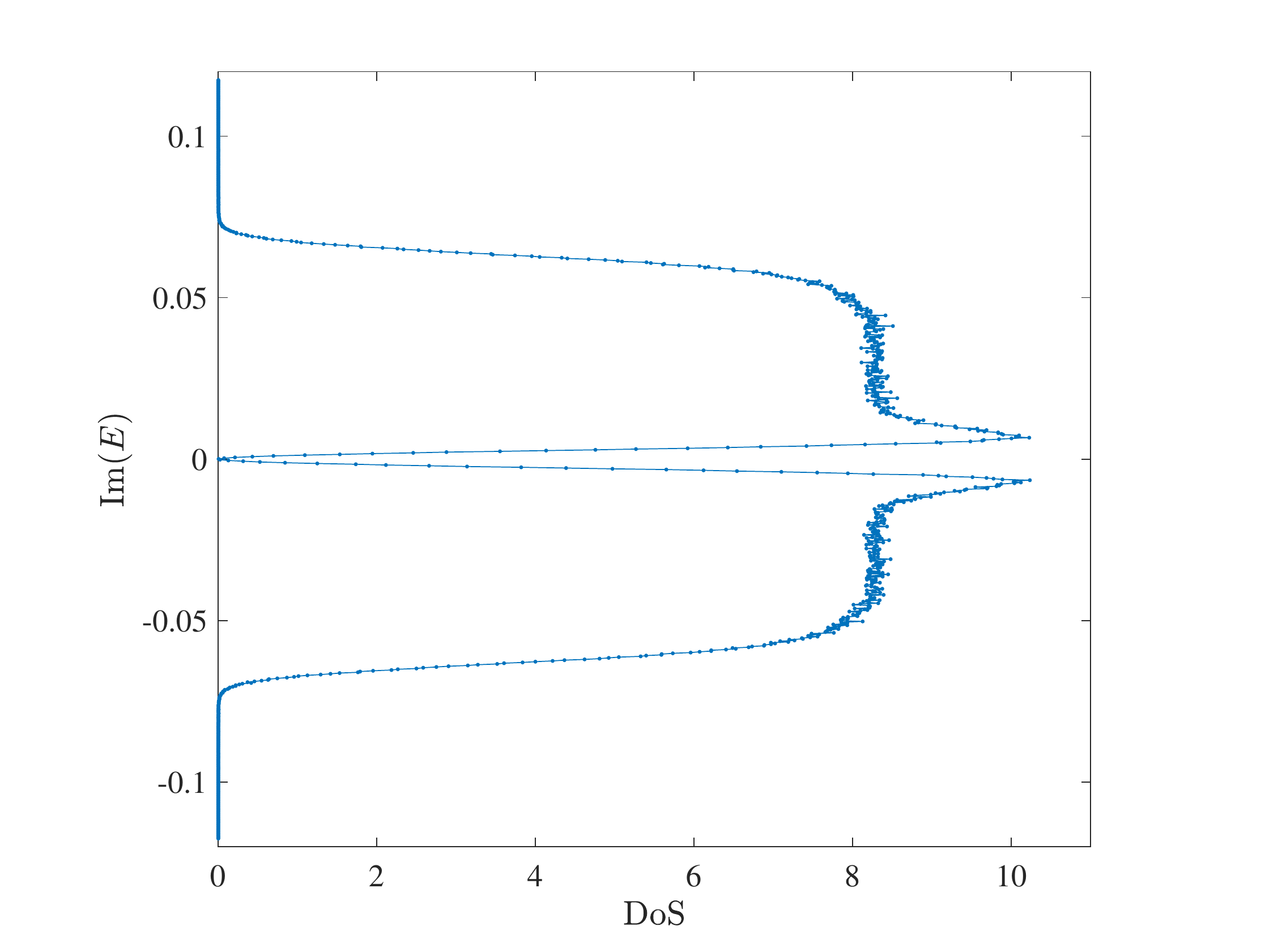}
		\end{minipage}%
	}%
	\caption{(a), (c)~Heat maps of eigenenergy density in the complex plane. 
		(b), (d)~Density of states (DoS)
		for the imaginary part of eigenenergies. 
		Eigenenergies are calculated 
		for the 2D and 3D SU(2) model in class AII
		under the periodic boundary conditions, 
		with $W_r=W_i=1$, 
		and 
		over the 640 samples with different disorder 
		realizations. The system sizes are $70 \times 70$ for 
		2D and $16\times 16 \times 16$ for 3D.}
	
	\label{AII_2D3D}
\end{figure}

\begin{figure}[bt]
	\centering
	\subfigure[2D]{
		\begin{minipage}[t]{0.5\linewidth}
			\centering
			\includegraphics[width=1\linewidth]{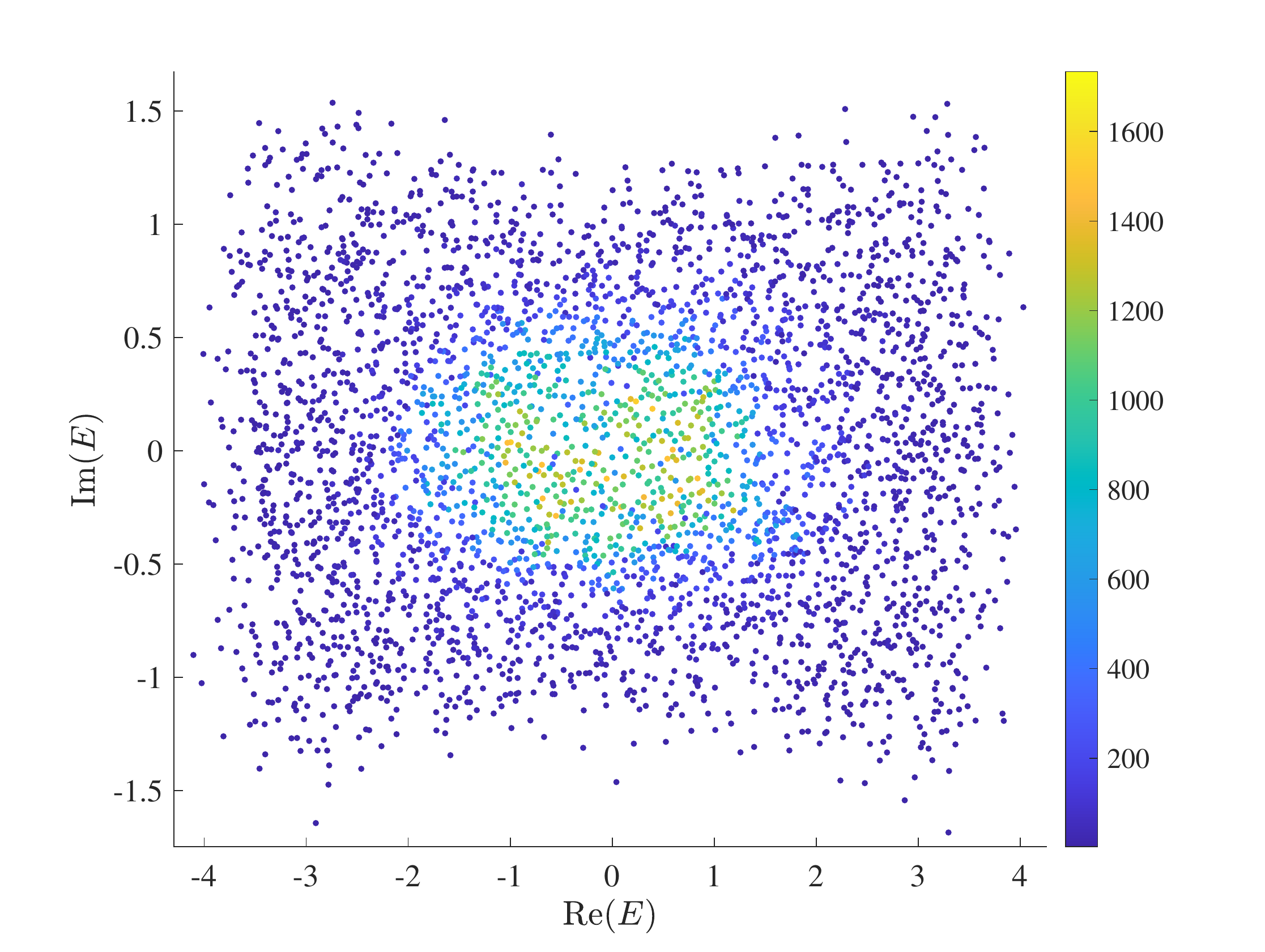}
		\end{minipage}%
	}%
	\subfigure[3D]{
		\begin{minipage}[t]{0.5\linewidth}
			\centering
			\includegraphics[width=1\linewidth]{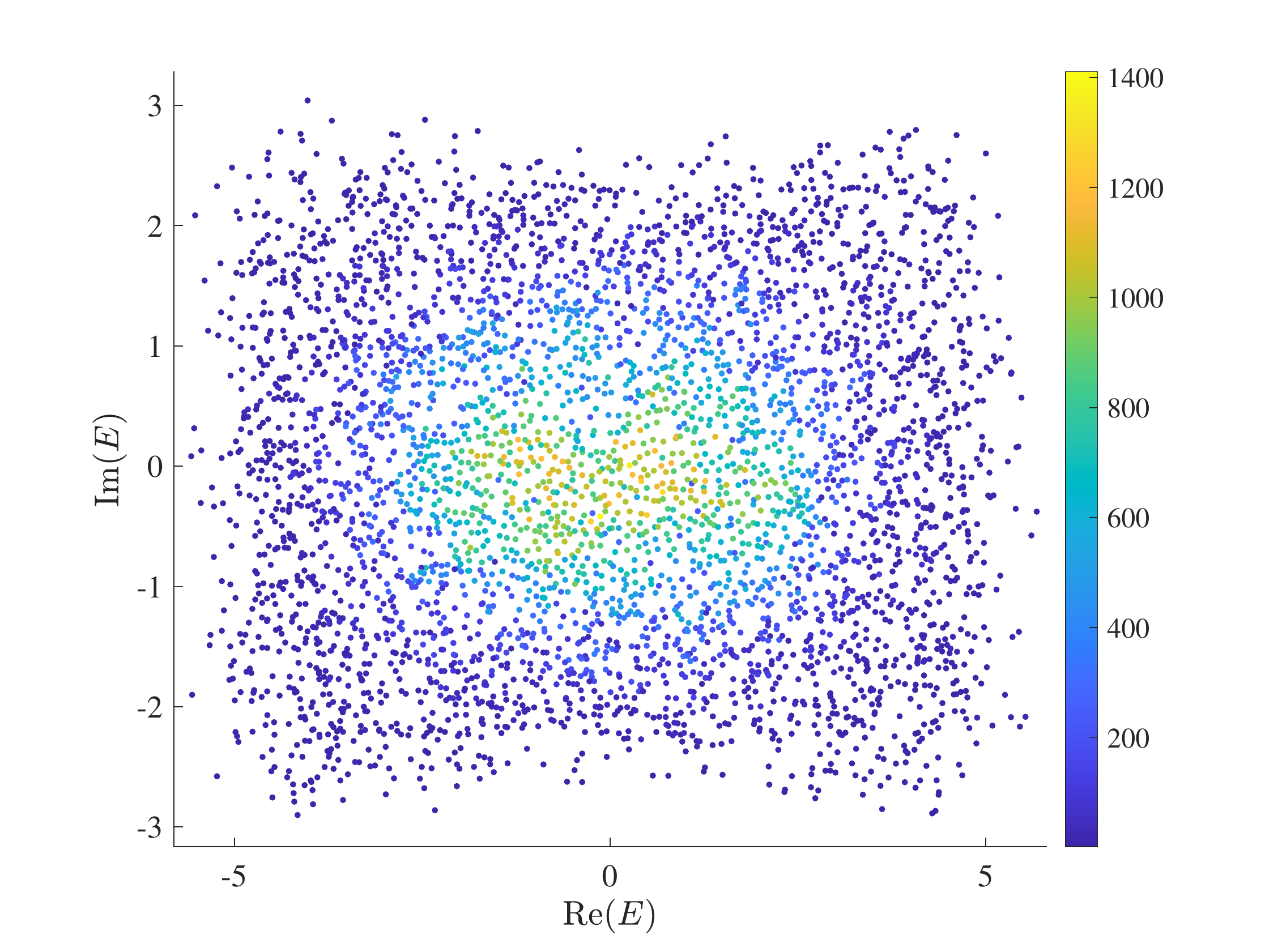}
		\end{minipage}%
	}%
	\caption{Eigenenergy distributions in the complex plane.
		Each point corresponds to each eigenenergy in one sample.
		The color of the points describes 
		$1/I$ with the inverse participation ratios $I$ for the corresponding eigenmodes.
		The eigenenergies 
		are calculated for the SU(2) models in class AII$^{\dagger}$ 
		with the system sizes $60 \times 60$ for 2D and $16 \times 16 \times 16$ for 3D.
		The periodic boundary conditions are imposed in all the directions for both 2D and 3D. 
		The disorder strength is set to 
		$W_r=W_i=4$ for 2D and $W_r=W_i=7$ for 3D.
	}
	\label{AII_dagger_2D3D}
\end{figure}

\begin{figure}
	\centering
	\subfigure[ Ginibre Orthogonal ensemble]{
		\begin{minipage}[t]{0.5\linewidth}
			\centering
			\includegraphics[width=1\linewidth]{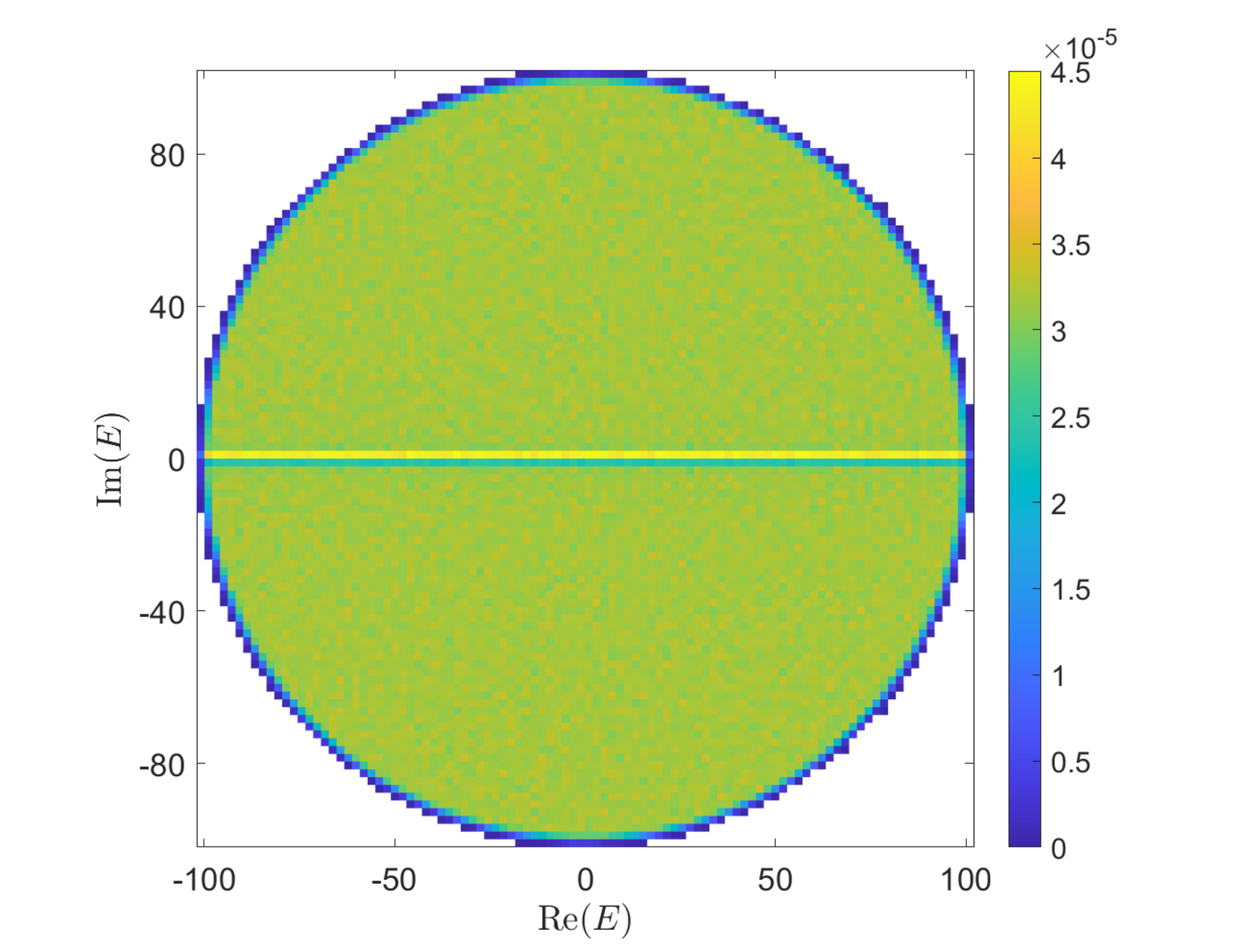}
		\end{minipage}%
	}%
	\subfigure[ Ginibre Orthogonal ensemble]{
		\begin{minipage}[t]{0.5\linewidth}
			\centering
			\includegraphics[width=1\linewidth]{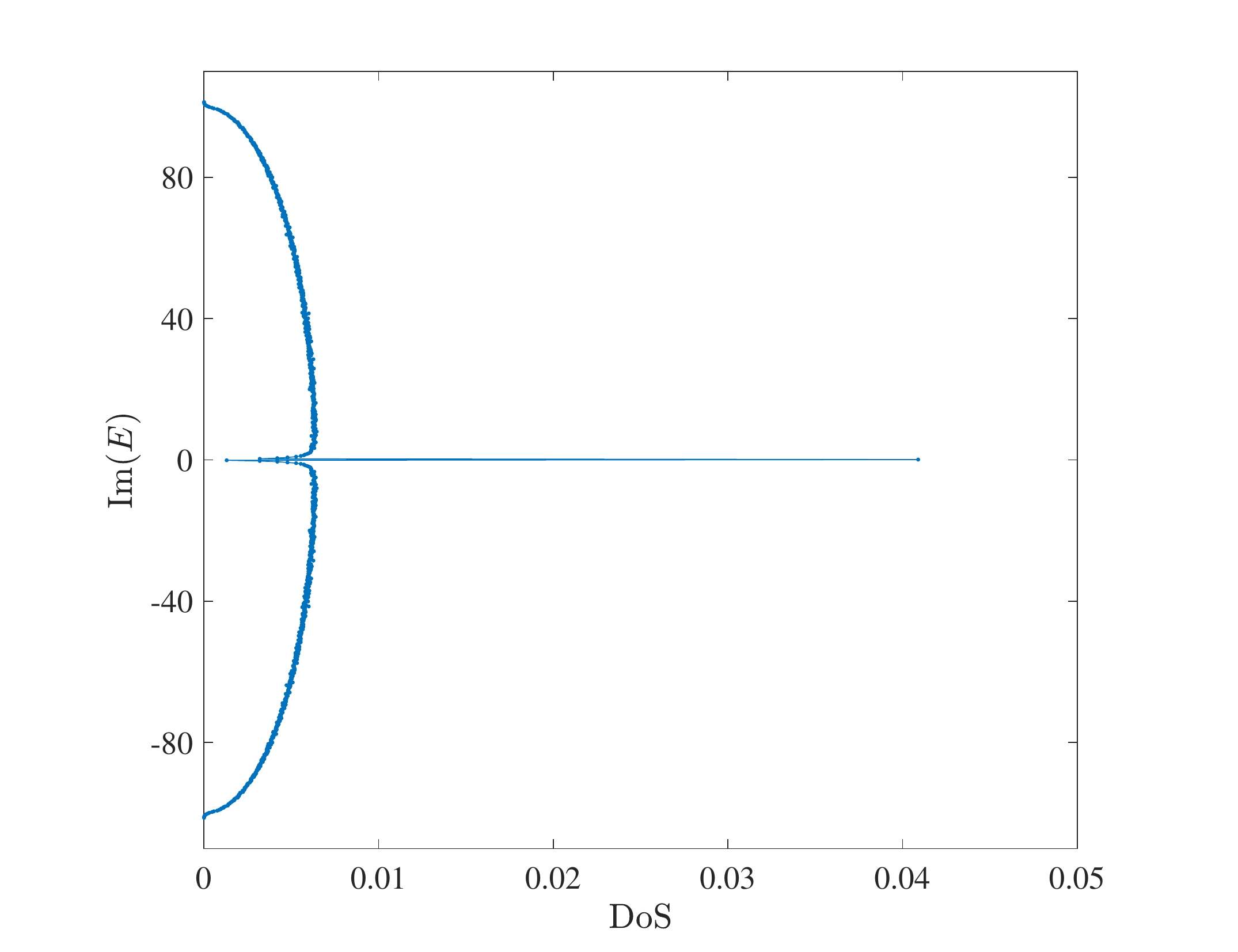}
		\end{minipage}%
	}%
	
	\subfigure[Ginibre Symplectic ensemble ]{
		\begin{minipage}[t]{0.5\linewidth}
			\centering
			\includegraphics[width=1\linewidth]{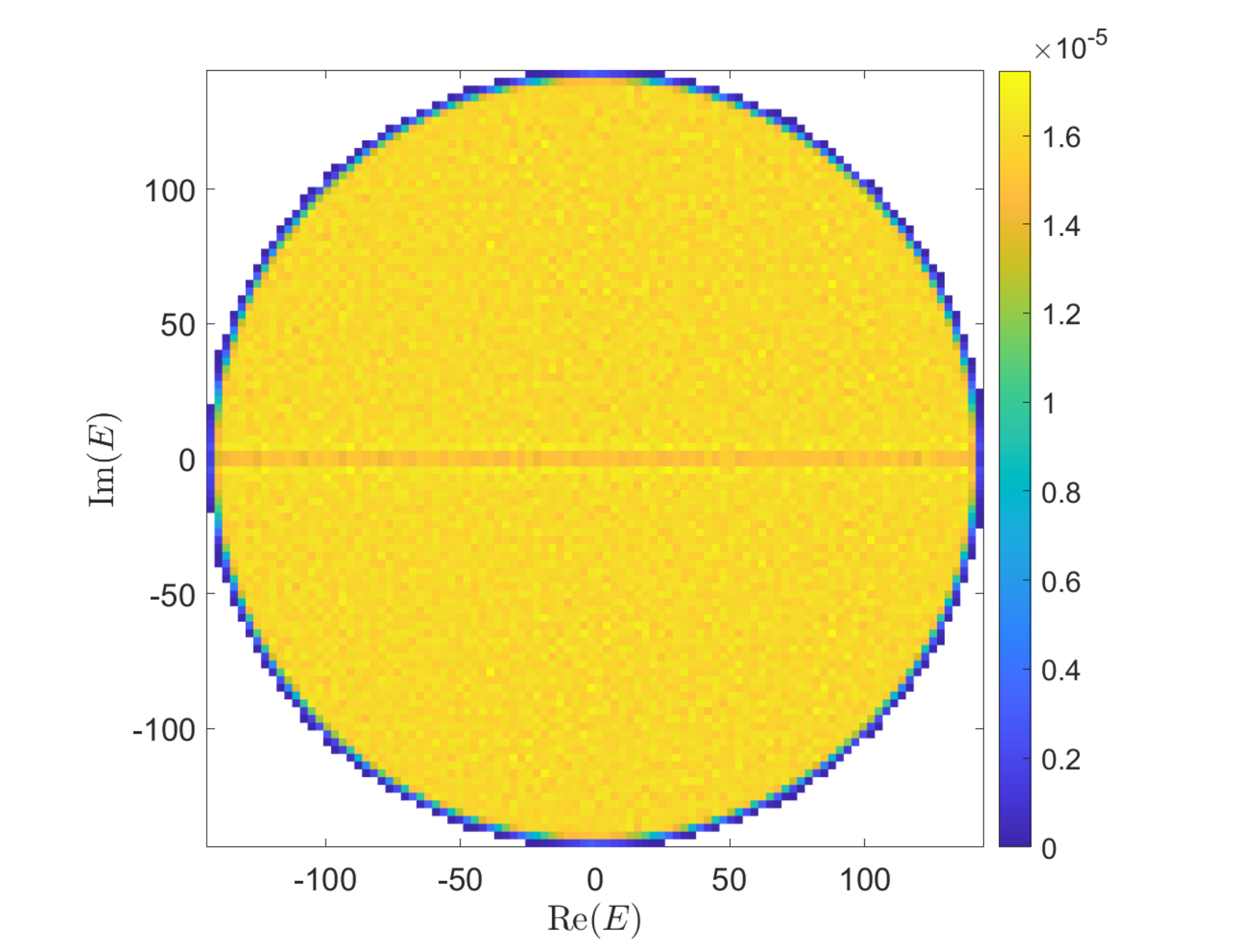}
		\end{minipage}%
	}%
	\subfigure[Ginibre Symplectic ensemble ]{
		\begin{minipage}[t]{0.5\linewidth}
			\centering
			\includegraphics[width=1\linewidth]{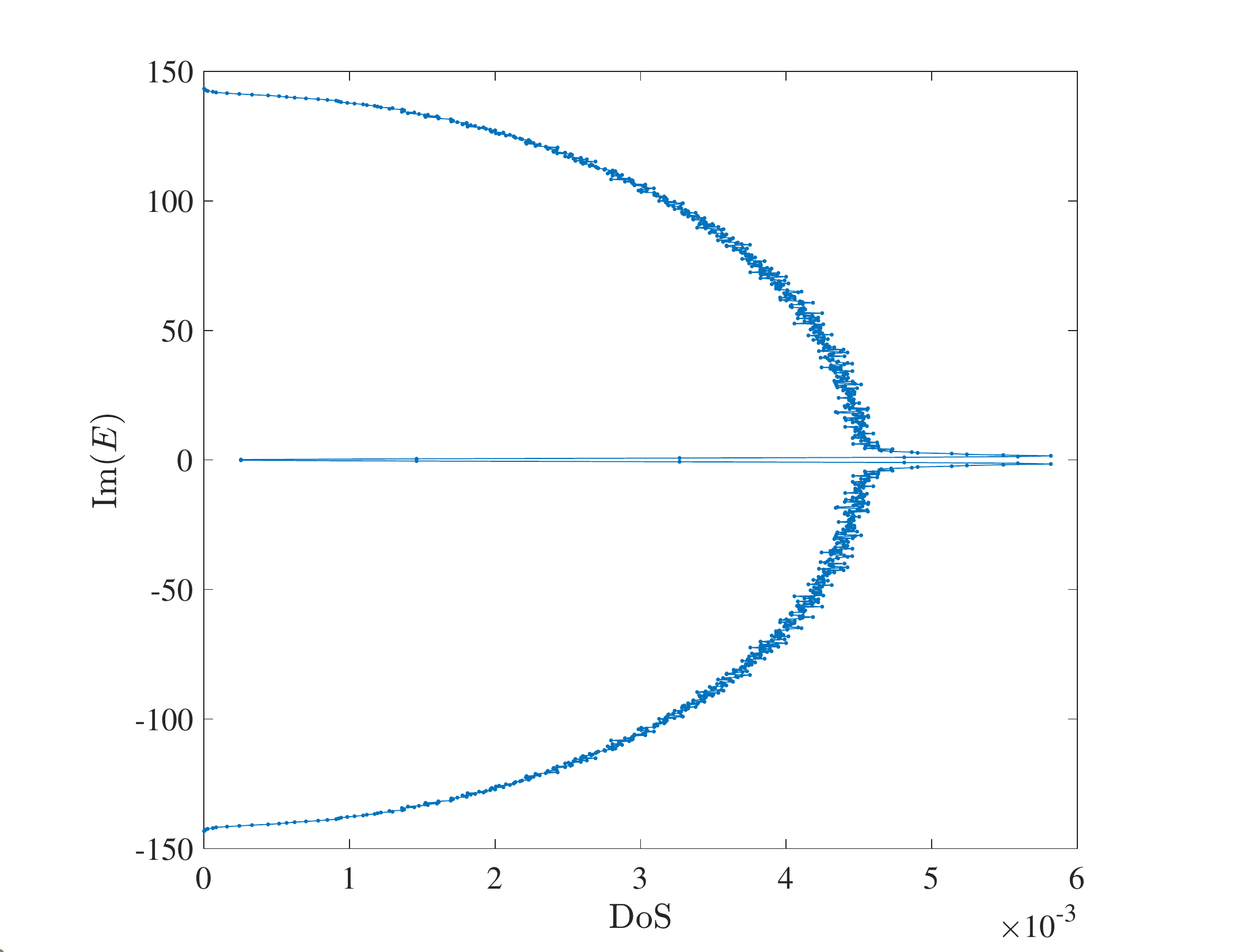}
		\end{minipage}%
	}%
	\caption{Heat maps of eigenenergy density in the complex plane for (a)~the Ginibre orthogonal ensemble and (c)~the Ginibre symplectic ensemble. The color describes the density of states (DoS) in the complex 
		plane.
		The DoS for the imaginary part of eigenenergies of (b)~the Ginibre orthogonal ensemble and (d)~the Ginibre symplectic ensemble.
		The numerical data come from the 640 realizations of random matrices with the size $10^4 \times 10^4$. 
	}
	\label{GinibreOSE}
\end{figure}

\end{widetext}
\end{document}